\documentclass[preprint,12pt]{elsarticle}

\newtheorem{les}{Lesson}
\usepackage{amssymb}
\usepackage{multirow}
\usepackage{hyperref} 
\usepackage{color}
\usepackage{amsmath}
\usepackage{makecell}
\usepackage{bm}
\usepackage{longtable}
\usepackage{caption}
\usepackage{subfigure}

\newcolumntype{P}[1]{>{\raggedright\arraybackslash\footnotesize}m{#1}}
\newcolumntype{A}[1]{>{\centering\arraybackslash\footnotesize}m{#1}}
\journal{Journal of Network and Computer Applications}

\begin{document}

\begin{frontmatter}

\title{A Survey of Secure Semantic Communications}

\author[label1]{Rui Meng}
\author[label1]{Song Gao}
\author[label1]{Dayu Fan}
\author[label1]{Haixiao Gao}
\author[label1]{Yining Wang}
\author[label1,label2]{Xiaodong Xu\corref{cor1}}
\author[label1]{Bizhu Wang\corref{cor1}}
\author[label3]{Suyu Lv}
\author[label1]{Zhidi Zhang}
\author[label1]{Mengying Sun}
\author[label1]{Shujun Han}
\author[label1]{Chen Dong}
\author[label4]{Xiaofeng Tao}
\author[label1,label2]{Ping Zhang}
\cortext[cor1]{Corresponding author. Email: xuxiaodong@bupt.edu.cn; wangbizhu\_7@bupt.edu.cn}

\affiliation[label1]{organization={State Key Laboratory of Networking and Switching Technology, Beijing University of Posts and Telecommunications}, 
city={Beijing}, 
postcode={100876}, 
country={China}} 

\affiliation[label2]{organization={Department of Broadband Communication, Peng Cheng Laboratory}, 
city={Shenzhen}, 
postcode={518066}, 
state={Guangdong}, 
country={China}} 

\affiliation[label3]{organization={School of Information Science and Technology, Beijing University of Technology}, 
city={Beijing}, 
postcode={100124}, 
country={China}} 

\affiliation[label4]{organization={National Engineering Laboratory for Mobile Network Technologies, Beijing University of Posts and Telecommunications}, 
city={Beijing}, 
postcode={100876}, 
country={China}} 

\begin{abstract}
Semantic communication (SemCom) is regarded as a promising and revolutionary technology in 6G, aiming to transcend the constraints of ``Shannon's trap" by filtering out redundant information and extracting the core of effective data. Compared to traditional communication paradigms, SemCom offers several notable advantages, such as reducing the burden on data transmission, enhancing network management efficiency, and optimizing resource allocation. Numerous researchers have extensively explored SemCom from various perspectives, including network architecture, theoretical analysis, potential technologies, and future applications. However, as SemCom continues to evolve, a multitude of security and privacy concerns have arisen, posing threats to the confidentiality, integrity, and availability of SemCom systems. This paper presents a comprehensive survey of the technologies that can be utilized to secure SemCom. Firstly, we elaborate on the entire life cycle of SemCom, which includes the model training, model transfer, and semantic information transmission phases. Then, we identify the security and privacy issues that emerge during these three stages. Furthermore, we summarize the techniques available to mitigate these security and privacy threats, including data cleaning, robust learning, defensive strategies against backdoor attacks, adversarial training, differential privacy, cryptography, blockchain technology, model compression, and physical-layer security. Lastly, this paper outlines future research directions to guide researchers in related fields.
\end{abstract}


\end{frontmatter}


\section{Introduction}
\subsection{Background}
As expected by academia, government, and industry, the sixth-generation communication (6G) is envisioned to open a new era of the ``Internet of Intelligence" with connected people, machines, things, and intelligence \cite{wang2023road}. Although the vision of what 6G should be is still an open issue, candidate key performance indicators, use cases and applications have already been summarized and forecasted by academia, industry, and standardization bodies. The International technology union (ITU) \cite{itur} provided the trends, usage scenarios, capabilities, and considerations of ongoing development of the future 6G networks. In addition to the extension of the three major scenarios of 5G, 6G will also support three new scenarios, including integrated sensing and communication, Artificial Intelligence (AI) and communication, and ubiquitous connectivity. Besides, six new capabilities are estimated targets for research of 6G, including coverage, sensing-related capabilities, applicable AI-related capabilities, sustainability, interoperability, and positioning. In conclusion, the emerging services in 6G not only rely on high-speed data transmission but also place higher demands on network intelligence and service diversity.
\subsection{Semantic Communication (SemCom)}
Researchers are currently reimagining how mobile communication systems transmit and utilize intelligence information to fully leverage the potential of AI in 6G. An increasing number of researchers are focusing on incorporating semantic information into wireless communications \cite{zhang2024intellicise} to address this challenge. Shannon and Weaver \cite{shannon2001mathematical} stated that communications could be categorized into syntactic, semantic, and pragmatic levels. The syntactic level focuses on accurately transmitting the symbols without considering the desired meaning at the semantic level or the intents at the pragmatic level. On the one hand, Semantic Communication (SemCom) enables the communication agents\footnote{In SemComs, ``communication agents" refer to the intelligent entities or communication entities that participate in the communication process. They leverage advanced AI technologies and specific functional modules to achieve accurate transmission and understanding of semantic information.} to extract essential semantic information from the original data, thereby reducing network transmission pressure and the processing delays for intelligent tasks \cite{xu2023task}. On the other hand, SemCom can be considered a form of AI-induced or brain-like communication mechanism, as the communication is based on a ``knowledge base" accumulated through continuous learning \cite{xu2023latent}.

According to recent efforts in SemCom, researchers have mainly focused on mathematical theory studies \cite{niu2024mathematical,shao2024theory,xin2024semantic}, infrastructure frameworks design \cite{zhang2023model,chaccour2024less,liang2024generative}, joint source-channel coding schemes \cite{xu2023deep,nemati2023vqvae,wu2024deep,huang2024joint,wang2024feature}, and advanced semantic extraction techniques for different communication modalities \cite{lo2022wireless,wang2022wireless,yao2022semantic,han2023semantic}. Niu et al. put forward the system framework of semantic information theory based on synonymous mapping for the first time, which revealed the key relationship between semantic information and grammatical information and expanded the scope of classical information theory in SemCom \cite{niu2024mathematical}. In addition, Zhang et al. proposed the model division multiple access technology, which can excavate the model information space based on SemCom to realize the purpose of distinguishing users, and can effectively improve the spectrum efficiency for semantic transmission \cite{zhang2023model}. Furthermore, Wu et al. introduce a vision transformer-based deep joint source and channel coding (JSCC) scheme for MIMO channels, named DeepJSCC-MIMO, which leverages self-attention to optimize feature mapping and power allocation, enhancing robustness against channel estimation errors and improving both distortion and perceptual quality across diverse scenarios \cite{wu2024deep}. Moreover, Wang et al. introduce the deep video semantic transmission, a novel method for end-to-end video transmission over wireless channels that leverages deep JSCC to adaptively transmit video semantic features and utilizes a nonlinear transform and conditional coding to optimize bandwidth use across video frames \cite{wang2022wireless}.


\subsection{Secure SemCom}
SemCom adaptively reshapes the core mode of information dissemination by integrating AI and communication technologies, and finally endows itself with endogenous intelligence and native conciseness. However, as SemCom evolves, a multitude of security and privacy concerns are surfacing. Notably, the knowledge base or semantic-related models cached for semantic extraction and recovery must continually adapt to various dynamics, including varying channel conditions, communication objectives, the content of the original information, the nature of communication recipients, network resource availability, and device capability limitations \cite{shen2023secure,yang2024secure,guo2024survey,du2023rethinking}. Additionally, attackers can launch model inversion attacks to deduce implicit information, such as the characteristics of data held by specific nodes, through interactions with targeted semantic-related models \cite{chen2023model}. Furthermore, by modeling the semantic transmission process, attackers can decipher the relationship between the original information, the transmitted core semantic content, and the reconstructed information, potentially leading to more severe privacy breaches compared to traditional bit-based communications \cite{wang2024privacy}.

In recent years, significant advancements have been made in secure SemCom, driven by various approaches aiming to enhance both privacy and robustness. Tung et al. \cite{10278612} proposed a framework that integrates encryption directly into the source-channel coding process to protect the privacy of semantic data. This method incorporates adversarial training to prevent attackers from reconstructing the original data under adverse network conditions, thereby enhancing the robustness and quality of semantic transmission. Building on this, Luo et al. \cite{10107616} combined adversarial training with encryption, enabling the encoder and decoder to accurately recover semantic information even in the presence of attackers. Additionally, Chen et al. \cite{10436996} introduced model inversion eavesdropping attacks, where attackers attempt to reconstruct original data by intercepting transmitted symbols. To defend against this, they suggested effective countermeasures such as random permutation and substitution to disrupt data continuity and protect semantic integrity. Furthermore, Chen et al. \cite{10547440} developed SemGuarder, a lightweight and robust wireless SemCom system that combines AES and RSA encryption for secure data transmission.

From a physical-layer security (PLS) perspective, Li et al. \cite{10662946} developed a deep neural network (DNN)-based secure SemCom system called DeepSSC, which uses a two-phase training strategy. Similarly, Mu et al. \cite{mu2024semanticcommunicationassistedphysicallayer} proposed another PLS approach that leverages semantic streams as beneficial noise to interfere with eavesdroppers, optimizing power allocation and decoding order to enhance secrecy. In the realm of learning-based techniques, Liu et al. \cite{liu2024learningbasedpowercontrolsecure} introduced a method using the soft actor-critic algorithm to optimize power control in covert SemCom. Qin et al. \cite{10279807} suggested generating a semantic key and employing subcarrier-level obfuscation to strengthen security in static channels. For image transmission, Tang et al. \cite{tang2024secure} employed signal steganography by embedding private image signals into non-sensitive host images to ensure that only legitimate receivers can extract private data. Lastly, Xu et al. \cite{10570800} proposed a covert SemCom framework for wireless edge networks that uses a full-duplex receiver to generate artificial noise, thereby interfering with eavesdroppers.



Representative overview/survey papers on secure SemCom are listed in Tab. \ref{tab1}. Won et al. \cite{won2024resource} comprehensively review resource management, security, and privacy in SemCom, introducing semantic-aware resource allocation and identifying open research challenges. This survey also emphasizes the importance of semantic-aware metrics and performance optimization techniques to enhance system resilience against eavesdropping, jamming, and semantic noise\footnote{Semantic noise can originate from the following aspects. (1) Semantic ambiguity is a common source of semantic noise in text. Minor changes in words or adjustments to sentence structures can potentially lead to misunderstandings of the transmitter's intent by the receiver. Additionally, different cultural backgrounds and language habits may result in diverse interpretations of the same information, thereby generating semantic noise. (2) Transmission errors, due to factors such as channel noise, can alter the semantic content of the information. (3) Data compression and encoding may introduce a certain level of semantic noise as they may lead to the loss or distortion of some details in the information. (4) Adversarial attacks refer to situations where malicious attackers may intentionally introduce semantic noise to disrupt or sabotage the communication process.}. Guo et al. \cite{guo2024survey} survey SemCom networks, addressing multi-layered architecture, security, and privacy issues while proposing future directions. E. Sagduyu et al. \cite{10328181} analyze deep learning (DL)-based vulnerabilities, particularly adversarial attacks, and propose novel models to enhance system resilience. Du et al. \cite{10183798} revisit wireless security in the Semantic IoT, proposing new metrics like semantic secrecy outage probability and highlighting key differences from traditional IoT. Shen et al. \cite{10292907} discuss secure SemCom challenges and propose defenses such as zero-knowledge proofs. Yang et al. \cite{10551876} examine SemCom security fundamentals, emphasizing protection mechanisms for semantic transmission and machine learning (ML) models. Luo et al. \cite{9679803} review SemCom theories and advancements, focusing on challenges in semantic extraction, compression, and transmission. He et al. \cite{he2024mixtureofexpertsenabledtrustworthysemantic} introduce a mixture-of-experts model for 6G SemCom, showcasing its success in mitigating heterogeneous attacks and preserving task performance.

\begin{longtable}{|A{0.6cm}|A{2.2cm}|P{9.5cm}|}  
\caption{\footnotesize{Representative Overview/Survey Papers on Securing SemComs}} 
\label{tab1}\\
\hline
\textbf{Ref.} & \textbf{Publication Year/Type} & \textbf{Major Contributions} \\
\hline
\endfirsthead

\hline
\textbf{Ref.} & \textbf{Publication Year/Type} & \textbf{Major Contributions} \\
\hline
\endhead

\hline
\endfoot

\hline
\endlastfoot
\cite{won2024resource} & 2024/Survey & Offers a comprehensive study focusing on resource management, security, and privacy in SemCom, introducing semantic-aware resource allocation methods and discussing open research challenges to advance the field.   \\
\hline
\cite{guo2024survey} & 2024/Survey & Provides an overview of SemCom networks, focusing on its architecture, security, and privacy issues. Proposes a multi-layered architecture from technical to semantic levels and outlines future research directions in enhancing system robustness and privacy protection.  \\
\hline
\cite{10328181} & 2023/Survey & Focuses on multi-domain vulnerabilities in DL-based SemCom, particularly adversarial attacks. Proposes adversarial attack models to test system resilience by introducing perturbations at input and channel levels, offering insights on defense strategies.\\
\hline
\cite{10183798} & 2023/Overview & Reviews traditional wireless security techniques like physical layer security and covert communication, and introduces new security metrics for Semantic IoT. Also examines semantic attacks and defense mechanisms, proposing targeted and training-free defenses. \\
\hline
\cite{10292907} & 2024/Overview & Comprehensive overview of security and privacy threats in SemCom, identifying five key requirements for secure communication (data confidentiality, integrity, availability, authenticity, and privacy protection) and suggesting countermeasures.  \\
\hline
\cite{10551876} & 2024/Overview & Discusses fundamental challenges in secure SemCom from the perspectives of information security and ML model security, presenting different methods for semantic information extraction (autoencoder, knowledge graph, etc.).  \\
\hline
\cite{9679803} & 2022/Overview & Provides a systematic overview of the basic principles and recent advancements in SemCom, highlighting key issues like semantic noise, knowledge base sharing, and semantic matching challenges. \\
\hline
\cite{he2024mixtureofexpertsenabledtrustworthysemantic} & 2024/Overview & Proposes a Mixture-of-Experts (MoE) model-based SemCom system framework to address diverse security threats in 6G, showcasing its effectiveness through case studies and simulations in vehicular networks.\\
\hline
\ / & Our Survey & Provide an overview of SemCom architecture from both link and network levels, outline the entire life cycle of SemComs, summarize various defense technologies, and point out the future research directions.\\
\hline
\end{longtable}

\subsection{Motivations}
Despite the fact that many researchers concentrate on ensuring the security of SemCom, it is surprising to find that a comprehensive understanding of the state-of-the-art in secure SemCom and its fundamental principles remains elusive. For example, \cite{guo2024survey} and \cite{won2024resource} do not provide the basic principle or detailed classification of defense technologies. \cite{10328181} only discusses adversarial attacks but does not provide a comprehensive overview of other attack threats and their corresponding defense methods. Therefore, the main objective of this paper is to offer an in-depth exploration of the characteristics and technologies that can be employed in the realm of secure SemCom. Initially, we introduce the architectures of SemCom at both the link and network levels for readers unfamiliar with the concept. Subsequently, we delineate the life cycle of SemCom and discuss the security and privacy concerns associated with each stage. Furthermore, we present various defensive techniques, some of which have already been implemented in SemCom, while others hold promise for its future, to facilitate a better understanding for readers without a background in security and privacy. Finally, we outline future research directions, highlighting research challenges and potential solutions. We hope that this work will provide researchers with a foundational understanding and inspiration to further advance the development of SemCom.


\subsection{Contributions}
The contributions of this paper are summarized as follows:
\begin{itemize}
\item Initially, we provide an overview of SemCom architecture from both link and network levels. The link-level architecture focuses on the joint optimization of each module, while the network-level architecture highlights the transmission process of deep JSCC models.
\item Following this, we outline the entire life cycle of SemComs, encompassing model training, model transfer, and semantic information transmission stages. We follow the measures of security and privacy in AI field, including confidentiality, integrity, and availability, to describe SemCom systems. The security and privacy challenges in these stages include: 1) model training, which faces knowledge base poisoning attacks, gradient leakage, server vulnerabilities, and attacks targeting communication bottlenecks; 2) model transfer, which is susceptible to model slice availability attacks and model slice forgery attacks; and 3) semantic information transmission, which is vulnerable to semantic adversarial attacks, semantic eavesdropping attacks, semantic inference attacks, and semantic jamming attacks.
\item To counteract these security and privacy threats in SemCom, we summarize various defense technologies. These include: 1) data cleaning to address dirty data that may introduce noise, missing values, or inconsistencies; 2) robust learning to enhance the resistance of DL models to noise or input disturbances; 3) defensive methods against backdoor attacks, provided from both data and model perspectives; 4) adversarial training, approached from sample generation and model optimization angles; 5) differential privacy, which introduces noise into the data to prevent attackers from accurately restoring individual information; 6) cryptography technology, further classified into homomorphic encryption, secure multi-party computation, trusted execution environments, secure aggregation, and quantum cryptography; 7) blockchain technology, which combines data blocks into a tamper-proof and unforgeable chain structure; 8) model compression, further categorized into model pruning, parameter quantization, low-rank decomposition, and knowledge distillation; and physical-layer security, which includes beamforming, artificial noise, relay cooperation, intelligent reflecting surfaces, physical-layer key generation, and physical-layer authentication.
\item Furthermore, we point out the future research direction, including dynamic and intelligent data cleaning, explainable robust learning, multi-strategy combined backdoor defense, differential privacy-based deep JSCC, efficient homomorphic encrypted SemCom, smart contract-enabled SemCom, and semantic channel fingerprint database-enabled PLS.
\end{itemize}

\begin{figure*}[t]
\centering
\vspace{-10mm}
\includegraphics[width=\linewidth]{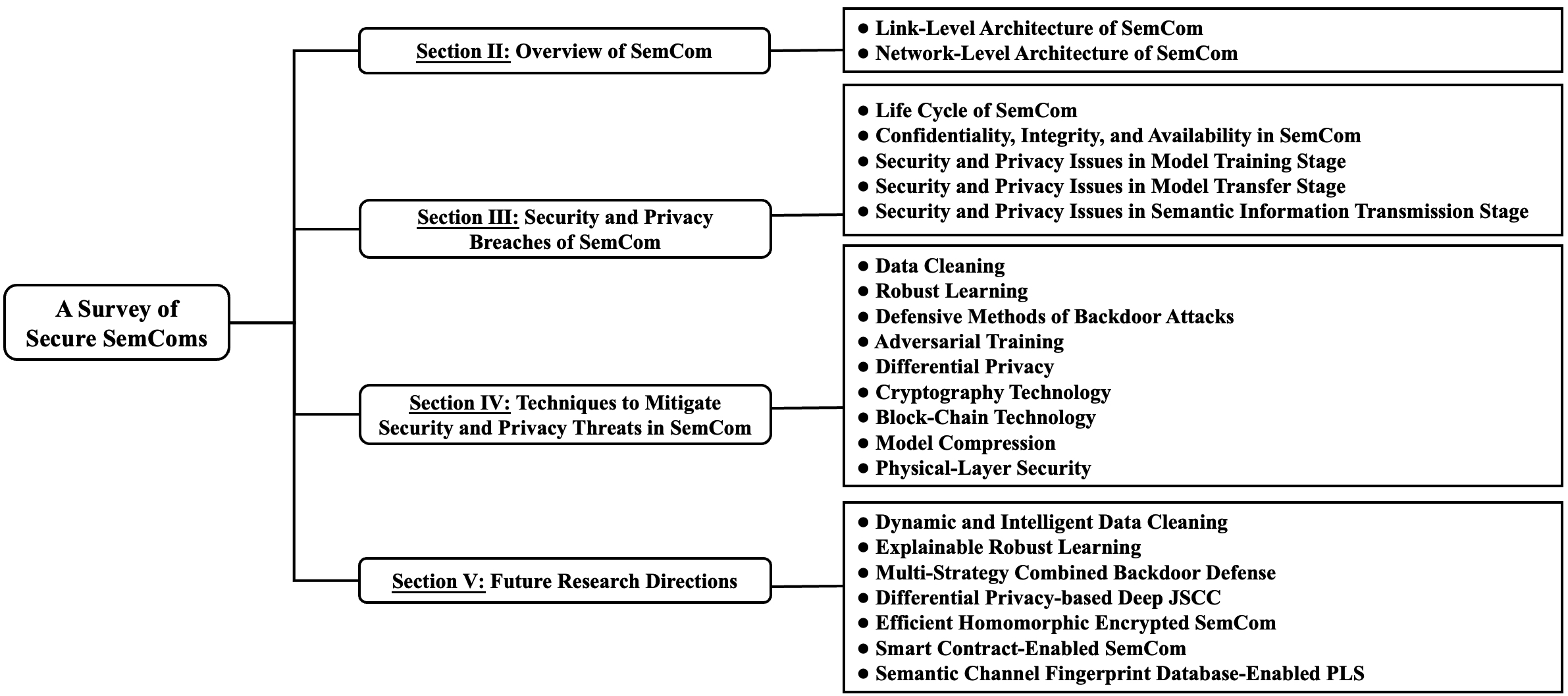}
\caption{Organizations of this paper.}
\label{fig11}
\end{figure*}

\subsection{Organizations}
As illustrated in Fig. \ref{fig11}, the rest of this paper is organized as follows. Sec. \ref{section 2} overviews the link- and network-level architectures of SemCom. Sec. \ref{section 3} provides the security and privacy violation issues during each stage of SemCom. Sec. \ref{section 4} provides insights into existing defense mechanisms to enhance the security and privacy of SemCom. Sec. \ref{section 5} looks forward to the future research direction. Finally, Sec. \ref{section 6} concludes this paper. The acronyms used in this paper are listed in Tab. \ref{Acronyms}.

\begin{longtable}{|A{2.4cm}|A{3.5cm}|A{2.4cm}|A{3.5cm}|}
\caption{\footnotesize{List of Acronyms Used in the Paper}} 
\label{Acronyms}\\
\hline
\textbf{Abbreviations} & \textbf{Full Name} & \textbf{Abbreviations} &\textbf{Full Name}
\\
\hline
\endfirsthead
\hline
\textbf{Abbreviations} & \textbf{Full Name} & \textbf{Abbreviations} &\textbf{Full Name}\\
\hline
\endhead

\hline
\endfoot

\hline
\endlastfoot
AI            & Artificial Intelligence        & MEC           & Mobile Edge Computing         \\ \hline
CNN          & Convolutional Neural Network              & MIMO           & Multi-Input Multi-Output   \\ \hline
DL      &  Deep Learning            &    Non-IID     &  Non-Independent and Identically Distributed  \\ \hline
DNN      &  DNN            &    PHE     &  Partially Homomorphic Encryption \\ \hline
 DP     &  Differential Privacy            &   PLA      &  Physical-Layer Authentication  \\ \hline
 FFT     &  Fast Fourier Transform            &  PLKG       & Physical-Layer Key Generation  \\ \hline
 FL     &  Federated Learning            &   PLS      &  Physical-Layer Security \\ \hline
 FHE     &  Fully Homomorphic Encryption            &     QoS    & Quality of Service   \\ \hline
  GAN    &  Generative Adversarial Network            &   SMPC      &  Secure Multi-Party Computation \\ \hline
   HE   &   Homomorphic Encryption           &    SemCom     & Semantic Communication  \\ \hline
  IRS    &  Intelligent Reflecting Surface            &     SNR    & Signal Noise Ratio  \\ \hline
  JSCC    &  Joint Source and Channel Coding            &     SGE    & Software Guard Extensions  \\ \hline
 ML     &  Machine Learning            &    TEE     &  Trusted Execution Environment \\ \hline
\end{longtable}

\section{Overview of SemCom}
\label{section 2}
This section first compares the traditional communication and SemCom systems from the perspective of link-level architecture, and then introduces the model generation and transmission in SemCom from the perspective of network level.
\subsection{Link-Level Architecture of SemCom}

\begin{figure*}[ht]
    \centering
		\subfigure[Traditional communication: each module operates independently.]{
			\includegraphics[width=0.9\textwidth]{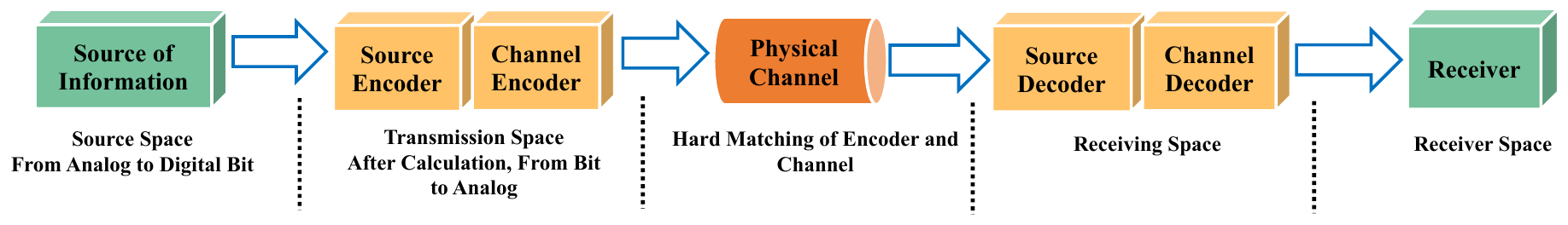}
			\label{figure1a}
		}
		\subfigure[SemCom: optimization of modules in the same semantic space.]{
			\includegraphics[width=0.9\textwidth]{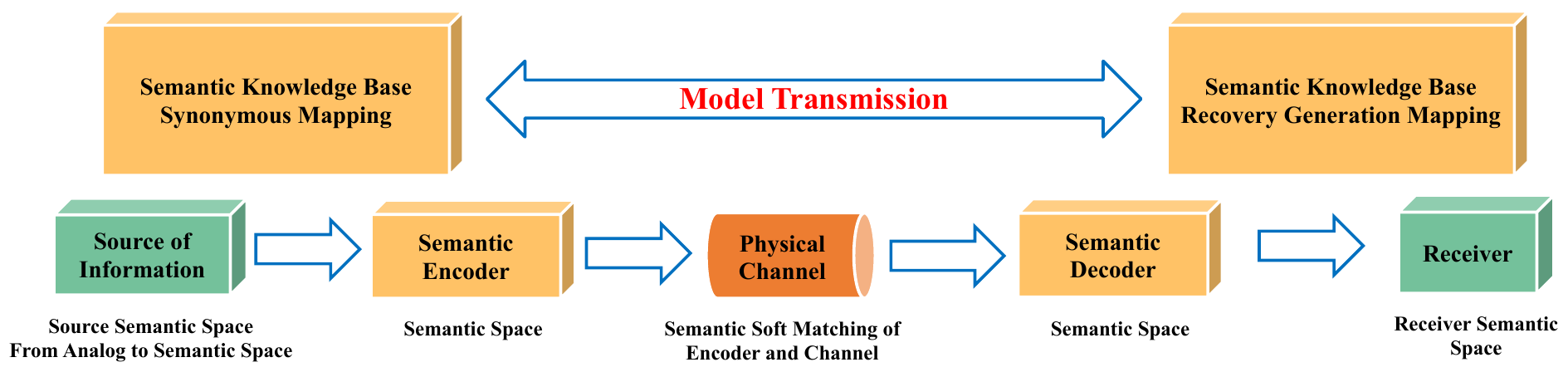}
			\label{figure1b}
		}
  \caption{Comparison of the link-level architecture between traditional communication and SemCom systems, where (a) shows the architecture of the traditional communication system, where the source and channel encoders operate independently to transmit and recover bits through the physical channel, focusing solely on accurate bit replication without considering semantic content; (b) illustrates the architecture of the SemCom system, where the semantic encoders and decoders are jointly trained to enable efficient transmission and recovery of semantically meaningful information, adapting to both the source and channel conditions. The above procedure is achieved by the transmission of models, which serve as semantic knowledge base for synonymous mapping, recovery and generation.}
\end{figure*}

Fig. \ref{figure1a} depicts the link-level architecture of the traditional communication system, where each module is separately designed and independently optimized. The information source is independently processed by the source encoder and the channel encoder, then transmitted via the physical channel, where a hard matching is occurred. After that, the noised signal is obtained at the receiver side and processed by the channel decoder and the source decoder. Traditional communication systems focus on the accurate replication of digital bits through the transmission of analog waves in the air, where the wireless channel is viewed as an opaque data pipe carrying messages. The contextual meaning and the validity of the messages are ignored, and the only requirement is to recover the transmitted bits as accurate as possible at the receiver space without considering the desired meaning or downstream actions. In this circumstance, a large amount of semantically irrelevant redundant data will be transmitted together, which causes unnecessary communication resource consumption.

Fig. \ref{figure1b} demonstrates the link-level architecture of the SemCom system, where AI models and abundant data samples are treated as the semantic knowledge base providing synonymous mapping between information source and semantics. As can be trained by distributed data samples and transmitted throughout the network (see Sec. \ref{sec:netlevel}), the AI model can perform efficient SemCom for semantic information generation and recovery on the transmitter and the receiver side \cite{0Intellicise}, thereby enabling the joint training and deployment of the semantic encoder and the semantic decoder.

As shown in Fig. \ref{figure1b}, the information source is directly processed by the semantic encoder in the semantic space, which aims to extract the contextual meaning of the input message. Then the extracted semantic symbols are passed into the physical channel, where the semantic soft matching \footnote{``Semantic soft matching" refers to a more flexible and tolerant matching approach adopted in the process of SemComs to process the semantic content of information. Compared to strict, rigid matching criteria in traditional communications, soft matching may allow for a certain degree of semantic differences or ambiguity, as long as these differences do not impair the overall understanding and transmission of the information.} is achieved to adapt to both the source information and the channel noise. At the receiver side, the noised semantic symbols are further processed by the semantic decoder, which recovers the source information under the joint consideration of channel noise and the received semantics. Since the semantic encoder and the semantic decoder are jointly trained under the semantic space and adapted according to the propagating channel environment, it eliminates the environment's effect on source information recovery facilitation under the receiver semantic space. It is noted that the multi-modal or heterogeneous original information requires combining other technologies such as semantic synthesis of multiple sources to deliver the desired meaning and match the transmission goal efficiently.

\subsection{Network-Level Architecture of SemCom}
\label{sec:netlevel}
Based on the SemCom framework at the link level, this subsection gives a detailed description of the semantic collaboration at the network level. As illustrated in Fig. \ref{fig:intellnetarc}, inspired by the cloud-edge-device architecture proposed to handle massive data transmission under different application scenarios (e.g., constrained energy availability and low latency tolerance limits) \cite{9539979}, the knowledge collaboration of the SemCom system applies the three-layer architecture as well. Unlike other SemCom systems proposed in \cite{SHIGuangming26,SHIGuangming91,chuanhong97}, which employ a cloud-based knowledge base to achieve semantic extraction and recovery remotely, the semantic collaboration framework aims at the replication of ``functions" via transmitting the AI models through the communication system to enable the ``flow of intelligence" \cite{yining2024intellicise}. The semantic collaboration procedure in each layer is illustrated as follows:
\begin{figure*}[!t]
\centering
\includegraphics[width=0.9\linewidth]{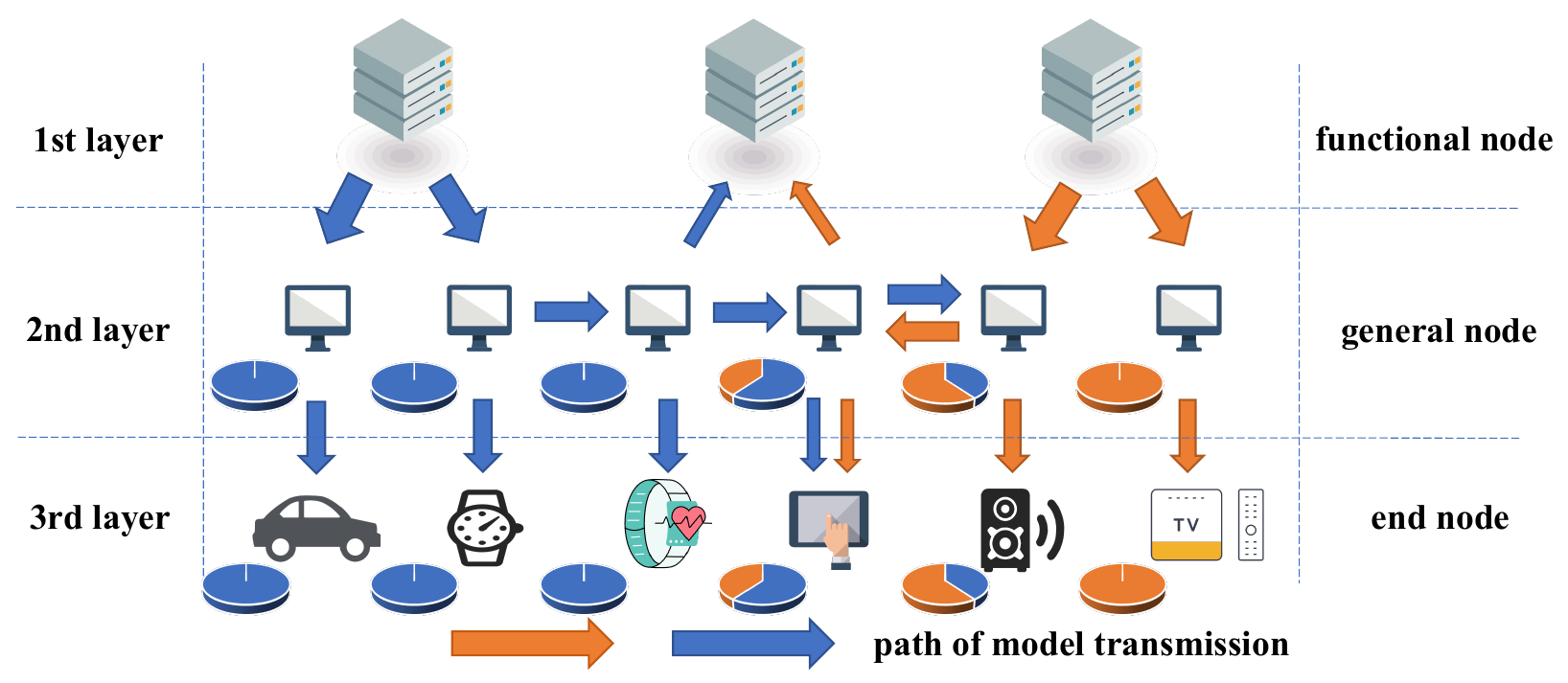}
\caption{Network-level architecture of SemCom systems, where the semantic collaboration spans across three layers: cloud, edge, and device. The first layer (cloud) hosts highly efficient models trained on vast distributed databases, providing intelligent semantic services and incentivizing user contributions. The second layer (edge) includes mobile edge computing nodes that retrain models using techniques like meta-learning and federated learning, collaborating globally to mitigate data bias and adapt to dynamic conditions. The third layer (device) consists of resource-constrained end nodes that retrieve and update models from the edge, ensuring synchronization for semantic recovery during transmission.}
\label{fig:intellnetarc}
\end{figure*}

\begin{itemize}
    \item The functional nodes on the first layer (cloud) consistently generate highly effective models trained on vast databases collected in a distributed manner, ensuring remarkable efficiency, accuracy, robustness, and reliability. For instance, Google’s servers host advanced models like the Switch Transformer, exemplifying state-of-the-art performance in natural language processing \cite{JMLR:v23:21-0998}.  Moreover, large language models (LLM) provided by AI companies, including OpenAI’s ChatGPT, Anthropic’s Claude, and the recently popular DeepSeek, can also be deployed in the cloud to offer intelligent semantic parsing and generation services \cite{fui2023generative}. Additionally, cloud platforms can utilize incentive mechanisms, such as blockchain-based reward schemes, to encourage user engagement and contributions to these large AI models. 

    \item The nodes with medium calculation and storage capability, such as mobile edge computing (MEC), can be placed on the second layer (edge). Upon receiving the model from the functional nodes in the first layer, the nodes on the second layer can retrain the model via the techniques such as meta-learning \cite{chen2024scalable}, transfer learning \cite{wu2022semantic}, knowledge distillation \cite{liu2023knowledge}, model distillation \cite{li2019fedmd} and federated learning (FL) \cite{10559618} for different purposes. For example, the edge nodes utilize distributed learning techniques to achieve global collaboration across different knowledge base. By leveraging data with heterogeneous semantic distributions\footnote{Information may originate from diverse sources such as text, images, audio, video, and more. These different information sources may adopt varying methods when expressing semantics, leading to heterogeneity in semantic distribution. Even for the same information source, different individuals or systems may utilize different modes of expression or linguistic styles to convey information. This variability in expression similarly results in heterogeneity in semantic distribution.} from geographically dispersed locations, the collaboration among edge nodes mitigates the impact of biased data samples on AI model performance, thereby constructing a globally consistent semantic knowledge base \cite{10559618}. Besides, the edge nodes can compress the large AI models from the first layer to varying degrees based on dynamic channel conditions and resource constraints, making it feasible to transmit the AI models to the nodes on the third layer within the limited bandwidth, storage and computation resources. 
    
    \item The resources-constrained end nodes on the third layer (device), such as sensors, cannot retrain the model. These nodes can send the model requirement messages to the edge nodes on the second layer or subscribe to the model update for model retrieval and pay the fees according to the model pricing. In the semantic collaboration framework, the semantic recovery model on the receiver side needs to be synchronized with the semantic encoding model held by the transmitter. Thus, the nodes on the third layer should be able to receive the models transmitted from the source directly and update the corresponding model based on the latest version. Additionally, it should also receive the newly retrieval model from the edge nodes or the nodes on the peer level when necessary.

\end{itemize}

\section{Security and Privacy Breaches of SemCom}
\label{section 3}
This section first introduces the full life cycle of SemCom, which includes model training, model transfer, and semantic information transmission phases. Then, it elaborates on the potential security and privacy threats that may arise in each of these three stages. The detailed security and privacy threats are listed in Tab. \ref{securitythreats}.

\begin{table*}[tbp] 
\centering
\caption{List of Security and Privacy Threats in SemCom.}
\label{securitythreats}
\renewcommand{\arraystretch}{0.9}  
\normalsize  
\begin{tabular}{
    |>{\centering\arraybackslash}m{0.1\textwidth}|>{\centering\arraybackslash}m{0.1\textwidth}|>{\centering\arraybackslash}m{0.1\textwidth}|>{\centering\arraybackslash}m{0.13\textwidth}|>{\arraybackslash}m{0.55\textwidth}|} 
    \hline
        \textbf{\scriptsize{Attacks}} & \textbf{\scriptsize{Positions}} & \textbf{\scriptsize{Phases}} & \textbf{\scriptsize{Properties}} & \textbf{\scriptsize{Effects}}\\
    \hline
        \tiny{Knowledge base poisoning attacks} & \tiny{Knowledge base} & \tiny{Model training} & \tiny{Integrity and Availability} & \tiny{It entails the premeditated introduction of harmful data into the knowledge base by an adversary, particularly in a publicly accessible one. The extraction and restoration of semantic characteristics hinge on the quality of the data in the knowledge base, and contaminated data undermines the authenticity and accessibility of the information. It is divided into data poisoning attacks, model poisoning attacks, and backdoor attacks.}\\
    \hline
        \tiny{Gradient leakage} & \tiny{Model transmission} & \tiny{Model training} & \tiny{Confidentiality} & \tiny{The attacker can update the dummy input and output to minimize the gradient distance. The dummy output is the same size as the dummy input during the deep JSCC model training instead of a class probability vector during the classification model training. Despite the difficulty of stealing data in realistic training regimes, the deep JSCC model's collaborative training procedure may lead to a deeper and easier privacy leakage.} \\

    \hline
        \tiny{Vulnerability of servers} & \tiny{Model transmission} & \tiny{Model training} & \tiny{Confidentiality, Integrity, and Availability} & \tiny{If the server's security is inadequate, attackers may directly access the global model from the server, further tampering with the model or extracting sensitive information, which poses a serious threat to the integrity and accuracy of the SemCom model. If the server is attacked during the aggregation process, it may lead to tampering with the model updates, thereby affecting the training effectiveness and performance of the global model.} \\

    \hline
        \tiny{Attacks against communication bottlenecks} & \tiny{Model transmission} & \tiny{Model training} & \tiny{Availability} & \tiny{Despite they infrequent occurrence, the consequences they trigger can be severe, potentially causing significant disruptions to the federated training environment of SemCom models. For exmaple, models have become increasingly intricate, with a substantial increase in the number of model parameters that change during each iteration, thereby significantly elevating communication overhead. Furthermore, limited network bandwidth can lead to delays for edge servers when uploading local model updates to the cloude server or downloading the global model, resulting in dropout.} \\

    \hline
        \tiny{Model slice availability attacks} & \tiny{Model transmission} & \tiny{Model transfer} & \tiny{Availability} & \tiny{Model slice availability attacks can corrupt the parameters of the model slices, directly impacting their SemCom performance once deployed.} \\

    \hline
        \tiny{Model slice forgery attacks} & \tiny{Model transmission} & \tiny{Model transfer} & \tiny{Integrity} & \tiny{Model slice forgery attacks involve deploying model parameters pre-trained by the attacker at the terminal, thereby steering the user's SemCom results towards the direction desired by the attacker.} \\

    \hline
        \tiny{Semantic adversarial attacks} & \tiny{Source of information and physical channel} & \tiny{Semantic information transmission} & \tiny{Availability} & \tiny{SemCom systems on DL models for the extraction, encoding, and transmission of semantic information. Semantic adversarial attacks are covert and efficient  Attackers can intentionally introduce imperceptible perturbations into the input data, causing the model to make erroneous predictions or exhibit unintended behaviors. Additionally, by introducing meticulously crafted perturbations into the channels, attackers can disrupt the transmission and reception of semantic information.} \\

    \hline
        \tiny{Semantic eavesdropping attacks} & \tiny{Physical channel} & \tiny{Semantic information transmission} & \tiny{Confidentiality} & \tiny{Semantic eavesdroppers can capitalize on openness of wireless channels by employing specialized receiving equipment to capture and analyze signals in transit. If eavesdroppers obtain access to the semantic decoder, they could potentially decode some or all of the private or sensitive semantic information, even when their channel conditions are substantially inferior to those of the legitimate receivers.} \\

    \hline
        \tiny{Semantic inference attacks} & \tiny{receiver and physical channel} & \tiny{Semantic information transmission} & \tiny{Confidentiality and Integrity} & \tiny{By analyzing model intermediate information exposed, semantic inference attackers can infer whether specific data belongs to the training set members and even further reconstruct sensitive semantic information. This attack method directly threatens the privacy security of each user participating in distributed semantic learning.} \\
 
    \hline
        \tiny{Semantic jamming attacks} & \tiny{physical channel} & \tiny{Semantic information transmission} & \tiny{Availability and Integrity} & \tiny{They focuse on interfering with the semantic content of transmitted data, thereby degrading the consistency and quality of the data, and hindering the receiver's accurate understanding of the intended message.} \\
        
    \hline
\end{tabular}
\end{table*}

\subsection{Life Cycle of SemCom}

\begin{figure*}[t]
\centering
\vspace{-10mm}
\includegraphics[width=\linewidth]{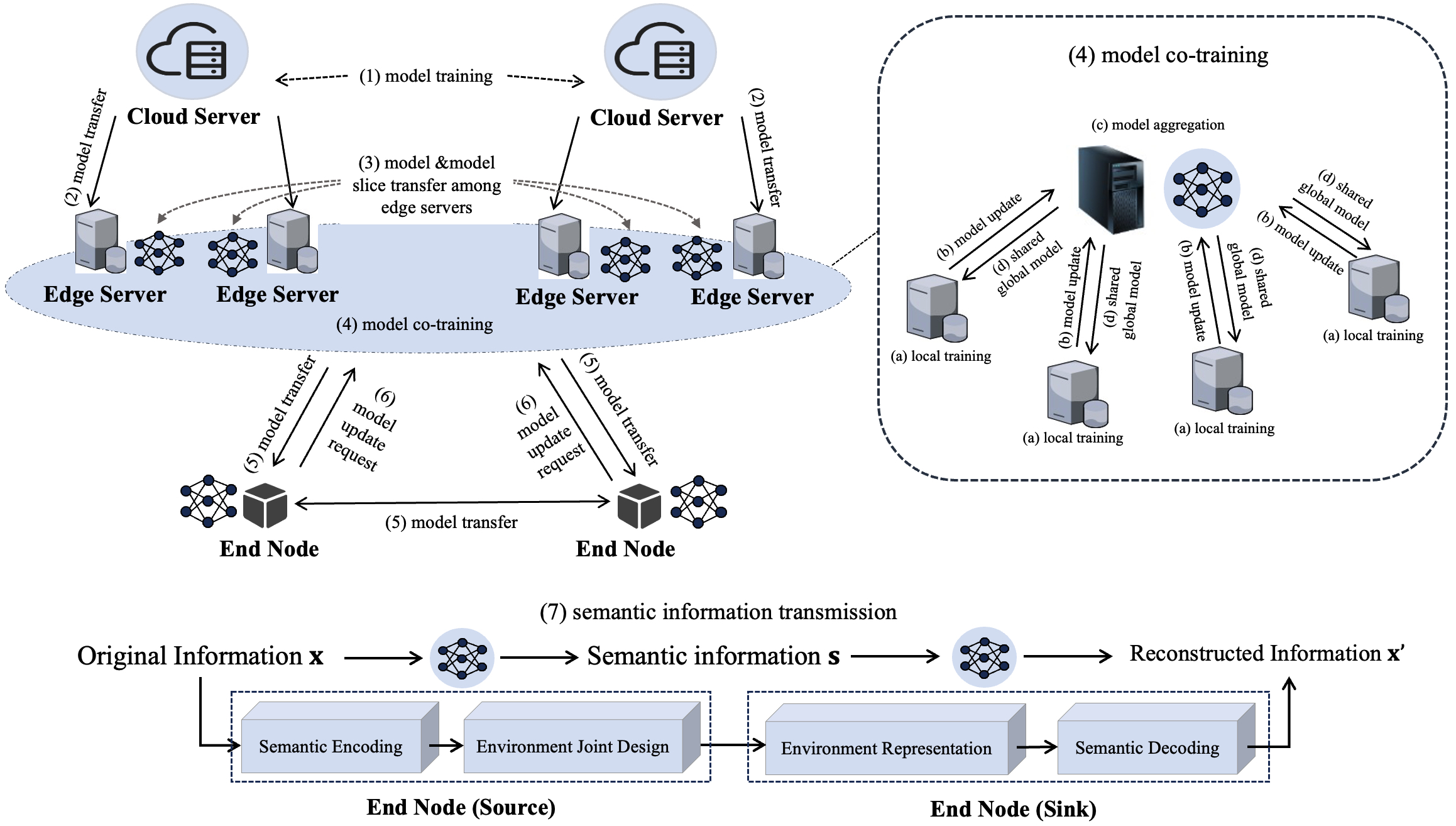}
\caption{Life cycle of SemComs, where a centralized model is first trained on the cloud and then transferred to edge servers. At the edge, the model and its slices are exchanged among servers to enable a robust co-training process that involves local training, model updates, aggregation into a refined global model and sharing it. This updated model is then delivered to the end node, and any further update requests trigger a new cycle of collaborative re-training, ensuring continuous optimization and efficient semantic information transmission throughout the network.}
\label{fig:intellcycle}
\end{figure*}

As illustrated in Fig. \ref{fig:intellcycle}, the whole life cycle of SemComs includes the following process.

\begin{enumerate}[{(1)}]
    \item Model training on cloud server.
    \item Model transfer from cloud server to edge server.
    \item Model and model slice transfer among edge servers.
    \item Model co-training among edge servers.
    \begin{enumerate}
        \item Local training on edge servers.
        \item Model update fragmentation and transfer.
        \item Model aggregation.
        \item Shared global model fragmentation and transfer.
    \end{enumerate}
    \item Model transfer from cloud server to end node.
    \item Model update request and model re-training (back to (3)).
    \item Semantic Information Transmission.
\end{enumerate}

The aforementioned process of SemComs 
life cycle can be summarized into the following three key phases.

\paragraph{Phase 1: Model Training}
In the model training phase, various methods such as FL and distributed learning can be utilized. In the case of FL, each edge server or device performs training locally and periodically uploads model parameters or training gradients to the cloud server for aggregation or model updates, facilitating collaborative training across multiple data sources. Uploading model parameters reduces the frequency of data exchanges, but unencrypted transmission of the model may also increase the risk of privacy breaches \cite{nguyen2024efficient}. On the other hand, uploading gradients can effectively protect data privacy, but it requires frequent data exchanges, thus increasing network load \cite{sun2020toward}. Additionally, distributed learning allows multiple computing nodes to independently process data and perform model training, thereby more effectively utilizing computing resources distributed across various locations to accelerate the training process \cite{cao2023communication}.
\paragraph{Phase 2: Model Transfer}
In the model transfer phase, sharing models between different network nodes can be optimized through techniques such as parameter passing, model splitting, or model compression. Model slicing, a common approach, involves dividing a large model into smaller segments for transmission across the network, which helps to reduce bandwidth requirements \cite{dong2022semantic}. Moreover, techniques such as model pruning, parameter quantization, low-rank factorization, and sparsification can be employed to reduce the model size and resource occupancy, thus enabling efficient model transmission \cite{li2023modelcompression}. Additionally, mechanisms for model exchange and model sharing between edge servers can also be used to expedite the model update process across nodes \cite{qi2024model}.
\paragraph{Phase 3: Semantic information transmission}
In the semantic information transmission phase, the model processes data through online inference, transmitting only the semantic features. This stage can utilize online inference methods based on attention mechanisms, knowledge distillation, or incremental learning to extract the most semantically significant information from the data. Attention mechanisms enhance the precision of extracting important information, while knowledge distillation helps reduce the complexity of inference, making it suitable for resource-constrained devices \cite{wheeler2023engineering,liu2023survey}. By employing methods of semantic information transmission, it is possible to reduce bandwidth while maintaining the effectiveness and contextual relevance of the transmitted information, meeting the demands of real-time communication.

\subsection{Confidentiality, Integrity, and Availability In SemCom}

\subsubsection{Confidentiality}
In information security, the break up of confidentiality refers to the release of unauthorized information \cite{samonas2014cia}. The confidentiality threat sometimes extends to the information content inference from the observed pattern during traffic analysis \cite{samonas2014cia}. In the AI-enabled traditional communication systems, confidentiality is roughly treated as the system's privacy, which protects sensitive information against misuse and authorizes access \cite{dilmaghani2019privacy}. As stated in \cite{dilmaghani2019privacy}, confidentiality-related attacks generally can be divided into two classes, namely, data breaches and model extraction. Data breaches are defined as disclosing sensitive data in unauthorized access \cite{dilmaghani2019privacy}. Take re-identification as an example. The authors in \cite{sweeney2000simple} presented the possibility of identifying medical records via exploiting the dataset from the public electoral rolls of the city of Cambridge. Model extraction refers to the attacks that aim to infer the property of the record used to train the model \cite{meminfer,properinfer} and even cloning the model directly by observing the input and the output pairs \cite{steal,Oh2018TowardsRB}.

As shown in Fig. \ref{fig:intellnetarc}, the model trading and model transmission among the functional, general nodes and end nodes in the SemComs enable the `flow of the intelligence’ and the data breaches and model extraction attacks like in the AI-enabled traditional networks. This article addresses that more considerations are necessary during the model transmission process to ensure the model confidentiality and data privacy in SemComs.


\subsubsection{Integrity}
In AI-enabled traditional networks, integrity refers to the consistency and accuracy of data through the AI system workflow against unauthorized modification \cite{dilmaghani2019privacy}. An attacker may modify the system towards misclassification and yet does not affect the performance of the systems, such as evasion attacks and backdoor attacks \cite{eykholt2018robust,moosavi2016deepfool,biggio2014poisoning,newell2014practicality,hinnefeld2018evaluating}.

Despite the attacks targeted on the complete model mentioned above, the attackers in the SemComs can also target the modifications of the transmitted model slices. For example, the malicious node can send the minimally modified models and parameters in a federated setup, such as using an appropriate linear layer and ReLU activation if it knows the cumulative distribution function of some quantity associated with user data \cite{fowl2021robbing}. Under such a case, the attacker can directly recover the part of the trained data via monitoring the gradient updates’ response from the victim \cite{fowl2021robbing}. Moreover, the semantic information, which considers the meaning and veracity of source information \cite{hinnefeld2018evaluating}, is a good choice for attackers to use.

\subsubsection{Availability}
In the AI era, availability can describe the system's power to perform to achieve the expected purpose-designed for the AI system with reliable outputs \cite{dilmaghani2019privacy}. For example, the data poisoning attack that alters the boundaries of the classifier such that the model becomes useless can be treated as a classic availability attack in the AI era \cite{dilmaghani2019privacy}.

In addition to the attacks in the AI-enabled traditional networks, the availability-related attacks in information security also threaten the security of SemComs. Data availability ensures the accessibility of the data and the system to the authorized users whenever they are needed \cite{samonas2014cia}. Typical availability issues include non-malicious issues like hardware failures, unscheduled software downtime, human error, or malicious issues like cyber-attacks and insider threats \cite{wang2022impact}.

In SemCom, model availability attacks disrupt model or model updates transmitted to authorized users. For instance, attackers can access many compromised nodes to set up attack armies, then send the meaningless model update to the model-assembling node, resulting in performance degradation. Besides, the malicious nodes can launch DDoS-alike attacks to disrupt legitimate users' access to the targeted models when the receiver attempts to get the latest model for semantic recovery.

\subsection{Security and Privacy Issues in Model Training Stage}

As detailed in Section~\ref{sec:netlevel}, the semantic recovery model on the receiver needs to be synchronized with the semantic encoding model held by the receiver. Thus, the receiving nodes on the second layer can receive the model transmitted from the source directly and update the model based on the latest model and the newly retrieval model sent from the functional nodes or the nodes on the peer level when necessary. As for the receiving nodes allocated on the third layer, the nodes can receive the well-trained model sent from the source or the edge nodes on the second layer.

By employing AI technology, such as meta-learning \cite{khodak2019adaptive}, transfer learning \cite{zhao2018federated}, knowledge distillation \cite{hinton2015distilling}, model distillation \cite{li2019fedmd} and FL \cite{pmlr-v54-mcmahan17a}, the model training and model update can be achieved through sharing the model parameters without revealing raw personal data. Nodes on the second layer can train their local model and send the local model updates to a server or a distributed ledger, followed by the global model aggregation and the dispensation of the latest updated model to the collaborative peer nodes.

However, similar to AI-equipped networks, the model training and update process also faces security and privacy issues even when there is no need for raw data collection \cite{Sun2694}. Besides, the close relationship between the original information, the semantic information extracted by the semantic encoding model and the recovered information by the semantic decoding model in SemCom systems, poses new challenges and opportunities for security and privacy detection, as follows.

\subsubsection{Knowledge Base Poisoning Attacks}
Each edge server has access to the training data during the collaborative training process, leading to a high possibility of adding the malicious model updates to the global model \cite{mothukuri2021survey}. In addition, no training samples and training process will be released to and checked by trustworthy authorities \cite{cao2019understanding}. Therefore, attackers can explore the lack of transparency in the agent updates to perform the data poisoning attacks or model poisoning attacks \cite{bhagoji2019analyzing}.

\paragraph{Data poisoning attacks}
Malicious clients can generate dirty samples to train the global SemCom model to produce falsified model parameters and send them to the server \cite{mothukuri2021survey}. The most commonly used data poisoning strategy is label-flipping \cite{blanchard2017machine}, where the label of training samples is modified to another category, resulting in taking control of the client’s local models and ultimately manipulating the global model \cite{mothukuri2021survey}. The objective of the data poisoning attacks is generally to degrade the accuracy of learning tasks, increase the convergence time of the global model, and the probability of erroneous learning results \cite{kang2020scalable}. For example, Peng et al. \cite{peng2024adversarial} present the targeted clean-label poisoning attack (TCLA) in multiuser SemCom, where an adversary meticulously selects and alters a subset of training data. This alteration introduces subtle perturbations that cause the SemCom model to misclassify specific instances during inference. Unlike conventional poisoning attacks that involve label flipping or the introduction of noisy data, TCLA achieves its objective without modifying the actual ground truth labels. Instead, it inserts meticulously crafted samples into the training dataset that resemble the target instances the attacker wishes to misclassify. By carefully selecting these instances, the attacker influences the model's decision boundaries for specific classes or instances, potentially leading to security or privacy risks.

\paragraph{Model poisoning attacks}
Unlike data poisoning attacks, model poisoning is designed to poison the global model in a targeted manner by compromising a small number of malicious agents \cite{bhagoji2019analyzing}. Instead of using fake data, the model poisoning attackers directly target the global and attempt to ensure the global model converges to a point with good performance on the test or validation data but misclassify a set of chosen inputs with high confidence \cite{bhagoji2019analyzing}. For example, the malicious classifier trained in \cite{gu2019badnets} identifies the stop signs like speed limits when a special sticker is added to the stop sign without affecting classification accuracy on other images.
Model poisoning attacks are more effective and challenging to explicate than data poisoning attacks \cite{bagdasaryan2020backdoor,bhagoji2019analyzing,fang2020local}. On the one hand, the poisoned model is equally accurate on the FL task, making it stealth to the detection algorithms that rely on the performance verification. Attackers can also incorporate evasion of anomaly detection into the attacker’s loss function to enable the maliciously trained model to evade even relatively sophisticated detectors, e.g., those that measure cosine similarity among submitted model updates. On the other hand, the researchers find that the poisoned model suffers from a longer-lasting performance degradation even when the model has later retrained for another task \cite{gu2019badnets}.

\paragraph{Backdoor attacks}
Backdoor attack is a special kind of backdoor attacks. According to the attacking scenarios, backdoor attacks can be divided into one-time (single-shot) and continuous (multiple-shot) poisoning settings \cite{xie2019dba}. The continuous poisoning attackers \cite{bhagoji2019analyzing} can train the models on backdoored inputs and change the local learning rate and the number of local epochs to maximize the overfitting of the backdoored data. As the global aggregation generally cancels out most of the backdoored model’s update, namely catastrophic forgetting, the multiple-shot attackers need to participate in every round of global aggregation to maintain the backdoor accuracy \cite{bagdasaryan2020backdoor}. While the one-time poisoning strategies are generally effective if staged by a single participant in a single round by applying the model replacement method to ensure that the attacker’s contribution survives averaging \cite{bagdasaryan2020backdoor}.
In addition, backdoor attacks can be divided into centralized and distributed backdoor attacks. The centralized attackers embed the same global trigger pattern in all malicious agent while the distributed attacker decomposes the global trigger pattern into local patterns and embeds them in the different parties \cite{xie2019dba}. Since the local trigger pattern is more insidious and easier to bypass the robust aggregation rules, distributed backdoor attacks are more effective and stealthy \cite{xie2019dba}.

\subsubsection{Gradient Leakage}
Zhu et al. \cite{zhu2019deep} found that it is possible to obtain the private training data from the publicly shared gradients. The normal participant calculates the gradient to update parameters using its private training data. In contrast, the malicious attacker updates its dummy inputs and labels with the goal of minimizing the gradients distance. When the optimization finishes, the evil user is able to steal the training data from the victims \cite{zhu2019deep}. Following the gradient leakage from gradients algorithm proposed in \cite{zhu2019deep}, several improved versions \cite{zhao2020idlg,yin2021see} are proposed to solve the convergence difficulties and relax the assumptions such as the batch size, the scale of the data set and the complexity of the target model. For example, Zhao et al. \cite{zhao2020idlg} demonstrate that when the targeted model uses cross-entropy as the objective function, the signs of the last-layer weights of the correct and wrong labels are opposite, enabling consistently discovering the ground-truth labels and facilitating the data extraction with increased fidelity.

Gradient leakage algorithms can also be applied to steal private data from the shared gradients in SemCom networks. The attacker can update the dummy input and output to minimize the gradient distance. The dummy output is the same size as the dummy input during the deep JSCC model training instead of a class probability vector during the classification model training. However, gradient leakage-alike attacks are limited in the scope of the batch size, the resolution of the data, and the model's complexity, leaving some to conclude that data privacy is still intact for realistic training regimes \cite{zhu2019deep,fowl2021robbing}. Despite the difficulty of stealing data in realistic training regimes, the deep JSCC model's collaborative training procedure may lead to a deeper and easier privacy leakage. Through modelling the relationship among the original information $x'$, the semantic information $x'$ extracted by the encoding module, and the reconstructed information $y'$ outputted from the decoding module, denoted as $f(x',s',y')$, the gradient approximation process can be expressed as
\begin{equation}
    \begin{aligned}
   & \mathcal D={\Vert \nabla W'-\nabla W \Vert}^2\\
    & s.t. \quad x',s',y' \sim f(x',s',y'). \\
    \end{aligned}
\end{equation}

The gradient approximation process is executed under the constraint of $f(x',s',y')$, narrowing the approximation space and simplifying the whole approximation process. Although the modelling of SemCom systems is still in its infancy, the deeper privacy leakage breaches in SemCom needs to be taken into account seriously.




\subsubsection{Vulnerability of Servers}
In the process of federated training for SemCom models, the edge server is responsible for securely aggregating the parameters updated by local users into the global model parameters, and then returning the updated parameters to the local users. If the server's security is inadequate, attackers may directly access the global model from the server, further tampering with the model or extracting sensitive information, which poses a serious threat to the integrity and accuracy of the SemCom model \cite{liu2022privacy}. The server is also responsible for receiving local model updates from various clients and aggregating them to generate a new global model. If the server is attacked during this process, it may lead to tampering with the model updates, thereby affecting the training effectiveness and performance of the global model. The server can control when each client accesses and manipulates the model during the FL training process. Therefore, a malicious server can predict the model's average or worst-case attack sensitivity, thereby designing the lowest-cost attack schemes. The security of network environments in which the server operates is also crucial. If the server operates in a dangerous network environment, the likelihood of being attacked will significantly increase \cite{mothukuri2021survey}.

\subsubsection{Attacks Against Communication Bottlenecks}
During the SemCom model federated training process, there is a frequent necessity to transmit model parameters or update information between servers. As DNNs evolve, models have become increasingly intricate, with a substantial increase in the number of model parameters that change during each iteration, thereby significantly elevating communication overhead. Furthermore, limited network bandwidth can lead to delays for edge servers when uploading local model updates to the cloude server or downloading the global model. When these delays become excessive, edge servers may fail to complete communication within the specified timeframe, resulting in dropout. The variability in data distribution and format across heterogeneous devices further complicates model training and increases communication overhead. During data transmission, additional processing steps may be required to ensure data consistency and accuracy, which can also prolong communication time. The aforementioned issues are collectively termed as attacks against communication bottlenecks \cite{mothukuri2021survey,liu2022privacy}. Despite their infrequent occurrence, the consequences they trigger can be severe, potentially causing significant disruptions to the federated training environment of SemCom models.

\subsection{Security and Privacy Issues in Model Transfer Stage}

As shown in Fig. \ref{fig:intellnetarc}, the model updates are shared among collaborative training participants while the latest model travels through the network to the receiving nodes for semantic recovery. Like the packet transmission, the communication systems impose very different upper limits on the size of packets that they can accept and transport \cite{1979Packet}. At the same time, the ongoing of artificial intelligence has led to surprising emergence that results from the scale \cite{bommasani2021opportunities}. For example, the introduction of transformer structure \cite{transformer}  made the DL model parameters exceeded 100 million in 2017. While the BERT network model \cite{devlin2018bert} was introduced, making the number of parameters exceed 300 million scales for the first time, the GPT-3 model \cite{brown2020language} exceeded 10 billion, and Pengcheng Pancake \cite{pengchengpancke} achieved a scale of 100 billion densities. The introduction of the Switch Transformer \cite{fedus2021switch} surpassed the trillion scale in one fell swoop.

Constrained by the network capacity, the AI model needs to be broken up into smaller slices for transport and then reassembled when entering the receiving node, a process usually known as fragmentation \cite{bommasani2021opportunities}. Inspired by the works to deal with AI-related tasks in real-time with bandwidth, computational capability, memory and delay limit via splitting the model \cite{sun2018slim,li2018jalad,teerapittayanon2016branchynet,tang2020joint}, model slices are adopted in SemCom systems. Slicing can divide the model into multiple smaller parts, thereby reducing the data volume per transmission and increasing transmission speed. During the model slice transfer stage, it becomes easier to detect which slice has failed or contains errors, and only those faulty slices, rather than the entire model, need to be retransmitted, thus saving time and bandwidth. Therefore, slicing allows for flexible adjustment and expansion of the model according to different application scenarios and requirements. We believe that different levels of security are needed to safeguard different model slices. For example, the attacker can directly identify the ground-truth labels based on the transmitted gradients on the last layer of the classification model trained with cross-entropy loss during the model update process \cite{zhao2020idlg}. In other words, the model slices should no longer be treated as non-differential bits when transmitted through the physical channel, and each model slice is with different vulnerability level. Therefore, it is reasonable that the model slice with high vulnerability, such as the last layer of the cross-entropy-trained model, needs extra protection when transmitted.

However, during the model slice transfer stage, it is susceptible to availability attacks and forgery attacks. Availability attacks can corrupt the parameters of the model slices, directly impacting their SemCom performance once deployed. Forgery attacks involve deploying model parameters pre-trained by the attacker at the terminal, thereby steering the user's SemCom results towards the direction desired by the attacker.

\subsubsection{Model Slice Availability Attacks}



During the semantic model slice transfer stage, each slice may contain specific parts of the model, such as parameters, structures, or training data. These slices must be strictly isolated and protected during transmission to prevent unauthorized access. If the isolation measures are not sufficiently strict, attackers may have the opportunity to steal sensitive information from these slices. Once attackers obtain this sensitive information, they can reconstruct the model or use it to launch targeted attacks. Additionally, if the slices contain training data or derived information, the leakage of these data may violate personal privacy, especially when dealing with sensitive data such as medical records and financial information. Besides isolation measures, resource allocation is crucial in the transmission of semantic model slices. Uneven resource allocation may lead to unfair treatment of certain slices during transmission, causing a series of problems. For example, Denial of Service (DoS) attackers can block slice transmission channels by consuming large amounts of resources (such as bandwidth and computing resources). When attackers successfully block the channels, the model slices cannot reach the receiving nodes as expected, resulting in the inability to fully reconstruct or update the model, and ultimately disrupting the operation of the entire system. Furthermore, uneven resource allocation may also lead to a decline in the overall system performance, with increased latency and reduced throughput. This reduces the availability and reliability of the system, making it susceptible to instability even under normal loads.

\subsubsection{Model Slice Forgery Attacks}

Semantic model slices, by aggregating data, computing power, and domain knowledge, can be directly applied to commercial services, such as semantic IoT, possessing clear economic value. This makes semantic models, as core intellectual property, a common practice for trading and sharing. However, attackers may attempt to obtain unauthorized semantic model slices by forging seemingly legitimate model slice requests. For instance, model slice forgery attackers can exploit loose access control policies or logical vulnerabilities in semantic slice models to bypass normal security mechanisms and directly access and download semantic model slices. Additionally, during the semantic model slices transfer stage, strong identity binding mechanisms (such as digital certificates, hardware fingerprints, etc.) are crucial for ensuring slice security and preventing illegal access. However, if the semantic slice model management system lacks these strong identity binding mechanisms, attackers can more easily forge identity information, impersonate legitimate users, and initiate illegal requests.

\subsection{Security and Privacy Issues in Semantic Information Transmission Stage}

As depicted in Fig. \ref{fig:intellcycle}, the extracted semantic information is communicated between the transmitter and the receiver in SemCom networks. Compared with raw data-based communication, privacy properties of SemCom degrade much more grace if the security mechanisms fail \cite{chuanhong97}.

Take asymmetric key encryption as an example. A crypto-system must make it impossible for a computationally-bounded adversary to derive meaningful information about a message from (plain text) using only its cipher text and the corresponding public encryption key \cite{bellare2012semantic}. In other words, the adversary chooses two plain texts  $m_0,m_1$, one of which is encrypted as $c\leftarrow E(k,m_i)$. Then the adversary has to guess which message is ciphered and the encryption is semantically secure if for all efficient adversaries, the advantage function is negligible \cite{bellare2012semantic}. The advantage function $A$ is expressed as 
\begin{equation}
Adv[A,E]=\lvert Pr(M_0-M_1)\rvert   
\end{equation}
where $M_i$ represents the event that the adversary A decides that $m_i$ is ciphered. In SemCom networks, the semantic information $s_i$ instead of the original data $m_i$ is transmitted through the network. Thus, the wiretapper has to steal the semantic model $model$ via model reverse engineering techniques to obtain $m_i$. Thus, in SemCom system, the ambiguity brought by the reconstructed model on the attacker side needs to be considered. And the advantage function mentioned above should be expressed as
\begin{equation}
Adv[A,E,model]=\lvert Pr(M_0-M_1)\rvert 
\end{equation}

\subsubsection{Semantic Adversarial Attacks}
Adversarial attacks mislead learning models by subtly altering input data \cite{huang2017adversarial}. These attacks are equally pertinent to SemCom systems, which rely on DL models for the extraction, encoding, and transmission of semantic information. Semantic adversarial attacks are covert and efficient \cite{guo2024survey}. Attackers can intentionally introduce imperceptible perturbations into the input data, causing the model to make erroneous predictions or exhibit unintended behaviors. These perturbations are barely noticeable to human senses but can exert significant impacts on DL models. Additionally, during the process of SemCom, wireless channels may be susceptible to various interferences and noises, creating vulnerabilities for attackers \cite{luo2022semantic}. By introducing meticulously crafted perturbations into the channels, attackers can disrupt the transmission and reception of semantic information. For example, Sagduyu et al. \cite{sagduyu2023semantic} present the process of semantic adversarial attacks as follows. During the inference stage, semantic adversarial attacks aim to tweak the input data fed into the adversary's model. This tweaking, often achieved through minor alterations or perturbations, renders the model unable to make accurate decisions for those particular samples. The effectiveness of these attacks is evaluated by examining the model's accuracy on the altered test inputs. The more the accuracy drops, the more successful the attack is deemed. The perturbations are chosen by minimizing their impact while ensuring that they cause the victim model to make errors. Additionally, the power of these perturbations is kept within a certain threshold. Adversaries can launch both targeted and non-targeted attacks. Targeted attacks aim to mislead the semantic model's outputs specifically for a set of non-target labels, redirecting them to target labels by minimizing the loss function related to those target labels. Conversely, non-targeted attacks strive to cause errors for samples from all labels by maximizing the loss function for all samples under attack.

\subsubsection{Semantic Eavesdropping Attacks}

Wireless channels, which serve as the conduit for signals traveling from the sender to the receiver, are characterized by their openness. This implies that during transmission, signals are not solely received by legitimate receivers but can also potentially be intercepted by nearby devices operating within the same frequency band. This inherent openness makes wireless channels vulnerable to a multitude of noises, interferences, and other unpredictable time-varying disruptions, including those arising from user mobility \cite{meng2024tdgcn}. Semantic eavesdroppers can capitalize on this openness by employing specialized receiving equipment to capture and analyze signals in transit. Such equipment may encompass high-sensitivity antennas, signal amplifiers, demodulators, and various other tools dedicated to receiving, amplifying, and dissecting wireless signals. In certain instances, semantic eavesdroppers may even deploy sophisticated signal processing techniques, such as signal separation and interference suppression, to boost the success rate and precision of their eavesdropping endeavors. If eavesdroppers obtain access to the semantic decoder, they could potentially decode some or all of the private or sensitive semantic information, even when their channel conditions are substantially inferior to those of the legitimate receivers \cite{tang2024secure}.

\subsubsection{Semantic Inference Attacks}

By analyzing intermediate information exposed during model training, semantic inference attackers can infer whether specific data belongs to the training set members and even further reconstruct sensitive semantic information. This attack method directly threatens the privacy security of each user participating in distributed semantic learning. Specifically, attackers can achieve privacy threats by directly exploiting shared information, poisoning the training process, and conducting collaborative attacks as follows.
\begin{itemize}
    \item Direct exploitation of shared semantic information: By observing the gradient distribution of the global model, attackers can identify the contribution of specific samples to the gradients. If the gradient direction of a sample is highly correlated with the global update, it can be inferred that the sample is likely part of the training set. Furthermore, semantic embeddings in shared knowledge bases may implicitly contain the distribution characteristics of training data. Attackers can use reverse engineering to infer original data from embedding semantic representations, for example, by discovering frequently occurring data categories through clustering analysis.
    \item Poisoning the training process: Attackers can upload local data containing adversarial samples, causing the model to overfit on specific semantic categories. This overfitting phenomenon amplifies the possibility of semantic inference because the introduction of adversarial samples alters the model's training trajectory. Additionally, during local training, attackers can inject malicious gradients, such as by scaling gradients or shifting their directions, to interfere with global model parameters. This interference can induce legitimate users to expose more sensitive semantic information during the inference phase. Moreover, attackers can modify the loss function during local training, such as by adding penalties for specific samples, causing the model to behave abnormally when processing these data. This abnormal behavior also aids attackers in subsequent semantic inference.
    \item Collaborative attacks: Multiple malicious participants can jointly launch semantic inference attacks, enhancing the accuracy of inference by cross-referencing gradient or parameter updates from different participants. This collaborative semantic inference attack method can further amplify the risk of privacy leakage because multiple attackers can share information and work together to improve the attack effectiveness.
\end{itemize}



\subsubsection{Semantic Jamming Attacks}

In traditional communications, jamming attackers disrupt or impede normal wireless communication processes by emitting targeted interference electromagnetic waves, leading to command failures or communication difficulties. These attackers can utilize specialized radio interference equipment, including jammers and disruptors, to emit such waves. Furthermore, they can exploit naturally occurring electromagnetic interference in the environment (like lightning and solar activity) or interference generated by human factors (such as power lines and industrial equipment) to execute interference attacks \cite{pirayesh2022jamming}. 

In contrast, semantic jamming attack focuses on interfering with the semantic content of transmitted data, thereby degrading the consistency and quality of the data, and hindering the receiver's accurate understanding of the intended message. For example, Tang et al. \cite{tang2023gan} proposes an intelligent jamming framework designed to analyze the security of SemCom systems. The framework introduces a semantic jammer aiming to alter the decoded content semantically from the original, while the receiver aims to correctly decode the semantic information despite the jamming attack. A game model is established between the jammer and receiver, incorporating a GAN-like strategy to optimize their interactions.

\begin{table*}[tbp] 
\centering
\caption{List of Defensive Techniques for Secure SemCom.}
\label{defensivetechniques}
\renewcommand{\arraystretch}{0.9}  
\normalsize  
\begin{tabular}{
    |>{\centering\arraybackslash}m{0.105\textwidth}|>{\arraybackslash}m{0.2\textwidth}|>{\centering\arraybackslash}m{0.17\textwidth}|>{\arraybackslash}m{0.425\textwidth}|} 
    \hline
        \textbf{\scriptsize{Methods}} & \textbf{\scriptsize{Sub-Class}} & \textbf{\scriptsize{Attacks}} & \textbf{\scriptsize{Effectiveness}} \\
    \hline
        \tiny{Data cleaning} & \tiny{Data cleaning for textual data, image data, audio data, and structured data} & \tiny{Knowledge base poisoning attacks} & \tiny{It simplifies the data set by removing redundant, repetitive and invalid data, thus reducing the data processing burden in the process of semantic model training. Additionally, it can identify and correct errors and abnormal values in data sets, thus improving the accuracy of data and reducing misunderstandings and ambiguities caused by data errors.}  \\
    \hline
        \tiny{Robust learning} & \tiny{Robust aggregation, data heterogeneity, the impact of noise and malicious clients, and balancing robustness and communication efficiency} & \tiny{Knowledge base poisoning attacks} & \tiny{Robust learning can enhance the semantic model's anti-interference ability to abnormal data and noise, so that the SemCom system can still maintain stable and reliable performance in the face of all kinds of interference and noise, and improve the semantic model's resistance ability in the face of malicious attacks or data tampering.} \\
    \hline
        \tiny{Defensive methods of backdoor attacks} & \tiny{Data processing and model identification and construction} & \tiny{Backdoor attacks} & \tiny{It can be achieved from data and model perspectives.}  \\
    \hline
        \tiny{Adversarial training} & \tiny{Adversarial sample generation, adversarial regularization, adaptive disturbance constraint, and matching pluggable modules} & \tiny{Semantic adversarial attacks and semantic inference attacks} & \tiny{Adversarial training is realized by adding adversarial attacks to the training samples to defend against semantic adversarial attacks. In SemCom, semantic adversarial samples are generated by adding noise to the semantic information to train the model together with the semantic information in the original training set.}  \\
    \hline
        \tiny{Differential privacy} & \tiny{Parameter value protection, objective function perturbation, and data generation and pre-processing} & \tiny{Gradient leakage and semantic inference attacks} & \tiny{In SemCom, the single semantic symbol cannot influence the meaning of the whole sentence. Therefore, the risk of privacy breaches arising from the addition of a piece of semantic symbol to a dataset is kept so small that an attacker cannot obtain accurate information by observing the results of the computation.}  \\
    \hline
        \tiny{Cryptography technology} & \tiny{Homomorphic encryption, secure multi-party computation, trusted execution environment, secure aggregation, and quantum cryptography} & \tiny{Vulnerability of servers, model slice availability attacks, model slice forgery attacks, semantic eavesdropping attacks, and semantic jamming attacks} & \tiny{Encryption algorithm can convert plaintext data into ciphertext, which enhances the confidentiality of semantic information. Integrity verification algorithms can ensure that the semantic data has not been tampered with during transmission and confirm the authenticity and integrity of the semantic data. Cryptography technology can also be used for identity authentication to ensure the true identity of both parties.}  \\
    \hline
        \tiny{Block-chain technology} & \tiny{Data security, prevent node leakage, and model framework design} & \tiny{Model slice availability attacks, model slice forgery attacks, semantic adversarial attacks, semantic eavesdropping attacks, semantic inference attacks, and semantic jamming attacks} & \tiny{Once the data is written into the blockchain, it cannot be easily changed or deleted, ensuring that the semantic information is not maliciously tampered with during transmission. In addition, the decentralized nature of blockchain makes semantic information no longer rely on a single central node for storage and transmission, which reduces the risk of data being single point of failure and improves the overall security of SemCom system. Blockchain can also be used to ensure that the identities of both communication parties are authentic, thus preventing impersonation and fraud.}  \\
    \hline
        \tiny{Model compression} & \tiny{Model pruning, parameter quantization, low-rank decomposition, and knowledge distillation} & \tiny{Attacks against communication bottlenecks, semantic inference attacks} & \tiny{By employing model compression techniques, devices can effectively decrease communication costs, conserve storage space, and eliminate parameter redundancy, thereby strengthening the SemCom system to defend against against multi-party cooperation attacks.} \\
    \hline
        \tiny{Physical-layer security} & \tiny{Beamforming, artificial noise, relay cooperation, intelligent reflecting surface, physical-layer key generation, and physical-layer authentication} & \tiny{Semantic eavesdropping attacks, semantic inference attacks, and semantic jamming attacks} & \tiny{It involves utilizing the randomness of wireless channels, such as interference, fading, and noise, to ensure secure transmission.}  \\
    \hline
\end{tabular}
\end{table*}

\section{Techniques to Mitigate Security and Privacy Threats in SemCom}
\label{section 4}
In response to the security and privacy threats detailed in Sec. \ref{section 3}, this section provides defensive techniques, including data cleaning, robust learning, defensive methods of backdoor attacks, adversarial training, differential privacy, cryptography technology, block-chain technology, model compression, and physical-layer security. The detailed techniques to mitigate security and privacy threats in SemCom are shown in Tab. \ref{defensivetechniques}.

 \subsection{Data Cleaning}
\begin{figure*}[htbp]
\centering
\vspace{0mm}
\includegraphics[width=\linewidth]{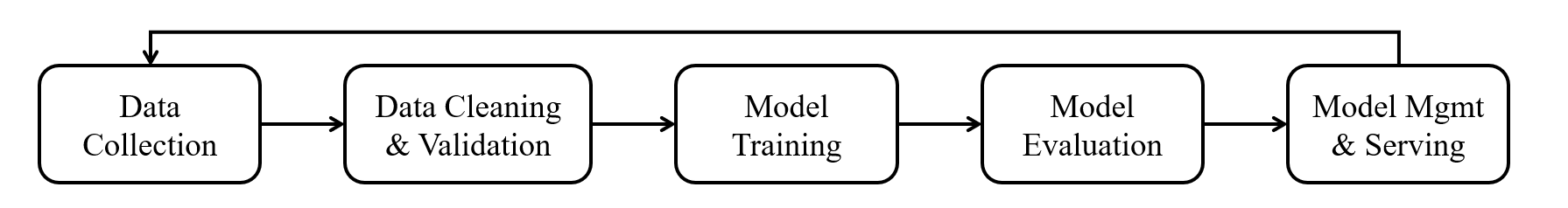}
\caption{Data cleaning in DL \cite{whang2020data}, where the process ensures the quality and consistency of data through collection, cleaning, validation, training, evaluation, and serving. Data cleaning, being a fundamental step, helps improve the accuracy, robustness, and reliability of the model, enabling better overall performance in real-world applications.}
\label{fig411}
\end{figure*}

Dirty data can introduce noise, missing values, or inconsistencies, directly causing semantic misunderstandings and even communication failure. Fig. \ref{fig411} illustrates the use of data cleaning in DL.

The below, as shown in Table \ref{datacleaning}, the following content, based on multiple academic studies, delves into various data cleaning methods for different data types and their potential applications in SemCom.

\begin{longtable}{|A{2.2cm}|A{0.6cm}|P{9.5cm}|}  
\caption{\footnotesize{List of Representative Data Cleaning Schemes}}
\label{datacleaning}\\
\hline
\textbf{Sub-class} & \textbf{Ref.} & \textbf{Descriptions, including} $\circledast$: \textbf{contributions; }

$\circleddash$: \textbf{trade-offs;} 
$\boxplus$: \textbf{performance metrics;}

$\Cap$: \textbf{applied in SemCom or promising for SemCom} \\
\hline
\endfirsthead

\hline
\textbf{Sub-class} & \textbf{Ref.} & \textbf{Descriptions, including} $\circledast$: \textbf{contributions; }

$\circleddash$: \textbf{trade-offs;} 
$\boxplus$: \textbf{performance metrics;} 

$\Cap$: \textbf{applied in SemCom or promising for SemCom} \\
\hline
\endhead

\hline
\endfoot

\hline
\endlastfoot
Textual Data Cleaning & \cite{rajan2024data} &
$\circledast$: Introduces an efficient deduplication method using hash algorithms and similarity matching (e.g., Jaccard similarity and cosine similarity). Also optimizes text standardization with lemmatization and stemming.

$\circleddash$: Accuracy and scalability.

$\boxplus$: F1-score and redundancy reduction.

$\Cap$: Promising for ensuring high-quality text data input in SemCom.
\\
\hline

Textual Data Cleaning & \cite{chu2016data} &
$\circledast$: Utilizes GANs for spelling correction and semantic enhancement, expanding datasets with diverse text expressions and improving semantic consistency.

$\circleddash$: Accuracy and computational efficiency.

$\boxplus$: Error detection rate and data quality improvement.

$\Cap$: Promising for ensuring high-quality text data input in SemCom.
\\
\hline

Textual Data Cleaning & \cite{lee2021survey} &
$\circledast$: Proposes a contextual reasoning-based text completion method using generative models (e.g., GPT) for missing information imputation.

$\circleddash$: Model training complexity and inference time.

$\boxplus$: Cleaning accuracy and model performance improvement.

$\Cap$: Promising for filling the missing text data in SemCom.
\\
\hline

Image Data Cleaning & \cite{tang2020data} &
$\circledast$: Develops a method for noise removal through Gaussian blur and bilateral filtering, maintaining edge details and removing random noise.

$\circleddash$: Data preprocessing complexity and computational cost.

$\boxplus$: Classification accuracy and generalization capability.

$\Cap$: Promising for enhancing signal interpretation and anomaly detection in SemCom.
\\
\hline

Image Data Cleaning & \cite{tae2019data} &
$\circledast$: Extends training datasets through data augmentation techniques like rotation and scaling to simulate real-world scene variations.

$\circleddash$: Accuracy and fairness.

$\boxplus$: Model accuracy and fairness ratio.

$\Cap$: Promising for autonomous - driving vision in SemCom.
\\
\hline





Audio Data Cleaning & \cite{li2021cleanml} &
$\circledast$: Proposes audio denoising using spectral subtraction combined with deep learning models to separate speech from background noise.

$\circleddash$: Cleaning thoroughness and computational efficiency.

$\boxplus$: Classification accuracy and F1-score.

$\Cap$: Promising for real-time voice in SemCom.
\\
\hline

Comprehensive Optimization of Structured Data & \cite{lotfi2025vmguard} &
$\circledast$: VMGuard introduces a reputation-based framework integrating  SemCom and dynamic trust evaluation to detect and mitigate data poisoning attacks in real-time across vehicular Metaverse and IoT ecosystems.

$\circleddash$: Computation Overhead and Real-time Response.

$\boxplus$: F1-score and Real-time Latency.

$\Cap$: Applied in filtering poisoned semantic metadata during transmission in semcom.
\\
\hline

\end{longtable}

\subsubsection{Intelligent Cleaning of Textual Data}
Textual data cleaning focuses on enhancing the accuracy and coherence of text used in SemCom by addressing issues like duplication, inconsistency, and errors. Techniques include deduplication, spelling correction, and missing information imputation, which improve the semantic clarity and reliability of textual datasets in applications like social media analysis and sentiment detection.

\textbf{Deduplication and standardization:} Rajan et al. \cite{rajan2024data} point out that duplicate data significantly impacts the learning efficiency and predictive performance of semantic models. The proposed efficient deduplication method combines hash algorithms and similarity matching (e.g., Jaccard similarity and cosine similarity). This study further optimized text standardization by lemmatization and stemming to eliminate semantic ambiguity, enhancing the model's robustness when facing diverse expressions. This method has been widely applied in social media data analysis and customer review processing, effectively improving data quality.

\textbf{Spelling correction and semantic enhancement:} Ilyas et al. \cite{chu2016data} developed a method for spelling correction and semantic enhancement using Generative Adversarial Networks (GANs). Their research demonstrated that GANs could significantly expand datasets by generating diverse text expressions, thus enhancing the model's adaptability to complex contexts. This method showed exceptional results in legal document processing and multilingual corpus analysis, providing strong support for semantic consistency.

\textbf{Missing information imputation:} Lee et al. \cite{lee2021survey} proposed a contextual reasoning-based text completion method that uses generative models (such as GPT) to predict and fill missing words or sentences. Their method performed excellently in sentiment analysis and dialogue systems, effectively improving data coherence and completeness, providing more reliable data for model training.

\subsubsection{Diverse Cleaning of Image Data}
Image data cleaning refers to the process of reducing noise, correcting artifacts, and enhancing data quality in images used for SemCom. This includes techniques like noise removal through Gaussian blur and bilateral filtering, which preserve critical features while eliminating distortions. Data augmentation methods, such as rotations and scaling, further improve adaptability to diverse visual scenarios, and GANs can generate synthetic data to address training data shortages.

\textbf{Noise filtering and enhancement:} Tang et al. \cite{tang2020data} designed an image cleaning method based on Gaussian blur and bilateral filtering. Gaussian blur effectively removes random noise, while bilateral filtering retains edge details while removing low-frequency interference. This method performed exceptionally well in processing medical images, providing high-quality inputs for disease diagnosis and treatment planning.

\textbf{Data augmentation:} Whang et al. \cite{tae2019data} extended training datasets using data augmentation techniques such as rotation, scaling, and color adjustments to simulate real-world scene variations. Their method was validated in autonomous driving visual systems, significantly improving model adaptability in complex environments.

\textbf{GANs:} Ilyas et al. \cite{chu2016data} explored the potential of GANs in generating image samples. Their method is particularly useful in situations with insufficient data, generating high-quality image samples to augment datasets and significantly improving model recognition performance for minority classes. This technology has been applied in ecological monitoring and remote sensing image analysis.

\subsubsection{Audio Data Denoising and Completion}
Audio data cleaning involves removing noise, filling gaps, and enhancing the clarity of audio signals to ensure effective SemCom. Methods such as spectral subtraction combined with DL models are used to isolate speech from background noise. WaveNet-based techniques are employed to restore missing segments and improve audio fidelity, ensuring reliable performance in real-time auditory applications.

\textbf{Denoising and separation:} Chu et al.\cite{li2021cleanml}  proposed an audio denoising method combining spectral subtraction and DL models. Their method excelled in voice assistant and multilingual speech recognition systems by separating speech signals from background noise, enhancing the clarity of audio data.

\textbf{Signal enhancement and completion:} Rajan \cite{rajan2024data} used the WaveNet model to complete missing audio segments and combined adaptive filtering technology to enhance the quality of low-sampling-rate audio signals. Their method was successfully applied in remote communication and real-time dialogue scenarios.

\subsubsection{Comprehensive Optimization of Structured Data}  
Structured data cleaning ensures logical consistency and integrity by detecting and correcting errors in datasets. Techniques like probabilistic reasoning tools identify and impute missing values, while entity matching algorithms remove duplicate records. These processes are essential for maintaining data reliability in critical fields such as healthcare and finance.

\textbf{SemCom for poisoning attack mitigation:} Lotfi et al. \cite{lotfi2025vmguard} proposed VMGuard, a reputation-based framework for detecting data poisoning attacks in vehicular Metaverse systems. By integrating SemCom and subjective logic modeling, VMGuard evaluates the trustworthiness of IoT devices, ensuring reliable data collection and mitigating adversarial tampering in real-time semantic data streams.

\textbf{Outlier and missing value handling:} Lee et al. \cite{lee2021survey} developed HoloClean, a probabilistic reasoning-based tool for outlier detection and missing value imputation. This tool is particularly suitable for financial and healthcare data analysis and significantly improves data consistency.

\textbf{Duplicate record removal:} Rajan \cite{rajan2024data} developed DeepMatcher, a DL-driven entity matching tool that efficiently identifies duplicate records in large datasets. This tool has been widely applied in e-commerce recommendations and customer relationship management.

\textbf{Logical consistency constraints:} Ilyas et al. \cite{chu2016data} proposed a dynamic constraint validation method that combines logical rules and ML. By dynamically verifying functional and inclusion dependencies, their method significantly enhances data's logical integrity.

\subsection{Robust Learning}
\begin{figure*}[htbp]
\centering
\vspace{0mm}
\includegraphics[width=\linewidth]{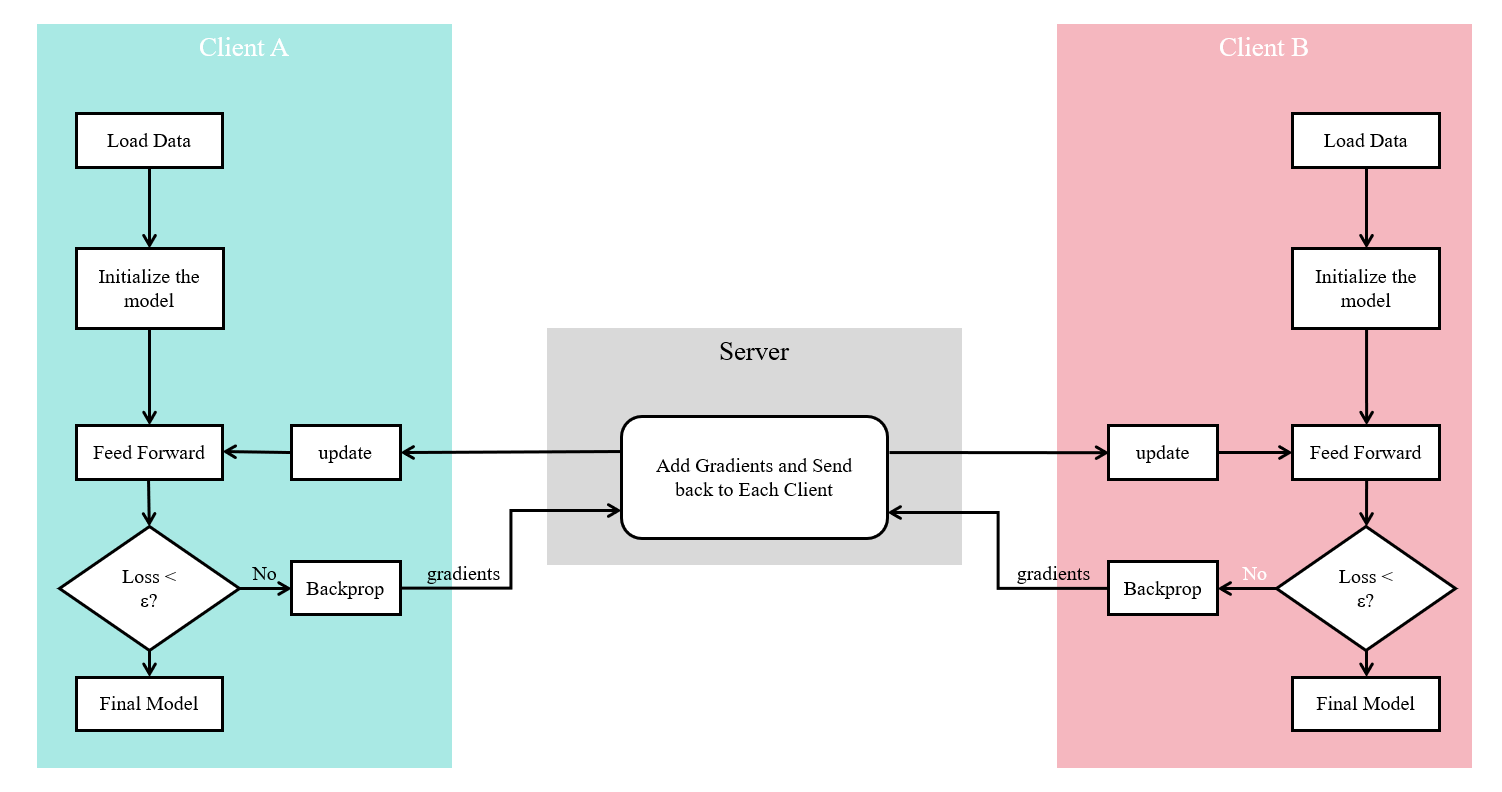}
\caption{Diagram of FL \cite{zhang2020enabling}, where data is processed in a decentralized manner by multiple clients, each performing forward propagation, model updating, and backpropagation before sending gradients to the server for aggregation. This decentralized approach ensures data privacy by allowing model training on local devices without the need to share raw data. The server then aggregates the gradients from the clients to update the global model, which is sent back for further iterations, leading to a more robust and generalized model.}

\label{fig423}
\end{figure*}

The design characteristics of FL, as shown in Fig. \ref{fig423}, make FL face numerous challenges in practical applications, such as noise interference, data heterogeneity, malicious clients, and low communication efficiency. These challenges not only hinder the improvement of model performance but also pose higher demands on system robustness and efficiency. The following text, some typical papers as shown in the table \ref{robustlearning}, reviews the latest advances in the robustness of federated learning, and presents possible directions for improvement as well as potential applications in SemCom.

\begin{longtable}{|A{2cm}|A{0.6cm}|P{9.5cm}|}
\caption{\footnotesize{List of Representative Robust Learning Schemes}} 
\label{robustlearning}\\
\hline
\textbf{Sub-class} & \textbf{Ref.} & \textbf{Descriptions, including} $\circledast$: \textbf{contributions; }

$\circleddash$: \textbf{trade-offs;} 
$\boxplus$: \textbf{performance metrics;}

$\Cap$: \textbf{applied in SemCom or promising for SemCom} \\
\hline
\endfirsthead
\hline
\textbf{Sub-class} & \textbf{Ref.} & \textbf{Descriptions, including} $\circledast$: \textbf{contributions; }

$\circleddash$: \textbf{trade-offs;} 
$\boxplus$: \textbf{performance metrics;} 

$\Cap$: \textbf{applied in SemCom or promising for SemCom} \\
\hline
\endhead
\hline
\endfoot
\hline
\endlastfoot

Robust Aggregation & \cite{pillutla2022robust} &
$\circledast$: Propose a geometric median-based aggregation method using the Weiszfeld algorithm to mitigate malicious client effects. 

$\circleddash$: Robustness and communication efficiency.

$\boxplus$: Model accuracy and convergence speed.

$\Cap$: Promising for distributed model training in SemCom.\\
\hline

Robust Aggregation & \cite{ghosh2019robust} &
$\circledast$: Introduce Trimmed Mean to exclude extreme updates, enhancing Byzantine attack resistance.

$\circleddash$: Robustness and communication efficiency.

$\boxplus$: Misclustering rate and estimation error.

$\Cap$: Promising for filtering model updates in SemCom.  \\
\hline

Data Heterogeneity & \cite{fang2022robust} &
$\circledast$: Develop a noise-tolerant loss function and dynamic weighting for Non-IID data.

$\circleddash$: Robustness and communication efficiency.

$\boxplus$: Classification accuracy and robustness against noise.

$\Cap$: Promising for personalized model optimization in SemCom.  \\
\hline









Noise \& Malicious Clients & \cite{lyu2022privacy} &
$\circledast$: Geometric median-based aggregation with trimming to filter malicious updates.

$\circleddash$: Privacy protection and model utility.

$\boxplus$: Privacy leakage resistance and model accuracy.

$\Cap$: Promising for privacy-preserving and robust collaborative learning in SemCom systems. \\
\hline

Robustness \& Communication & \cite{sattler2019robust} &
$\circledast$: Sparse Ternary Compression reduces communication costs via parameter sparsification and quantization.

$\circleddash$: Compression efficiency and model convergence speed.

$\boxplus$: Model accuracy and communication overhead reduction.

$\Cap$: Promising for bandwidth-efficient collaborative learning in SemCom. \\
\hline

Robustness \& Communication & \cite{ang2020robust} &
$\circledast$: Worst-case optimization with noise-aware regularization for communication robustness.

$\circleddash$: Robustness and convergence speed.

$\boxplus$: Promising for ensuring reliable model updates in noisy SemCom.

$\Cap$: Promising for optimizing noisy environments in SemCom, improving model robustness under low-quality communication conditions. \\
\hline
\end{longtable}

\subsubsection{Robust Aggregation Methods}
The model aggregation process in FL is critical as it determines the performance of the global model. However, in the presence of noise, malicious clients, or device failures, traditional methods such as weighted averaging or arithmetic mean are highly susceptible to outliers, leading to significant degradation in the global model's performance. To address this, researchers have proposed a series of robust aggregation methods.

\textbf{Geometric median aggregation:} Geometric median aggregation is a classical approach to enhance robustness in high-dimensional spaces. Pillutla et al. \cite{pillutla2022robust} introduced a robust federated aggregation method based on the geometric median, leveraging the Weiszfeld algorithm to compute the geometric median efficiently. This approach successfully mitigates the adverse effects of malicious updates on the global model. Experimental results demonstrate that the proposed method maintains high performance even with a high proportion of malicious clients, showcasing significant advantages in robustness and model accuracy over traditional FedAvg, especially in highly polluted environments.

\textbf{Trimmed mean strategy:} The Trimmed Mean method offers a simplified yet effective aggregation strategy by excluding extreme values from model updates, thereby reducing the impact of outliers. Ghosh et al. \cite{ghosh2019robust} demonstrated that the Trimmed Mean method performs exceptionally well against Byzantine attacks. By discarding a fixed proportion of maximum and minimum values, this method significantly enhances robustness in adverse environments while avoiding the risk of extreme values disrupting global model updates.

\subsubsection{Addressing Client Data Heterogeneity}
The highly non-independent and identically distributed (Non-IID) nature of client data poses a significant challenge to global model training in FL. To accommodate such data heterogeneity, researchers have proposed personalized optimization strategies and client grouping methods.

\textbf{Robust heterogeneous FL framework:} Fang et al. \cite{fang2022robust} developed a Robust Heterogeneous FL framework tailored for heterogeneous data environments. By adaptively adjusting the local training process, this framework enables each client's model to better align with its unique data distribution. Notably, a noise-tolerant loss function is introduced during local training to alleviate the adverse effects of noisy data on the global model. Moreover, a dynamic weighting mechanism balances the contributions of different clients to global aggregation, thereby improving the robustness of the global model.

\textbf{Cluster-based client grouping:} Ghosh et al. \cite{ghosh2019robust} proposed a clustering-based method to address challenges associated with heterogeneous data. By analyzing the similarities among local client data, clients with similar distributions are grouped into clusters. Within each cluster, clients share and optimize their local models, effectively mitigating the impact of data heterogeneity. Experimental results indicate that this clustering-based grouping method significantly enhances global model performance in Non-IID data environments.

\subsubsection{Mitigating the Impact of Noise and Malicious Clients}
\begin{figure*}[htbp]
\centering
\vspace{0mm}
\includegraphics[width=0.85\linewidth]{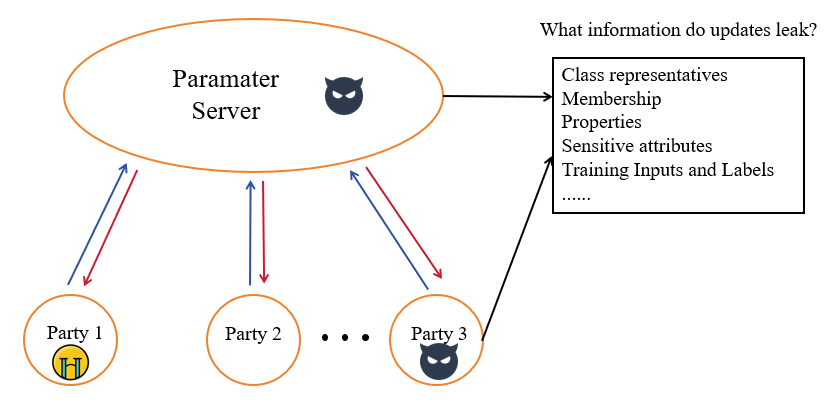}
\caption{Malicious client attacks in FL \cite{lyu2022privacy}, where malicious clients attempt to leak sensitive information through their updates, affecting the parameter server and potentially compromising the privacy of the training process. These attacks could expose critical data such as class representatives, membership details, properties of the data, sensitive attributes, or even training inputs and labels, ultimately undermining the security and trustworthiness of the FL system.}
\label{fig422}
\end{figure*}

To counter malicious clients and noisy data, several mechanisms have been developed to dynamically adjust client contributions, filtering out noise or malicious updates. Fig. \ref{fig422} illustrates common malicious client attacks in FL. 

\textbf{Client confidence re-weighting mechanism:} Fang et al. \cite{fang2022robust} introduced a Client Confidence Re-weighting mechanism to dynamically adjust each client's weight in model aggregation. By evaluating the training quality of each client's local model on its private data, this mechanism  quantifies the actual contribution of each client to global model training. For underperforming clients, their weights are reduced to mitigate the adverse effects of noise and malicious clients. This mechanism not only enhances the robustness of the RHFL framework but also maintains strong model performance in high-noise environments.

\textbf{Geometric median-based aggregation and filtering:} Lyu et al. \cite{lyu2022privacy} proposed robust aggregation methods based on the geometric median, effectively mitigating the impact of outliers on the model by replacing traditional mean aggregation. Further developments included trimming and filtering mechanisms, such as magnitude clipping of model updates and geometric median-based aggregation, to filter out malicious updates. These strategies significantly improve global model performance and robustness, especially in environments with high proportions of malicious clients.

\subsubsection{Balancing Robustness and Communication Efficiency}
Communication costs are a critical issue in FL applications. Frequent communication, particularly in scenarios involving numerous devices, can significantly increase bandwidth and energy consumption. Consequently, some studies focus on developing solutions that are both communication-efficient and robust.

\textbf{Sparse ternary compression:} Sattler et al. \cite{sattler2019robust} proposed Sparse Ternary Compression, a method specifically designed to reduce communication costs. In this method, model updates are sparsified by selecting only essential parameters for transmission and further compressing data using quantization techniques. This method not only significantly reduces communication overhead per training round but also demonstrates high robustness in Non-IID data environments. Compared with traditional FedAvg, this method effectively balances model accuracy and bandwidth demands, making it highly suitable for resource-constrained settings.

\begin{figure*}[htbp]
\centering
\vspace{0mm}
\includegraphics[width=\linewidth]{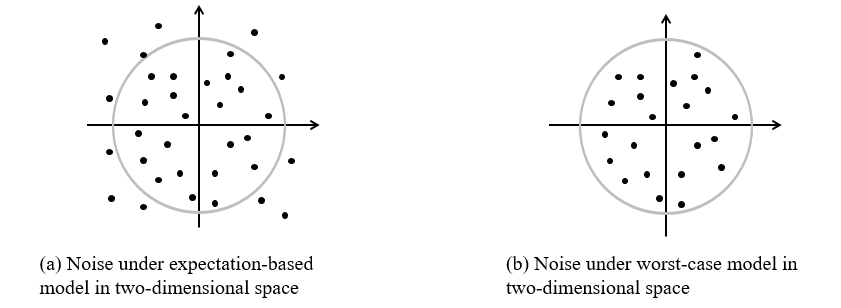}
\caption{Comparison of noise distributions under the expectation-based model and the worst-case model \cite{ang2020robust}, where panel (a) visualizes the noise characteristics under an expectation-based model in two-dimensional space, and panel (b) depicts noise under the worst-case model in the same space. The analysis demonstrates that the expectation-based model effectively reduces the noise effects in the loss function, enhancing the global model's robustness against communication noise. Experimental results show that the worst-case model also achieves comparable performance while maintaining strong model accuracy even in challenging communication scenarios.}
\label{fig421}
\end{figure*}

\textbf{Worst-case optimization for communication robustness:} Ang et al. \cite{ang2020robust} analyzed an expectation-based model that incorporates noise into statistical modeling. They designed a regularization approach that introduces noise effects into the local loss function, enhancing the global model's robustness against communication noise. For worst-case scenarios, researchers adopted sampling and successive convex optimization techniques to address non-convex optimization challenges. Fig. \ref{fig421} visualises the noise characteristics under two different noise models in 2D space. Experimental results reveal that these designs outperform traditional methods in reducing loss function values and improving prediction accuracy. Additionally, convergence analysis demonstrates that their convergence rates are comparable to centralized learning while maintaining strong model performance under worst-case communication noise.

\begin{les}
Improving robustness in FL requires a multifaceted approach, encompassing optimization of aggregation methods \cite{pillutla2022robust,ghosh2019robust}, addressing data heterogeneity \cite{fang2022robust,ghosh2019robust}, filtering noise and malicious clients \cite{fang2022robust,lyu2022privacy}, and balancing communication efficiency with model performance \cite{sattler2019robust,ang2020robust}. These methods enhance the practicality of FL and lay a solid foundation for its deployment in complex application scenarios. In SemCom systems, these technologies further enhance the efficiency and reliability of distributed model training, supporting the development of more intelligent communication networks.
\end{les}

 \subsection{Defensive Methods of Backdoor Attacks}
\begin{figure*}[t]
\centering
\vspace{0mm}
\includegraphics[width=0.7\linewidth]{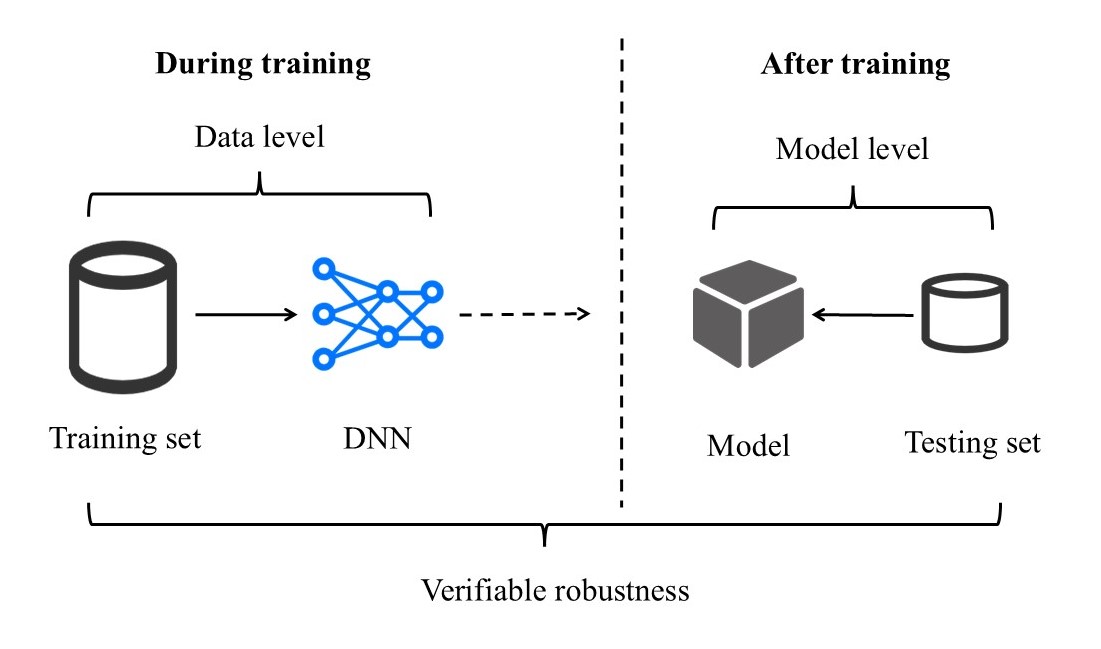}
\caption{Classification of neural network backdoor
defense \cite{Jiang2024OverviewofDeepNeuralNetworkBackdoorDefense}, where the process of data collection and model training is referred to as model training, while the stages of model testing and model deployment are considered subsequent to model training. The phases of training and post-training correspond to two distinct types of neural network backdoor attacks: one in which the user acquires training data from a third party, and the other in which the user obtains a model from a third party.}
\label{fig431}
\end{figure*}

As illustrated in Fig. \ref{fig431}, backdoor attacks of DL models are mainly achieved in two ways: one is to adulterate the training data with malicious samples, and the other one is to directly modify the model structure. These operations embed a backdoor in the model, causing the model unusually sensitive to the presence of specific triggers. Extend to SemCom, the attacker can embed triggers to alter the semantic symbols transmitted by the transmitter and modify the symbols in the receiver's dataset to align with targets and labels of its own choosing, which shows the capability to manipulate the semantics of reconstructed symbols in SemCom.To defend against backdoor attacks, researchers have proposed different defense methods of various focuses at different stages of the model. This section will analyze these methods both on the data level and model level.

The below, as shown in Table \ref{backdoorattacks}, the following content, based on multiple academic studies, delves into various defensive methods of backdoor attacks in different stages and their potential applications in SemCom.

\begin{longtable}{|A{2cm}|A{0.6cm}|P{9.5cm}|}
\caption{\footnotesize{List of Representative Backdoor Defense Schemes}}
\label{backdoorattacks}\\
\hline
\textbf{Sub-class} & \textbf{Ref.} & \textbf{Descriptions, including} $\circledast$: \textbf{contributions;}

$\circleddash$: \textbf{trade-offs;} $\boxplus$: \textbf{performance metrics;} 

$\Cap$: \textbf{applied in SemCom or promising for SemCom} \\
\hline
\endfirsthead

\hline
\textbf{Sub-class} & \textbf{Ref.} & \textbf{Descriptions, including} $\circledast$: \textbf{contributions;} 

$\circleddash$: \textbf{trade-offs;} $\boxplus$: \textbf{performance metrics;} 

$\Cap$: \textbf{applied in SemCom or promising for SemCom} \\
\hline
\endhead

\hline
\endfoot

\hline
\endlastfoot
\textbf{Data Processing} & \cite{tran2018spectral} &
$\circledast$: Proposes a detection algorithm based on PCA of spectral signatures to detect poisoned samples. Uses singular value decomposition to calculate anomaly scores for each sample. 

$\circleddash$: Detection precision and false positive rate, model performance and defense capability.

$\boxplus$:  Norm of mean, covariance matrix. 

$\Cap$: Promising for detecting poisoned data in SemCom systems.
\\
\hline

\textbf{Data Processing} & \cite{chen2018detecting} &
$\circledast$: Activates clustering defense method through analyzing network activations to detect and classify toxic data using dimensionality reduction and clustering.

$\circleddash$: Dimensionality reduction and computational complexity.

$\boxplus$: Silhouette score and ExRe score.

$\Cap$: Promising for filtering toxic data in large-scale SemCom systems.
\\
\hline

\textbf{Data Processing} & \cite{geiping2021doesn} &
$\circledast$: Introduces Poison Immunity, extending adversarial training for data augmentation, creating poisoned samples dynamically to help the model recognize backdoor attacks during training.

$\circleddash$: Potential for slower training times and reduced model convergence.

$\boxplus$: Detection accuracy, detection rate.

$\Cap$: Promising for improving model resilience in SemCom systems under adversarial conditions.
\\
\hline

\textbf{Data Processing} & \cite{gao2019strip} &
$\circledast$: STRong Intentional Perturbation (STRIP) performs perturbations on input data during inference to distinguish Trojan attacks through entropy analysis.

$\circleddash$: Computational cost and detection effectiveness.

$\boxplus$: Entropy distribution, F1 score, FAR, and FRR.

$\Cap$: Promising for real-time filtering of toxic data in SemCom systems.
\\
\hline

\textbf{Model Identification and Construction} & \cite{huang2020one} &
$\circledast$: One-Pixel Signature model uses a meta-classifier to identify whether a model has been poisoned by comparing the one-pixel signature of clean and backdoored models.

$\circleddash$: Computational cost and model complexity.

$\boxplus$: Detection Rate.

$\Cap$: Promising for identifying and repairing poisoned models in SemCom.
\\
\hline

\textbf{Model Identification and Construction} & \cite{liu2018fine} &
$\circledast$: Fine-Pruning method combines pruning inactive neurons and fine-tuning to eliminate backdoors while preserving model performance.

$\circleddash$: Pruning and fine-tuning operation and computation cost.

$\boxplus$: Classification accuracy, utility matrix.

$\Cap$: Promising for reconstructing models post-backdoor in SemCom.
\\
\hline

\textbf{Model Identification and Construction} & \cite{zhou2024backdoor} &
$\circledast$: Proposes a split learning framework to secure data transmission and prevent backdoor injections during training.

$\circleddash$: Model Performance and defense capability.

$\boxplus$: PSNR, F1-score, and difference dc. 

$\Cap$: Applied for securing SemCom systems from backdoor attacks during transmission.
\\
\hline

\end{longtable}

\subsubsection{Data Processing}


For the training phase, the importance of data protection focuses on how the defender obtains training data from external sources. In these cases, an attacker may attack the defender by contaminating the training dataset with malicious samples triggers. If the defender fails to identify these malicious samples and directly uses the data for training, it may lead to backdoor implants in the model. To against the attackers, there are two most direct and effective method based on the training dataset: the first one is to identify and remove the malicious samples in the dataset, while the second one is to reduce the impact of the toxic data, which can prevent the backdoor attack effectively. For the inference Phase, the importance of data protection focuses on how to recognize and pre-process the obtained data. 

\textbf{Filtering Toxic Data:} To filter toxic data, there are three main methods focusing on spectral signatures, clustering and gradient as below.

\begin{itemize}
\item Spectral Signatures based Defense. One of the effective ways to recognize a backdoor attack is based on its spectral signature. Spectral signature refers to the fact that the samples of a backdoor attack have a distributional difference from the normal samples in the feature representation space, and this difference will be reflected in the spectrum of the covariance matrix. 
Based on the above theoretical foundation, Tran \cite{tran2018spectral} proposed a detection algorithm based on principal component analysis of spectral signatures. In the trained neural network, a learned representation of each input for each label is extracted. The covariance matrix of these representations is then subjected to a singular value decomposition to obtain a series of eigenvalues and corresponding eigenvectors. The results of the decomposition are used to calculate the anomaly score for each sample, which means the size of the projection of each sample in the direction of the top singular vector of the covariance matrix. The feature representation of a contaminated sample will have a larger projection on one of the principal components of the covariance matrix. In the other words, the samples with the highest scores are the contaminated data points. Then, removing the toxic data and retraining the network can effective against backdoor attacks.
However, the above methods have some disadvantages in some aspects. The defense can only be detected when one of the spectral features of the poisoned sample is large enough, otherwise it is not effective to identify the poisoned sample. Hayase \cite{hayase2021spectre} proposed a new method to defense against general backdoor attacks, called Spectral Poison ExCision Through Robust Estimation. It estimates the mean and covariance of clean data through robust statistical methods, which makes it possible to accurately estimate the statistical characteristics of normal data even when the data is contaminated. Then the spectral features are amplified by whitening process, and the contaminated samples are removed based on quantum entropy score. Comparing with principal component analysis, the proposed method is able to adapt to different types of backdoor attacks, including the hidden triggers used in some case. It works effectively with a small number of contaminated samples and removes the backdoor completely, which ensures the high efficiency and accuracy in multiple attack scenarios.
\item Clustering based Defense. For contaminated datasets, toxic data can be separated from normal data according to cluster analysis. Activation Clustering is a defense method proposed by Chen \cite{chen2018detecting} et al. The key idea is to analyze the activation mode of network for the training data, to determine whether the data is poisoned and which data points are toxic. For each sample in the training set, AC records the activation output of the hidden layer. These activation outputs are flattened into one-dimensional vectors and classified based on their labels. After dimensionality reduction, these vectors are clustered and analyzed into toxic and normal patterns. The data is analyzed by exclusionary reclassification, silhouette score, size comparison and many other ways to filter out the toxic data. Based on the labeled toxic data, the system can further train the model until convergence to achieve the purpose of backdoor repairing.
One more clustering approach is Heatmap Clustering proposed by Schulth et al. \cite{schulth2022detecting}. This method utilizes Explainable Artificial Intelligence techniques, specifically Layer-wise Relevance Propagation, to generate heatmaps. The heatmap shows the correlation between each pixel point of the input sample and the network predictions, which can be visualized as a two-dimensional array where the color shades indicate the strength of the relevance. By calculating the distance between heatmaps, the Differences between different data points in network decisions are quantified, which helps the model to identify the location of backdoor triggers. The processed heatmaps are categorized using clustering methods to distinguish normal and manipulated data points. Comparing these two clustering methods, Heatmap clustering has a significantly higher computational cost than activation clustering, which is caused by Gromov-Wasserstein to calculate the distances between heatmaps. However, it is more adaptable and robust to different attack patterns and data characteristics.
\item Gradient based defense. For the gradient defense approach, Chan et al. \cite{chan2019poison} use the gradient of the model's loss function with contaminated inputs to extract signals associated with backdoor attacks. The model assumes that there are specific neurons model that are activated only in the presence of the backdoor trigger. As a result, the weights of these neurons are typically much larger than the weights of the normal one, resulting in a relatively large absolute value of the input gradient at the trigger location having . In this way, toxic and clean samples can be effectively separated in the training model.
\end{itemize}

\textbf{Reducing the impact of toxic data:} Despite of filtering toxic data, backdoor attacks can be resisted by reducing the impact of toxic data, which can be achieved by data augmentation and gradient.

\begin{itemize}
\item Data augmentation based defense. Geiping et al. \cite{geiping2021doesn} use adversarial training to achieve the aim of data augmentation, extending the adversarial training framework to the defense of poison data. It proposes a defense strategy named Poison Immunity, where adversarial data augmentation involves generating poisoned samples dynamically. These samples intend to imitate the attacks that an attacker might employ. In this way, the model learns to recognize and defend these attacks during the training process. Also, it uses data augmentation including mixup and cutout to create new samples and improve the generalization ability of the model.
\item Gradient based defense. Hong et al. \cite{hong2020effectiveness} mitigate backdoor attacks through gradient shaping. In other words, when facing the threat of data poisoning, it can be observed that there are significant differences in the gradient characteristics, including the magnitude and direction of the gradient, between contaminated and uncontaminated data. To solve this problem, a gradient shaping technique is employed based on the differential privacy stochastic gradient descent method. Specifically, the technique first limits the magnitude of the gradient through gradient cropping to avoid any individual sample interfering the update of the model parameter too much. Subsequently, it further increases the difficulty for the attacker to manipulate the direction of the gradient by introducing noise into the gradient. According to these measures, even if the training data contains poisoned samples, the parameter updating of the model remains, which cannot be affected significantly by these abnormal samples.
\end{itemize}


\textbf{Filtering toxic data in inference phase:} These defense mechanisms are also capable of screening out toxic samples, but the process takes place in the inference phase of the model rather than in the training phase. Gao et al. \cite{gao2019strip} proposed a defense system called STRong Intentional Perturbation in response to Trojan attacks. This system performs multiple intentionally perturbations of the input and calculates the prediction of the perturbed input. If the entropy value based on the perturbed inputs is low, it means that the input sample is classification inducing, belonging to a part of Trojan attack, whereby malicious samples can be distinguished from normal samples. Subedar et al. \cite{subedar2019deep} proposed two methods to distinguish clean and poisoned samples by quantifying the uncertainty estimation associated with the training model. The strategy is based on adapting a category-specific probability density function to a deep representation of a DNN, thus constructing a generative model capable of characterizing the deep feature space. During the model testing phase, uncontaminated data samples are distinguished by evaluating the log likelihood values of the deep features of the test samples with respect to these probability distributions. In addition, this study employs the mean-field variational inference technique to infer the posterior probability distributions of the model weights and combines it with the Bayesian inference framework to quantify the uncertainty of the model predictions. Specifically, it utilizes the uncertainty measure in Bayesian active learning to assess the informational reciprocity between the posterior probability distributions of the parameters and the predicted output distributions, which provides a quantitative indicator of the model prediction uncertainty.

\textbf{Adding pre-processing module:} Filtering attack data can also be done by introducing a pre-processing module. This approach aims to adjust the trigger properties in the test samples, so that they can no longer match the activation conditions of the backdoor attack. With this modification, the potential threat of toxic samples can be effectively neutralized. At the earliest, Liu et al. \cite{liu2017neural} proposed the scheme of connecting a trained self-encoder to the input data as a pre-processor. The trained self-encoder can correctly recognize normal data and generate ``failure" commands for toxic data. Qiu et al. \cite{qiu2021deepsweep} proposed the DeepSweep framework to automatically evaluate and generate defense methods against backdoor attacks. The framework performs image transformation by Augmentation Library, and utilizes Inference Transformation Policy in the inference phase to pre-process the samples to disrupt the judgment of the triggers, which in turn corrects the outputs of the toxic samples.

\subsubsection{Model Identification and Construction}

The model-level based defense approach is mainly applied to determine whether the model is poisoned or not and repair, which is applied after model training. Also, some researchers are denoted to build reliable cyber-security architectures to resist backdoor attacks.

\textbf{Identifying poisoning models:} To identify the poisoning  model, there are two mainstream method based on trigger and meta-classifier.

\begin{figure*}[t]
\centering
\vspace{-10mm}
\includegraphics[width=\linewidth]{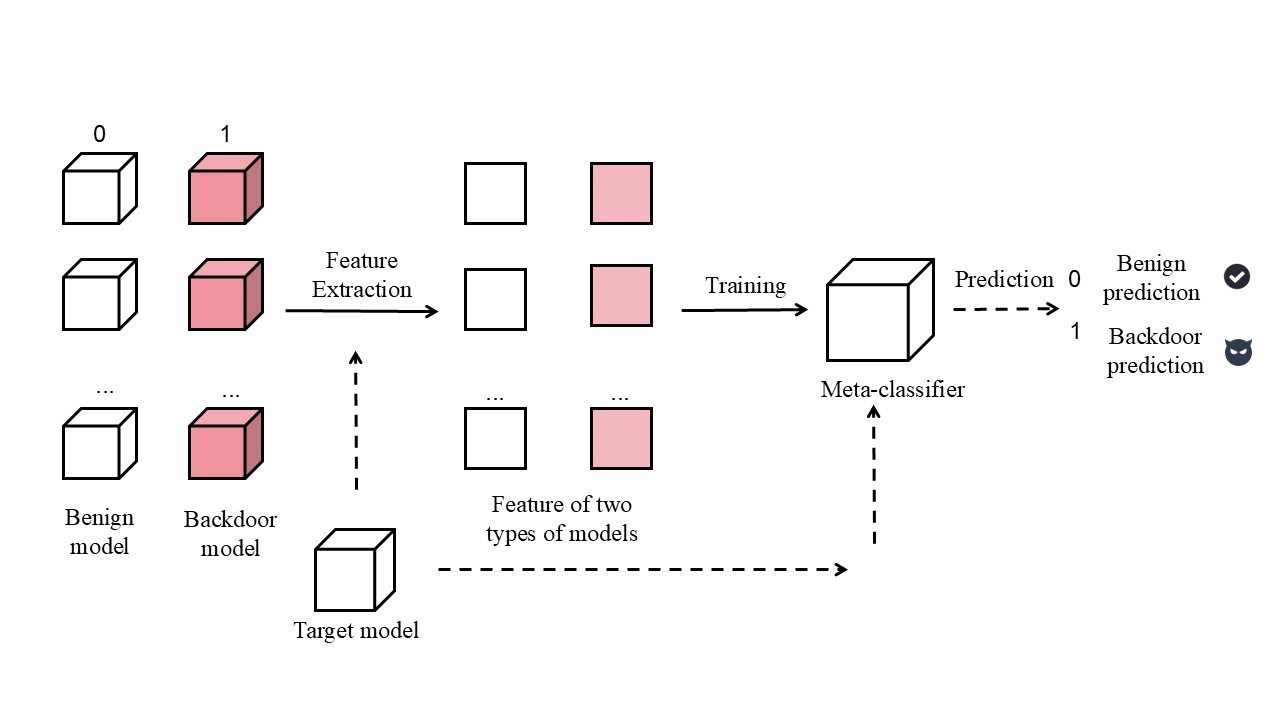}
\caption{The workflow of meta-classifier defense \cite{Jiang2024OverviewofDeepNeuralNetworkBackdoorDefense},
where the data that reflect the characteristics of both benign and backdoored models are extracted firstly; this data includes key information such as the models' internal parameters and output behaviors. Subsequently, these features are utilized to train a meta-classifier, which is a supervised learning model capable of learning to distinguish between benign and backdoored models based on these features. Once trained, the meta-classifier is employed to predict the nature of new model samples, determining whether they are benign or backdoored, with 0 representing a benign model and 1 representing a backdoored model. The entire process forms a closed loop, commencing with feature extraction, proceeding through training and prediction, and ultimately applying to the security detection of the target model. In the event that a backdoored model is detected, the system can implement corresponding measures, such as rejection, retraining, or model repair.}
\label{fig432}
\end{figure*}

\begin{itemize}
\item Trigger based defense. The trigger based defense is mainly used to detect and remove malicious triggers hidden in DNNs instead of removing the backdoor directly. Wang \cite{wang2019neural} proposed a Neural Cleanse approach to design an optimization scheme, which measures the minimum amount of perturbation that converts an arbitrary input sample into a target classification. Based on the measured value of L1 norm, the model finds the malicious triggers and remove them. Based on the idea of trigger defense, researchers also proposed a method to reversely generate the attacker's triggers with the help of Generative Model. Zhu et al. \cite{zhu2020gangsweep} proposed a framework called GangSweep to generate perturbation mask generation to misclassify the input image using GAN. By simulating the behavior of malicious triggers, GangSweep can identify perturbation masks that can trigger backdoor behaviors, for the  specific perturbations could generate specific changes in statistics.
    
\item Meta-classifier based defense. As illustrated in Fig. \ref{fig432}, meta-classifier identifies abnormal pattern by monitoring and analyzing the output of the main classifier. Huang \cite{huang2020one} proposed One-Pixel Signature model based on the principle of meta-classifier. Comparing the one-pixel signature of the clean model and the backdoor model, it can train an outlier detector as a classifier to detect whether a CNN model has been implanted with a backdoor. The meta-classifier makes decisions based on multiple base classifiers to improve the overall classification accuracy through integrated learning methods. It also helps to reduce the risk of overfitting based on the wide range of input data from different models.
\end{itemize}

\textbf{Reconstructing model:} Model reconstruction methods directly target suspicious models to eliminate malicious backdoor. Liu et al. \cite{liu2018fine} propose a defense method called “Fine-Pruning”, which consists of two methods: pruning and fine-tuning. The principle of utilizing the pruning is to remove neurons that are inactive under normal input data, for these neurons may be utilized by attackers to implant a backdoor. Fine-tuning is based on the principle of tuning the pre-trained model, which depends on adjusting the weights of the local model to mitigate the impact on backdoor attacks. Combining these two approaches, the strategy in this paper can reconstruct the model to effectively resist the backdoor attack. It can not only maintains the accuracy against clean inputs, but also significantly reduces the success rate of the backdoor attack, especially when complex pruning-aware attacks occur. To purify DNNs contaminated by backdoor attacks, Wu et al. \cite{wu2021adversarial} proposed a model reconstructing method of Adversarial Neuron Pruning. The core idea is to add perturbations to neurons and prune sensitive neurons to eliminate backdoor attacks. This method can reshape the model based on a small amount of clean data, and has wide applications in resisting backdoor attacks.

\textbf{Reliable model framework in Semcom:} Zhou et al. \cite{zhou2024backdoor} explored backdoor attacks that specifically target deep learning-based SemCom systems. Given that current studies on backdoor attacks are not adapted to the unique context of SemCom, a novel backdoor attack model, referred to as backdoor attack on semantic symbols, is introduced. Based on this model, appropriate countermeasures are proposed. In particular, a training framework is developed to mitigate the risks of Backdoor Attack on Semantic Symbols. Furthermore, defense strategies utilizing reverse engineering and pruning techniques are formulated to safeguard SemCom systems from backdoor attacks. Simulation results validate the effectiveness of both the proposed attack model and the corresponding defense mechanisms.

\begin{les}
Backdoor attacks on DL models are primarily carried out by contaminating training data with malicious samples or altering the model structure, resulting in abnormal sensitivity to specific triggers. Extending this to SemCom, attackers can embed triggers to manipulate the semantics of transmitted symbols. To counter these attacks, researchers have proposed various defense methods targeting different stages of the model from the data and model levels. The data processing techniques can be achieved during training \cite{tran2018spectral,hayase2021spectre,chen2018detecting,schulth2022detecting,chan2019poison,geiping2021doesn,hong2020effectiveness} and inference stages \cite{gao2019strip,subedar2019deep,liu2017neural,qiu2021deepsweep}. The model identification and construction techniques include identifying poisoning models \cite{wang2019neural,zhu2020gangsweep,huang2020one}, pruning \cite{liu2018fine,wu2021adversarial}, and reliable model framework \cite{zhou2024backdoor},
\end{les}

\subsection{Adversarial Training}
\begin{figure*}[t]
\centering
\vspace{-10mm}
\includegraphics[width=\linewidth]{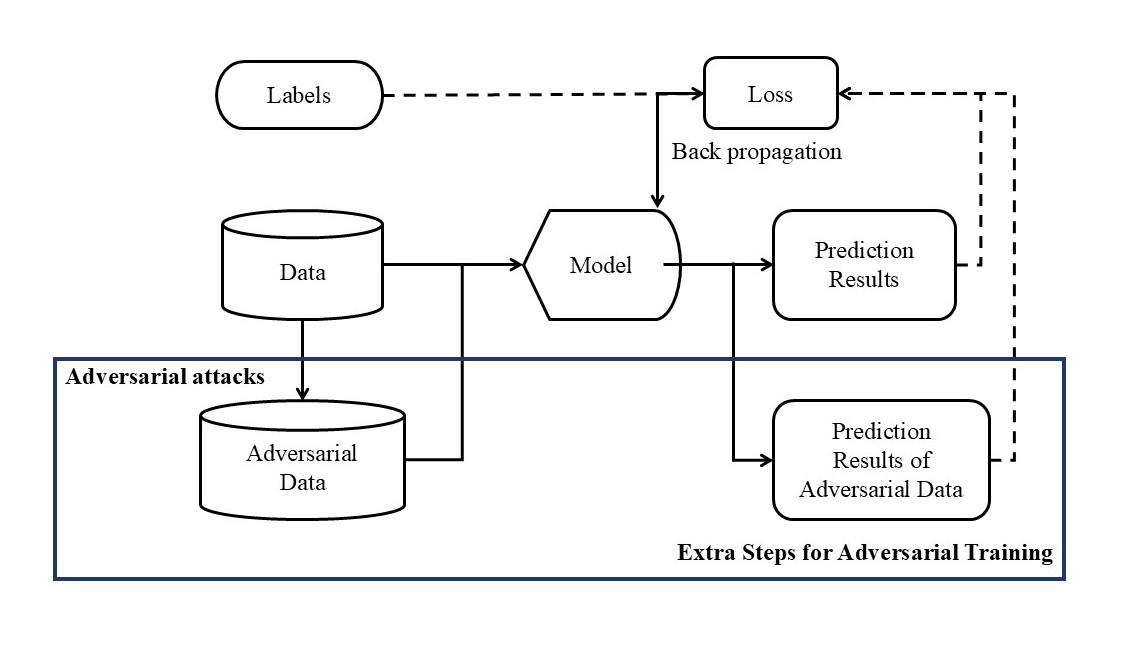}
\caption{Process of adversarial training \cite{zhao2022adversarial},
where adversarial training enhances traditional training methods by incorporating additional steps into the training process, including the generation of adversarial data and the prediction of results.Initially, data is fed into the model alongside their corresponding labels to produce prediction results, which are then compared against the true labels to calculate the loss.  This loss is subsequently used to perform backpropagation, adjusting the model's parameters to improve its accuracy on the given data.  Additionally, the model undergoes an extra step where it is trained with adversarial data, which has been intentionally modified to challenge the model's robustness.  This adversarial data is generated through adversarial attacks designed to deceive or mislead the model.  The model's performance on both standard and adversarial data is assessed, with the prediction results of adversarial data being a critical component in evaluating the model's resilience against such attacks.  This dual training process aims to enhance the model's ability to generalize well and maintain accuracy under both normal and adversarial conditions, thereby strengthening its overall robustness against potential threats in real-world applications.}
\label{fig441}
\end{figure*}

To resist adversarial attacks, adversarial training is favored by researchers for its remarkable performance. As illustrated in Fig. \ref{fig441}, adversarial training is realized by adding adversarial attacks to the training samples. In SemCom, semantic adversarial samples are generated by adding noise to the semantic information to train the model together with the semantic information in the original training set.

The below, as shown in Table \ref{adversarialtraining}, the following content, based on multiple academic studies, delves into various adversarial training methods for data level and model level, and their potential applications in SemCom.

\begin{longtable}{|A{2cm}|A{0.6cm}|P{9.5cm}|}  
\caption{\footnotesize{List of Representative Adversarial Training Schemes}} 
\label{adversarialtraining}\\
\hline
\textbf{Sub-class} & \textbf{Ref.} & \textbf{Descriptions, including} $\circledast$: \textbf{contributions; }

$\circleddash$: \textbf{trade-offs;} 
$\boxplus$: \textbf{performance metrics;}

$\Cap$: \textbf{applied in SemCom or promising for SemCom} \\
\hline
\endfirsthead

\hline
\textbf{Sub-class} & \textbf{Ref.} & \textbf{Descriptions, including} $\circledast$: \textbf{contributions; }

$\circleddash$: \textbf{trade-offs;} 
$\boxplus$: \textbf{performance metrics;} 

$\Cap$: \textbf{applied in SemCom or promising for SemCom} \\
\hline
\endhead

\hline
\endfoot

\hline
\endlastfoot
Adversarial Sample Generation & \cite{hu2022robust} &
$\circledast$: Introduces a robust framework for SemCom systems using the iterative Fast Gradient Symbol Method to generate semantic noise and improve robustness to adversarial samples. The method incorporates weighted perturbation adversarial training.

$\circleddash$: Robustness and transmission efficiency, semantic noise and system stability.

$\boxplus$: Classification accuracy and SNR.

$\Cap$: Applied in enhancing robustness in real-world SemCom systems.
\\
\hline

Adversarial Sample Generation & \cite{kang2023adversarial} &
$\circledast$: Proposes the Semantic Distance Minimization mechanism to dynamically generate adversarial samples by adding semantic noise to minimize the semantic distance from original data, optimizing robustness using Kullback-Leibler divergence.

$\circleddash$:  Robustness and transmission efficiency.

$\boxplus$: Model accuracy.

$\Cap$: Applied in improving security and reliability in SemCom systems.
\\
\hline

Adversarial Sample Generation & \cite{nan2023physical} &
$\circledast$: Develops the SemAdv framework and SemMixed training strategy to generate semantically-oriented adversarial samples using a multilayer perceptron and hybrid perturbation generator.

$\circleddash$: Model size and inference time, transmission efficiency and reconstruction performance.

$\boxplus$: SSIM and PSNR.

$\Cap$: Applied in robust SemCom systems in wireless environments.
\\
\hline

Adding adversarial regularization & \cite{luo2023encrypted} &
$\circledast$: Introduces adversarial regularization with encryption for SemCom systems to improve robustness while preserving privacy. Defines distinct loss functions for encrypted and unencrypted modes.

$\circleddash$: Model complexity and encryption accuracy.

$\boxplus$: BLEU score.

$\Cap$: Applied in privacy-preserving SemCom systems.
\\
\hline

Adding adversarial regularization & \cite{wang2019improving} &
$\circledast$: Proposes Misclassification Aware Adversarial Training (MART) that distinguishes between correctly and incorrectly classified adversarial samples, offering robustness against adversarial examples.

$\circleddash$: Complexity in optimizing adversarial risk and accuracy.

$\boxplus$: Accuracy.

$\Cap$: Applied in task-oriented SemCom applications.
\\
\hline

Adaptive Disturbance Constraint & \cite{ding2018mma} &
$\circledast$: Develops the Max-Margin Adversarial Training method that adapts perturbation values for each sample to improve robustness and accuracy.

$\circleddash$: Computational complexity and performance.

$\boxplus$: l2 norm.

$\Cap$: Promising for improving classification accuracy and robustness in SemCom systems.
\\
\hline

Adaptive Disturbance Constraint & \cite{cheng2020cat} &
$\circledast$: Introduces Customized Adversarial Training (CAT) to adaptively update perturbation levels for each sample, enhancing robustness against adversarial attacks.

$\circleddash$: Computational complexity and performance.

$\boxplus$: Accuracy.

$\Cap$: Applied in task-oriented SemCom applications.
\\
\hline

Matching Pluggable Modules & \cite{he2024secure} &
$\circledast$: Proposes pluggable modules for SemCom systems to counter adversarial attacks by employing a trainable Adversarial Residual Network and corresponding mitigation network at both transmitter and receiver sides.

$\circleddash$: Computational complexity and performance.

$\boxplus$:  MSE and accuracy.

$\Cap$: Applied in enhancing security in SemCom systems.
\\
\hline
\end{longtable}

\subsubsection{Adversarial Sample Generation}

Since the introduction of adversarial examples by Szegedy and Goodfellow et al.'s demonstration \cite{goodfellow2014explaining} that the high-dimensional linearity of neural networks is a fundamental cause of their vulnerability to adversarial attacks, a series of adversarial example generation methods have emerged. These methods are generally characterized by low computational cost and effective performance. 
The typical process for generating adversarial examples follows a series of key procedures. First, a network classifier is trained using normal data. Next, perturbations are introduced into the original samples to create adversarial examples. These perturbed samples are then input into the classifier, and the resulting classification error is computed. Subsequently, adversarial training is conducted to assess the classifier's robustness against adversarial examples. Finally, the classification process is repeated to further refine the model's resilience.

\textbf{Iterative fast gradient symbol method:} Hu et al. \cite{hu2022robust} presents a robust framework for developing end-to-end SemCom systems that can effectively withstand semantic noise. The approach utilizes an iterative Fast Gradient Symbol Method to dynamically generate semantic noise, allowing for realistic modeling of noise during training. To address this challenge, a novel weighted perturbation adversarial training method is introduced, which incorporates noise-affected samples into the training dataset. This method solves a complex optimization problem, enhancing the model's ability to resist semantic noise. By integrating these techniques, the proposed framework significantly improves the model's robustness, making it more reliable in real-world communication environments.

\begin{figure*}[t]
\centering
\vspace{-10mm}
\includegraphics[width=0.7\linewidth]{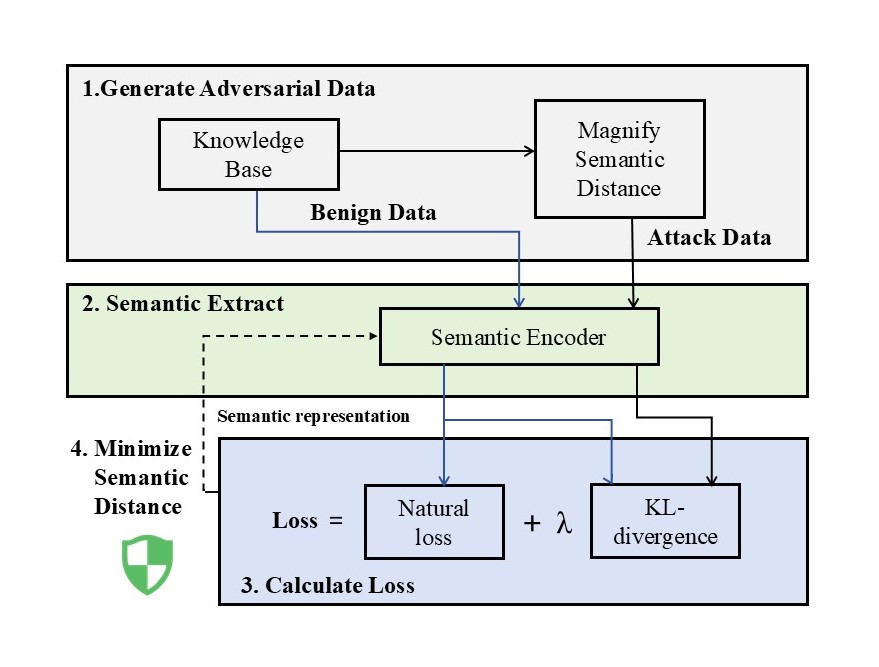}
\caption{Semantic distance minimization mechanism \cite{kang2023adversarial}, where lines of the same color represent the same data stream, while the dashed line indicates that the semantic encoder is updated based on the loss function.The diagram outlines a four-step process for generating and utilizing adversarial data to enhance a model's robustness. In step 1, adversarial data is produced by leveraging a knowledge base and benign data to magnify the semantic distance, creating attack data intended to challenge the model. In step 2, a semantic encoder is employed to extract semantic representations from both the benign and adversarial data. Step 3 involves calculating the loss, which combines the natural loss of the model with a Kullback-Leibler (KL) divergence term, weighted by a factor $\lambda$, to quantify the difference between the semantic representations of the benign and adversarial data. Step 4 aims to minimize this semantic distance, thereby reducing the disparity between the representations and enhancing the model's ability to generalize across both normal and adversarial conditions.  This process is crucial for improving the model's resilience against potential semantic attacks and ensuring its reliability in various scenarios.}
\label{fig442}
\end{figure*}

\textbf{Semantic distance minimization mechanism:} As shown in Fig. \ref{fig442}, Kang et al. \cite{kang2023adversarial} proposes a defense strategy called the Semantic Distance Minimization mechanism to enhance the resilience of SemCom systems against adversarial attacks. The proposed method incorporates adversarial samples during training, enabling the model to learn how to counteract carefully crafted perturbations. The generation of these adversarial samples is based on maximizing the semantic distance from the original data, achieved by adding imperceptible semantic noise to the original samples. the proposed method employs Kullback-Leibler divergence to measure the semantic distance between the original and adversarial samples and optimizes model performance through a loss function that combines natural and robust losses. This approach increases the model’s robustness to adversarial samples while preserving accuracy on normal data. By dynamically generating and handling adversarial samples during training, the proposed method ensures stable performance across varying SNR conditions, effectively improving the security and reliability of SemCom systems in the face of adversarial attacks.

\textbf{Defense Framework in multi-user Semcom:}
Peng et al. \cite{peng2024adversarial} proposes an effective attack-defense game framework, specifically designed to defend against data poisoning attacks during image transmission. The proposed method used reinforcement learning to determine the optimal attack strategy for each attack type, aiming to strengthen the attack while evading detection. To generate corresponding adversarial samples, we design an Adversarial Sample Generator based on Conditional Generative Adversarial Networks. Next, an Attack Defender is introduced to detect data poisoning attacks, exclude the poisoned samples from the learning process of the target model, and generate confrontation samples guided by the optimal attack strategy to enhance the robustness of the defender. Simulation results demonstrate that the framework successfully identifies the optimal attack strategy, which further degrades the target model’s accuracy while maintaining a high evasion rate. Additionally, the generator produces effective adversarial samples, and the defender outperforms five state-of-the-art detectors in terms of Top-1 accuracy on three widely used image datasets, tested under the AWGN channel.

\textbf{Hybrid adversarial training strategy:} Nan et al. \cite{nan2023physical} introduce MobileSC, an innovative SemCom framework designed to optimize both computational and memory efficiency in wireless environments. Building on this framework, we propose SemAdv, a novel adversarial perturbation generator operating at the physical layer to produce adversarial samples. SemAdv is implemented via a multilayer perceptron neural network, which is specifically designed to generate semantically-oriented, imperceptible, input-agnostic, and controllable adversarial perturbations. These perturbations are aimed at misleading the interpretation of a specific semantic target without disturbing other semantic elements, all while not requiring prior knowledge of the input data during an attack. The proposed method seeks to create semantic adversaries that align with these criteria, offering a powerful tool for testing and improving the robustness of SemCom systems. Also, it proposes a the SemMixed training method, an innovative hybrid adversarial training strategy that integrates SemAdv and perturbation generator model techniques to generate both semantically-oriented and content-oriented adversarial samples. This approach enables the model to simultaneously address the challenges posed by different attack types. During each training stage, the system randomly decides whether to introduce adversarial samples and from which sources, thereby enhancing the model's resilience to unknown attacks. Additionally, SemMixed combines clean and adversarial training to maintain the model's accuracy while actively seeking out and defending against perturbations that could degrade performance. This is achieved through the optimization of a worst-case loss function, which further strengthens the model's robustness against a wide range of potential adversarial threats.

\subsubsection {Adversarial Regularization}

Adversarial training enhances a model's resilience by teaching it to identify both regular data and adversarial samples—intentionally crafted inputs meant to deceive the model. By iteratively improving its ability to counter these attacks, the model gains robustness against real-world noise and malicious interference. This process requires balancing performance on standard and adversarial data while refining the model's decision boundaries to bolster generalization and defense.

Adversarial regularization is a technique used to improve the robustness of machine learning models, particularly in the context of adversarial training. It works by incorporating additional terms into the loss function that help the model to better defense adversarial attacks, including class consistency, semantic consistency, or distribution consistency. These terms typically focus on ensuring consistency between normal samples and adversarial examples in various dimensions to improve the robustness of adversarial training.

\textbf{Combination adversarial training and encryption}. Luo et al. \cite{luo2023encrypted} proposes an adversarial training-based Encrypted SemCom System designed to protect privacy while enabling secure SemCom. To ensure the accuracy of SemCom in both encrypted and un-encrypted modes, distinct loss functions are defined for each mode, and the transmitter, receiver, and attacker are trained simultaneously using adversarial training. In this setup, a regularization term, controlled by the hyperparameters, balance the performance of the decryptor and the effectiveness of the attacker’s efforts. Specifically, the loss function combines two components: LKd, the loss for the decryptor, and LA, the loss for the attacker. The training process consists of two main stages: first, training the channel coder and decoder, and then alternating the training of the attacker and the transmitter/receiver. This sequential training approach effectively mitigates the risk of the attacker eavesdropping on semantic information, enhancing the security of the communication system.

\textbf{Application for task-oriented SemCom.} Wang et al. \cite{wang2019improving} introduces a novel defense algorithm, Misclassification Aware Adversarial Training, which applies adversarial regularization. The key idea behind MART is to distinguish between correctly and incorrectly classified samples during adversarial training and to apply distinct handling strategies for each. MART redefines adversarial risk by incorporating incorrectly classified samples as a regularization term, thereby encouraging the neural network to remain stable against adversarial examples derived from misclassified samples. This regularization enhances the model’s robustness to these misclassified samples. MART formulates adversarial training as a min-max optimization problem, where adversarial samples are generated by maximizing the internal loss, while the model is trained by minimizing the external loss to improve robustness. Additionally, MART proposes a semi-supervised extension that leverages unlabeled data to further bolster robustness. 

\textbf{Trade off between the accuracy and robustness}. Zhang et al. \cite{zhang2019theoretically} proposes Tradeoff-inspired Adversarial Defense via Surrogate-loss minimization, a defense strategy designed to enhance the adversarial robustness of DL models. The core idea behind TRADES is to incorporate a regularization term during training that balances the model's accuracy on natural samples with its robustness against adversarial examples. TRADES is grounded in a theoretical framework that decomposes robust error into two components: natural error and boundary error. The optimization goal of TRADES is to minimize a loss function composed of two terms: one that reduces natural error by improving performance on clean data, and another that mitigates boundary error by increasing the distance between the decision boundary and both natural and adversarial samples, thereby enhancing robustness. The TRADES algorithm is implemented through alternating gradient descent and gradient ascent to approximate the minimax optimization problem. Specifically, adversarial samples are generated by adding small perturbations, and model parameters are updated by minimizing the combined loss of natural and boundary errors.

\subsubsection {Adaptive Disturbance Constraint} 
Parameter configuration plays a critical role in enhancing the robustness of a model. Typically, the parameters of the attack including the number of iterations K, the attack step size $\alpha$ with the perturbation constraint $\varepsilon$, which are artificially predefined during training. However, some researchers argue that individual data points may have different intrinsic robustness, which means they have different distances to the decision boundary of the classifier. If a fixed perturbation tolerance is imposed on all samples at the decision boundary, samples close to the decision boundary will be misclassified when the perturbation value is relatively large. The model is forced to adjust the decision boundary in order to adapt to these perturbed samples, resulting in a distortion of the decision surface of the entire model, which affects the model's classification performance for other samples, especially samples far from the decision boundary. However, traditional adversarial training uses a uniform treatment for all samples. To address this problem, a series of adaptive methods can be used when setting such parameters, which can better improve the model robustness.

\textbf{Max-margin adversarial training}. Ding et al. \cite{ding2018mma} proposed a Max-Margin Adversarial training method, the core idea of which is to adaptively choose a “correct” perturbation $\varepsilon$ for each sample instead of using a fixed $\varepsilon$. The proposed method separates correctly categorized samples from misclassified samples. correctly categorized samples and misclassified samples separately. For correctly classified samples, the algorithm aims to directly maximize the margins in the input space by optimizing the objective function. Specifically, for each data point, Max-Margin Adversarial training finds the optimal $\varepsilon$ value that maximizes its margin. For misclassified samples, the goal of this training is to minimize the classification loss of these samples.
With this differential treatment, the proposed training not only improves the robustness of the model for correctly classified samples, but also improves the accuracy for misclassified samples.

\textbf{Application for task-oriented SemCom}. Based on the training labels, Cheng et al. \cite{cheng2020cat} proposed a new algorithm called Customized Adversarial Training (CAT) to improve the traditional training approach, which achieves this target of adaptively customizing the perturbation level for each training sample. The CAT algorithm updates the perturbations by iterative way to update the perturbations, keeping the current $\varepsilon$ if a perturbation that makes the classifier's prediction wrong is found, while, on the other hand, increasing $\varepsilon$. Meanwhile, the CAT algorithm proposes an algorithm for adaptive label smoothing, which correlates the smoothing parameter with the level of perturbations and helps the model to learn smoother decision boundaries. This ensures that the samples are not under the control of absolute one-hot coding under adversarial attacks. In the absence of adversarial perturbations, the output of the model approaches a more reasonable probability distribution of [0.5,0.5] as the samples approach the decision boundary.

\subsubsection {Matching Pluggable Modules}
Pluggable modules offer substantial benefits for SemCom systems by significantly enhancing security while maintaining both efficiency and flexibility.  These modules are designed with high compatibility, allowing them to be easily integrated into existing system architectures without the need for extensive modifications. By providing such integration, pluggable modules ensure that security enhancements can be implemented with minimal disruption to the overall system performance. This adaptability not only helps in safeguarding sensitive communication but also preserves the system’s operational efficiency, making it an ideal solution for evolving communication environments where flexibility is essential.

He et al. \cite{he2024secure} investigate the advantages of secure-aware SemCom systems in countering adversarial attacks. They proposes installing a matched pair of pluggable modules—one positioned after the semantic transmitter and another before the semantic receiver. The transmitter module employs a trainable Adversarial Residual Network to generate adversarial examples, while the receiver module uses a corresponding trainable network to mitigate adversarial attacks and channel noise. To counter semantic eavesdropping, these trainable network are jointly optimized to minimize a weighted sum of the adversarial attack's strength, mean squared error in SemCom, and the eavesdropper’s confidence in accurately retrieving private information. Numerical results demonstrate that this approach effectively misleads eavesdroppers while preserving high-quality SemCom.

\begin{les}
Adversarial training, which involves adding adversarial attacks to training samples, is favored by researchers for enhancing the robustness of models in SemCom by enabling them to recognize semantic adversarial samples generated through noise addition. Adversarial training approaches can be divided into adversarial sample generation, adversarial regularization, adaptive disturbance constraint, and matching pluggable modules. The adversarial sample generation schemes includes iterative fast gradient symbol method \cite{hu2022robust}, semantic distance minimization mechanism \cite{kang2023adversarial}, and hybrid adversarial training strategy \cite{nan2023physical}. The adversarial regularization includes combination adversarial training and encryption \cite{luo2023encrypted}, application for task-oriented SemCom \cite{wang2019improving}, and trade off between the accuracy and robustness \cite{zhang2019theoretically}. The adaptive disturbance constraint methods are applied in max-margin adversarial training \cite{ding2018mma} and Semcom \cite{cheng2020cat}. Matching pluggable module is uesd in SemCom system to defense adversarial attacks \cite{he2024secure}.
\end{les}

\subsection{Differential Privacy}

\begin{figure*}[t]
\centering
\vspace{-10mm}
\includegraphics[width=\linewidth]{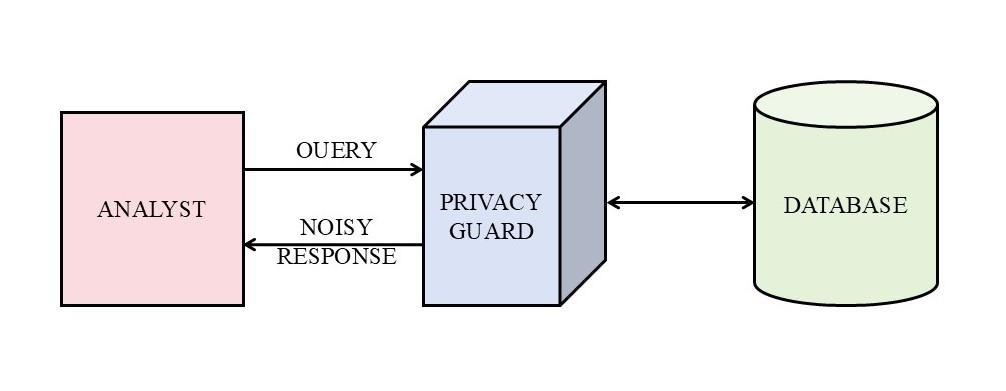}
\caption{Differential privacy mechanism \cite{jain2018differential}, where depicts a privacy-preserving system architecture where an Analyst submits a query to a Privacy Guard, which acts as an intermediary between the Analyst and the Database.   The Privacy Guard is responsible for processing the query in a way that protects sensitive information within the database.   It does this by adding noise to the query results before sending them back to the Analyst, thus providing a Noisy Response.   This mechanism ensures that while the Analyst can still receive useful information, the privacy of the data in the Database is maintained by preventing the Analyst from gaining access to exact or potentially sensitive data points.}
\label{fig451}
\end{figure*}

Differential privacy is a new definition of privacy proposed by Dwork \cite{dwork2008differential} in 2006 in response to privacy breaches in statistical databases. As shown in Fig. \ref{fig451}, under this definition, the result of computational processing of a dataset is insensitive to a specific data change, which means that the presence or absence of a single piece of data in the dataset has a negligible effect on the computational result. Extend to SemCom, the single semantic symbol cannot influence the meaning of the whole sentence. Therefore, the risk of privacy breaches arising from the addition of a piece of semantic symbol to a dataset is kept so small that an attacker cannot obtain accurate information by observing the results of the computation.

Assume two of semantic symbol sets are adjacent if they differ in a single entry. Then a randomized mechanism $\mathcal{M}: \mathcal{D}_{C} \rightarrow \mathcal{R}$ with domain $\mathcal{D}_{C}$ and range $\mathcal{R}$, satisfies $(\varepsilon, \delta)$-differential privacy if for any two adjacent inputs $d_{C}$ and $d'_{C}$, and for any subset of outputs $S_{C} \subseteq \mathcal{R}$, it holds that
\begin{equation}
    Pr[\mathcal{M}(d_{C})\in \mathcal{S_{C}}]\leq Pr[e^{\varepsilon} \mathcal{M}(d'_{C})\in \mathcal{S_{C}}]+\delta
\end{equation}
where $\delta$ denotes the probability of $\varepsilon$-differential privacy being broken.

Prior to its development, existing privacy-preserving algorithms, such as k-anonymity, had significant limitations against homogenization attacks. Differential privacy is able to address two shortcomings of traditional privacy-preserving models. First, the differential privacy-preserving model assumes that the attacker has access to semantic information about all other records except the target record, and the sum of this information can be understood as the maximum background semantic knowledge that the attacker can have. Under this semantic knowledge base assumption, differential privacy preservation does not need to consider any possible background knowledge possessed by the attacker. Second, it is built on a solid mathematical foundation that provides a rigorous definition of privacy protection and a quantitative evaluation method, which makes the level of privacy protection provided by datasets under different parameter treatments comparable.

\begin{figure*}[htbp]
\centering
\vspace{-10mm}
\includegraphics[width=\linewidth]{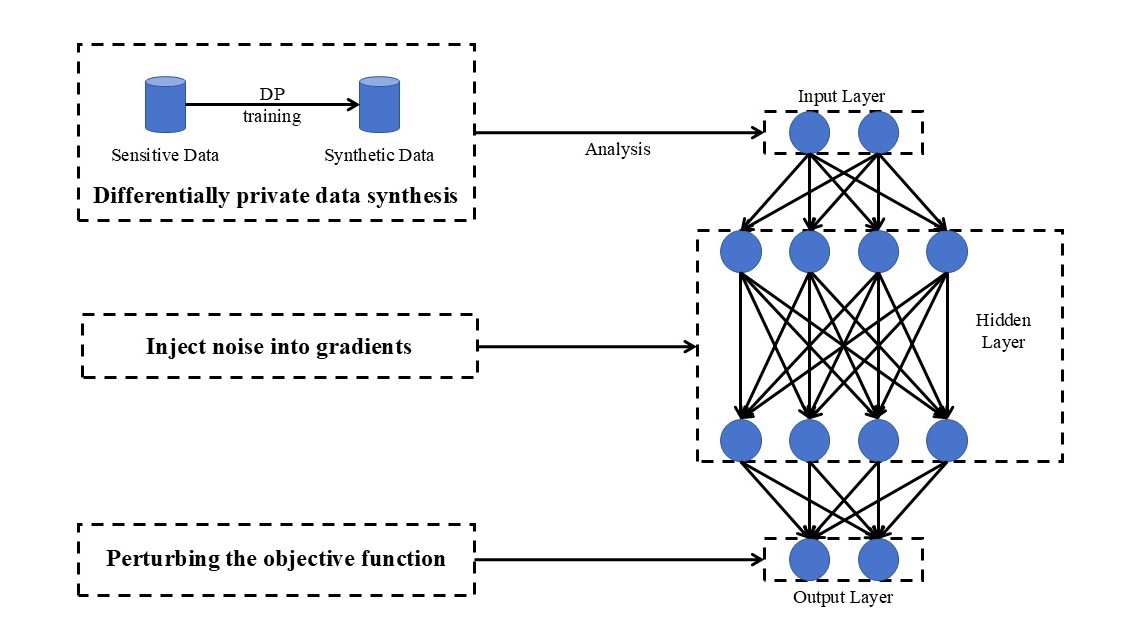}
\caption{The three phase of differential privacy deployed in DL model \cite{zhao2019differential}, where the phase of the mechanism is divided it into three types: input layer, hidden layer, and output layer. Firstly, sensitive data undergoes differential privacy training to generate synthetic data, ensuring that the privacy of individuals within the dataset is maintained. This synthetic data is then used for analysis within a neural network composed of an input layer, one or more hidden layers, and an output layer. To further enhance privacy, noise is injected into the gradients during the training process, and the objective function can be perturbed. These measures are designed to prevent the reconstruction of the original data from the model, thereby safeguarding sensitive information.}
\label{fig452}
\end{figure*}

Privacy-preserving ML requires the learner to be able to get hold of information about the distribution of data in a private dataset, while not revealing too much information about any individual in the dataset. Differential privacy can fulfill the above requirements well. Also differential privacy is a promising confidentiality technique in SemCom. Therefore, the use of differential privacy in the model training phase would be analyzed in the following three aspects: parameter values, objective function and data, which are illustrated in Fig. \ref{fig452}.

The below, as shown in Table \ref{differentialprivacy}, the following content, based on multiple academic studies, delves into various differential privacy methods in different stages and their potential applications in SemCom.

\begin{longtable}{|A{2cm}|A{0.6cm}|P{9.5cm}|}  
\caption{\footnotesize{List of Representative Differential Privacy Schemes}}
\label{differentialprivacy}\\
\hline
\textbf{Sub-class} & \textbf{Ref.} & \textbf{Descriptions, including} $\circledast$: \textbf{contributions; }

$\circleddash$: \textbf{trade-offs;} 
$\boxplus$: \textbf{performance metrics;}

$\Cap$: \textbf{applied in SemCom or promising for SemCom}
\\
\hline
\endfirsthead

\hline
\textbf{Sub-class} & \textbf{Ref.} & \textbf{Descriptions, including} $\circledast$: \textbf{contributions; }

$\circleddash$: \textbf{trade-offs;} 
$\boxplus$: \textbf{performance metrics;} 

$\Cap$: \textbf{applied in SemCom or promising for SemCom} \\
\hline
\endhead

\hline
\endfoot

\hline
\endlastfoot
Parameter Value Protection & \cite{wei2020federated} &
$\circledast$: Introduces the Noising before Aggregation FL framework to prevent information leakage by adding random Gaussian noise to model weights.

$\circleddash$: Privacy and model accuracy, number of clients and convergence performance.

$\boxplus$: Training loss and number of communication rounds.

$\Cap$: Promising for privacy-preserving model updates in multi-hop SemCom.
\\
\hline

Parameter Value Protection& \cite{shokri2015privacy} &
$\circledast$: Introduces differential privacy for gradient leakage prevention by selectively sharing model parameters.

$\circleddash$: Complexity in parameter selection and privacy.

$\boxplus$: Accuracy.

$\Cap$: Promising for secure model training in SemCom.
\\
\hline

Parameter Value Protection& \cite{liu2024adaptive} &
$\circledast$: Develops the the Adaptive Privacy Budget-based Differential Privacy method adjusting noise levels as models converge to ensure consistent privacy protection throughout training.

$\circleddash$: Privacy protection and performance efficiency.

$\boxplus$: PSNR, loss, and PLR.

$\Cap$: Applied in enhancing privacy in gradient-based optimization in SemCom.
\\
\hline

Objective Function Perturbation & \cite{chaudhuri2011differentially} &
$\circledast$: Designs a differential privacy framework for perturbing the objective function in classifier optimization.

$\circleddash$: Data size and model accuracy.

$\boxplus$: Accuracy.

$\Cap$: Promising for privacy-preserving machine learning in SemCom.
\\
\hline

Objective Function Perturbation & \cite{phan2017adaptive} &
$\circledast$: Proposes an Adaptive Laplace Mechanism for adaptive noise addition in the objective function for differential privacy.

$\circleddash$: Noise levels and privacy protection.

$\boxplus$: Accuracy.

$\Cap$: Promising for adaptive privacy mechanisms in SemCom.
\\
\hline

Data Generation and Pre-processing & \cite{acs2018differentially} &
$\circledast$: Introduces differential privacy in generative models to protect privacy during data generation.

$\circleddash$: Data utility and privacy.

$\boxplus$: Average relative error, clustering accuracy, and epsilon.

$\Cap$: Promising for privacy-preserving data generation in SemCom.
\\
\hline

Data Generation and Pre-processing & \cite{li2015differentially} &
$\circledast$: Introduces distance-based sampling for dynamic datasets under differential privacy.

$\circleddash$: Data dynamics and privacy efficiency.

$\boxplus$: Absolute error, query accuracy.

$\Cap$: Promising for dynamic data privacy in SemCom.
\\
\hline
\end{longtable}

\subsubsection{Parameter Value Protection}
To protect the privacy of the training data, it is possible to focus on the final parameters produced by the training process without focusing on the specific internal operations, considering the process as a black box. However, the correlation between these parameters and the input data is difficult to measure by a specific and reliable standard. The operation of adding noise to the parameters is conservative, which cannot guarantee the security of the learning model reliably. Therefore,the model is assumed to be vulnerable to malicious attacks on the parameters, assuming that the adversary fully understands the training mechanism and has access to the model's parameters. Based on the above problems, an effective method is to use differential privacy to protect the model parameters uploaded by the client. Model parameters are mainly divided into two types: model weight and gradient.

\textbf{Weight of Model Protection:} 
Wei et al. \cite{wei2020federated} added random Gaussian noise at the aggregated global model to hide any single client’s update in the centralized pattern. Specifically, Kang Wei et al. propose a new framework based on differential privacy, Noising before Aggregation FL, to effectively prevent information leakage. Each client locally perturbs its training parameters by purposefully adding noise to ensure that the uploaded parameters satisfy a specific level of privacy protection and uploads the updated model weights to the central server before uploading its training parameters to the server for aggregation. The system achieves an effective balance between privacy protection and model performance by sharing key parameters of the model to improve learning accuracy while protecting the privacy of participants' data.

\textbf{Gradient of Model Protection:}  Shokri et al. \cite{shokri2015privacy} chose to share the model parameter selectively. The parameter selection rate determines the number of parameters updated in iterations for each parameter subset. During the parameter subset selection process, the gradients whose median values are larger than a specific threshold are randomly selected for uploading, and this approach is similar to the idea of differential privacy. Song et al. \cite{song2013stochastic} introduced differential privacy in Stochastic gradient descent. They theoretically derived the gradient descent with differentially private algorithm and verified the feasibility of the algorithm. Abadi et al. \cite{abadi2016deep} developed the algorithm on this basis and proposed the differentially private stochastic gradient descent algorithm, which is reflected in the proposal of an effective algorithm for calculating the single training examples' gradient, the efficient algorithm for subdividing the task into smaller batches to reduce memory footprint, and the application of the differentially private principal projection at the input layer. Specifically, the algorithm adds Gaussian noise to the gradient to preserve privacy, where the size of the noise is proportional to the sensitivity of the gradient.

\textbf{Parameter Protection in Semcom:}
Liu et al. \cite{liu2024adaptive} introduces the Adaptive Privacy Budget-based Differential Privacy (APB-DP) method. APB-DP leverages differential privacy with an adaptive privacy budget, adjusting noise levels as models converge to ensure consistent privacy protection throughout training. Additionally, APB-DP accounts for wireless channel effects to reduce interference. Experimental results demonstrate that APB-DP decreases privacy leakage by 13 percent and minimizes performance loss by 71 percent compared to leading differential privacy-based training methods, providing a more secure and efficient solution for SemCom. In SemCom systems, the sender can add the noise to the model update before model-sharing, which is helpful for privacy-preserving, especially in multi-hop scenarios. Several researchers \cite{wei2020federated,mothukuri2021survey,10.1145/3133956.3134012,8737416} have demonstrated that the obfuscated parameters in differential privacy lead to a trade-off between the model training-accuracy and security. Therefore, adding well-designed differential privacy and the proper way to add noise matters to the model-driven networks, especially in the multi-hop manners.

\subsubsection{Objective Function Perturbation}

Differential privacy is one of the methods of target perturbation, which plays a crucial role in model training. The application of differential privacy in objective perturbation is mainly reflected in the design of privacy-preserving algorithms for security frameworks. Specifically, the objective perturbation approach adds random noise to the objective function by using differential privacy in the optimization process instead of injecting noise directly into the result. This approach puts privacy preservation into consideration in algorithm design. The optimization process is interfered by adding noise to the objective function to improve robustness, thereby protecting the privacy of individuals in the training data process. 

\textbf{Classifier optimization:} Chaudhuri et al. \cite{chaudhuri2011differentially} designed an empirical risk minimization framework that uses differential privacy to perturb the objective function, which is the combination of the loss function and the regularization term, and then optimizes the classifier based on this situation. Compared to the direct perturbation of the output result, perturbation of the objective function reduces the prediction error under the condition of preserving privacy. It achieves the goal that balances the trade-off between privacy preservation and learning performance. 

\textbf{Linear and logistic regression analyzation:} Zhang et al. \cite{winslett2012functional} proposed a concept of Functional Mechanism to describe the operation of perturbing the optimization goal function of regression analysis. The noise is injected into the objective function using differential privacy on the two most commonly used regression models, linear regression and logistic regression. Regularization and spectral pruning techniques are used to ensure the objective function to get a meaningful solution. Experimental results show that the accuracy of functional mechanism differential privacy in linear regression and logistic regression tasks significantly outperforms existing differential privacy preservation methods, including Filter-Priority and so on. In particular, the accuracy advantage of FM is more obvious when the data dimension increases. Meanwhile, the low-order approximation module of FM makes its computational efficiency much higher than that of ordinary baseline methods.


\textbf{Combination with Auto-Encoder:} Phan et al. \cite{phan2016differential} propose a Deep Private Auto-Encoder model, achieving $\varepsilon$ differential privacy preservation by introducing perturbations into the objective function traditional deep self-encoder. To make the objective function satisfy differential privacy, this paper approximates the cross-entropy error functions of the data reconstruction and softmax layers into polynomial forms using Taylor Expansion, and injects noise into these polynomial forms. The auto-encoder preserves most of the valid information of the input data by minimizing the reconstruction error between the input and the output results through an objective function. The autoencoder learns the important features of the input data to hide the sensitive information of the input data in the output data, which achieves accurate prediction of human behavior.

\textbf{Adaptive laplace mechanism:} The article as we mentioned before adds the same amount of noise in the model, which shows great limitations in the application of different application scenarios. To solve this problems, Phan et al. \cite{phan2017adaptive} also proposed a new mechanism of Adaptive Laplace Mechanism. It utilizes the Layer-wise Relevance Propagation algorithm to calculates the relevance between the input features and output feature of the model. The Adaptive Laplace Mechanism adds more noise into features that have less influence on the model output while adding less noise in significant feature. This scheme promotes the wide application of differential privacy techniques in DL.

\subsubsection{Data Generation and Pre-processing}

In terms of data generation, differential privacy is applied to protect data privacy in the generative model of the input layer, which can be regarded as a pre-processing of the training dataset. The data manager first generates synthetic data with the same statistical features as the original training dataset under differential privacy. The synthetic data and generated models are then released without compromising privacy, which can be used for a variety of analyses. 

\textbf{Generative model:} Acs et al. \cite{acs2018differentially} introduce differential privacy techniques during the training process of generated models to protect the privacy of the individual data used to train these models. Specifically, the dataset is divided into k clusters using differential privacy and k-mean clustering algorithms. A generative neural network is assigned to each cluster and cluster is trained by differential privacy gradient descent respectively. To solve the problem of complex data privacy protection, Zhang et al. \cite{zhang2018differentially} proposed to propose dp-GAN, a generalized private publishing framework for rich semantic data, in which the differential privacy mechanism is integrated based on the existing GAN models. The random noise is injected into the training step of discriminator to perform differential privacy constraints. Various optimization strategies, such as Parameter Grouping, Adaptive Clipping and Warm Starting, are used to generate high-quality synthetic data. Chen et al. \cite{chen2024enhancing} introduces a novel approach to secure SemCom by combining differential privacy with a GAN-based inversion technique to protect sensitive image data transmitted over insecure channels. By extracting and safeguarding essential semantic features, the proposed method reduces privacy risks while enabling accurate data reconstruction at the intended receiver. Simulation results confirm that this approach effectively prevents eavesdroppers from accessing sensitive information while preserving high-quality image reconstruction, highlighting its potential as a robust solution for privacy-preserving SemCom.


\textbf{Adding noise in tree structure:} Cormode et al. \cite{cormode2012differentially} proposed a framework called Private Spatial Decompositions, which utilizes the Laplace mechanism to add noise to each node in the dataset on a tree structure. Non-uniform assignment of noise parameters to data nodes ensures that both the tree structure and node counts satisfy differential privacy. 

\textbf{Dynamic dataset publishing:}  Li et al. \cite{li2015differentially} focus on differential privacy in dynamic dataset publishing. Two methods,distance-based sampling with fixed threshold and distance-based sampling with adaptive threshold, are proposed for handling the dynamics of dynamic datasets under user-level differential privacy. Specifically, the fixed threshold method uses a fixed threshold to publish a new differential privacy histogram only when the distance between the current dataset and the previously published dataset exceeds a predefined threshold. While the adaptive threshold method further improves fixed threshold by dynamically adapting the threshold using a feedback control mechanism to capture data dynamics.

\begin{les}
In differential privacy, user privacy in published semantic datasets is protected by introducing random noise. Noise-attachment mechanism is a method of protecting data by perturbing it through a predefined mechanism. The addition of noise makes it possible to add or remove any record from the input data without significantly affecting the output of the algorithm. This approach ensures that even if an attacker has some information, he or she cannot reliably infer sensitive information. Differential privacy is realized from the following perspectives: parameters (model weights \cite{wei2020federated,wei2020federated,mothukuri2021survey} and gradients \cite{shokri2015privacy,song2013stochastic,abadi2016deep,liu2024adaptive}), objective functions \cite{chaudhuri2011differentially,winslett2012functional,phan2017adaptive}, and data \cite{acs2018differentially,zhang2018differentially,cormode2012differentially,li2015differentially}. Generative models \cite{phan2016differential,acs2018differentially,zhang2018differentially,chen2024enhancing} are generally combined to protect differential privacy.
\end{les}

\subsection{Cryptography Technology}
In traditional communication systems, encryption technologies ensure data confidentiality during transmission by converting plaintext into randomized ciphertext. Methods such as AES and RSA rely on complex mathematical algorithms, making it difficult for unauthorized users to decrypt data. However, with the evolving demands of communication, particularly in SemCom, the limitations of traditional encryption methods have become increasingly apparent.

SemCom significantly enhances communication efficiency by extracting and transmitting semantic features. However, the randomized ciphertext generated by traditional encryption methods makes it difficult to extract semantic information and disrupts the semantic correlation of the data. Moreover, the ``avalanche effect" of traditional encryption and its limited ability to directly operate on ciphertext further constrain its applicability in SemCom. Against this backdrop, innovative encryption technologies such as Homomorphic Encryption (HE), Secure Multi-Party Computation (SMPC), Trusted Execution Environment (TEE), and secure aggregation have garnered significant attention.

These emerging encryption technologies offer a balance between data protection and semantic processing. For instance, SMPC enables collaborative data computation without compromising privacy; HE allows addition and multiplication directly on ciphertext, facilitating complex data processing without decryption; TEE ensures secure execution of sensitive data through hardware-level isolation; and Secure Aggregation provides an efficient mechanism for data merging in distributed SemCom. The integration of these technologies not only addresses the shortcomings of traditional encryption but also presents novel solutions for enhancing the efficiency and security of SemCom systems. These new technologies are described below, and some typical articles are shown in Table \ref{cryptography}.

\begin{longtable}{|A{2cm}|A{0.6cm}|P{9.5cm}|}
\caption{\footnotesize{List of Representative Cryptography Schemes}} 
\label{cryptography}\\
\hline
\textbf{Sub-class} & \textbf{Ref.} & \textbf{Descriptions, including} $\circledast$: \textbf{contributions; }

$\circleddash$: \textbf{trade-offs;} 
$\boxplus$: \textbf{performance metrics;}

$\Cap$: \textbf{applied in SemCom or promising for SemCom} \\
\hline
\endfirsthead
\hline
\textbf{Sub-class} & \textbf{Ref.} & \textbf{Descriptions, including} $\circledast$: \textbf{contributions; }

$\circleddash$: \textbf{trade-offs;} 
$\boxplus$: \textbf{performance metrics;}

$\Cap$: \textbf{applied in SemCom or promising for SemCom} \\
\hline
\endhead
\hline
\endfoot
\hline
\endlastfoot

 HE & \cite{aono2017privacy} &
$\circledast$: Propose a privacy-preserving DL framework using additive HE (Paillier) to encrypt gradients and prevent data leakage in cloud servers.

$\circleddash$: High computation and communication overhead.

$\boxplus$: Model accuracy and communication cost.

$\Cap$: Promising for secure and privacy-preserving collaborative learning in SemCom systems. \\
\hline

 HE & \cite{fang2021privacy} &
$\circledast$: Combine additive HE with FL using an optimized Paillier algorithm, reducing encryption time by 60\%.

$\circleddash$: Privacy preservation and computational overhead.

$\boxplus$: Model accuracy and encryption time.

$\Cap$: Promising for secure collaborative training in SemCom. \\
\hline





HE & \cite{lee2022privacy} &
$\circledast$: Implement FHE-based DL inference (ResNet-20 on CIFAR-10) with CKKS and polynomial approximations for non-linear functions.

$\circleddash$: Security and computational efficiency.

$\boxplus$: Prediction accuracy and training time.

$\Cap$: Promising for encrypted semantic feature processing in SemCom. \\
\hline

HE & \cite{meng2025secure} &
$\circledast$: Propose FHE-enabled SemCom with homomorphic deep JSCC, replacing ReLU with square functions for HE compatibility.

$\circleddash$: Security and computational complexity.

$\boxplus$: Classification accuracy, compression ratio, and computation time.

$\Cap$: Applied in semantic correlations in ciphertext for secure and efficient SemCom. \\
\hline













SMPC & \cite{goldreich1998secure} &
$\circledast$: Lay theoretical foundations for SMPC with semi-honest/malicious security models.

$\circleddash$: Security and computational complexity.

$\boxplus$: Communication overhead, computational cost, and security guarantees.

$\Cap$: Promising for designing SemCom protocols with formal privacy guarantees. \\
\hline

SMPC & \cite{mugunthan2019smpai} &
$\circledast$: Integrate SMPC with FL using cryptographic masking and differential privacy.

$\circleddash$: Privacy and model accuracy.

$\boxplus$: Classification accuracy and computational overhead.

$\Cap$: Promising for SemCom to protect semantic model updates. \\
\hline









TEE & \cite{zhang2021shufflefl} &
$\circledast$: ShuffleFL combines TEE with random grouping to prevent gradient inversion attacks.

$\circleddash$: Privacy and computational efficiency.

$\boxplus$: Aggregation time, communication overhead, and model accuracy.

$\Cap$: Promising for SemCom to secure semantic model aggregation. \\
\hline









Secure Aggregation & \cite{kadhe2020fastsecagg} &
$\circledast$: FFT-based FastSecAgg reduces computation costs via multi-secret sharing.

$\circleddash$: Privacy and computational efficiency.

$\boxplus$: Aggregation accuracy, computation time, and communication overhead.

$\Cap$: Promising for secure and efficient model aggregation in SemCom. \\
\hline













Quantum Cryptography & \cite{chehimi2024quantum} &
$\circledast$: Propose a quantum SemCom framework integrating QKD to secure semantic key transmission.

$\circleddash$: Resource efficiency and semantic fidelity.

$\boxplus$: Quantum semantic fidelity and resource consumption.

$\Cap$: Applied in resource-efficient SemCom systems \\
\hline

Quantum Cryptography & \cite{nunavath2024towards} &
$\circledast$: Explore QDS for signing semantic messages, ensuring authenticity and integrity in distributed SemCom networks.

$\circleddash$: Communication robustness and resource efficiency.

$\boxplus$: Quantum semantic fidelity and F1 score.

$\Cap$: Applied in enhancing knowledge transmission in SemCom \\
\hline

Quantum Cryptography & \cite{khalid2023quantum} &
$\circledast$: Investigate QSDC for direct semantic message transmission without pre-shared keys, reducing latency.

$\circleddash$: Privacy and computational complexity.

$\boxplus$: Quantum semantic fidelity and system latency.

$\Cap$: Applied for secure and resource-efficient communication in SemCom \\
\hline

\end{longtable}

\subsubsection{Homomorphic Encryption}
HE is an advanced cryptographic technique that allows mathematical operations to be performed directly on encrypted data without decryption. This characteristic is particularly valuable in privacy-preserving scenarios, where computations need to be performed without exposing the underlying data. The core concept of HE ensures that operations performed on ciphertext yield results that, when decrypted, match those obtained if the same operations were applied to the plaintext. This innovative feature has opened up new research areas, especially in cloud computing, FL, and privacy-preserving ML applications.

HE can be categorized into two types: Partially Homomorphic Encryption (PHE) and Fully Homomorphic Encryption (FHE). PHE supports a single type of operation, such as multiplication in RSA or addition in Paillier, while FHE supports both addition and multiplication operations an unlimited number of times, making it the most powerful encryption scheme in theory. In 2009, Craig Gentry \cite{gentry2009fully} introduced the first practical FHE scheme based on ideal lattices, providing a groundbreaking tool for privacy preservation in academia and industry.

Despite its strong privacy-preserving capabilities, the significant computational and communication overhead associated with HE remains a major challenge for its adoption in real-world applications. In recent years, researchers have focused on optimizing encryption algorithms to improve efficiency and reduce the computational cost of ciphertext operations. Below, we discuss some typical applications and optimization methods of HE in the field of DL.

\begin{figure*}[htbp]
\centering
\vspace{10mm}
\includegraphics[width=0.8\linewidth]{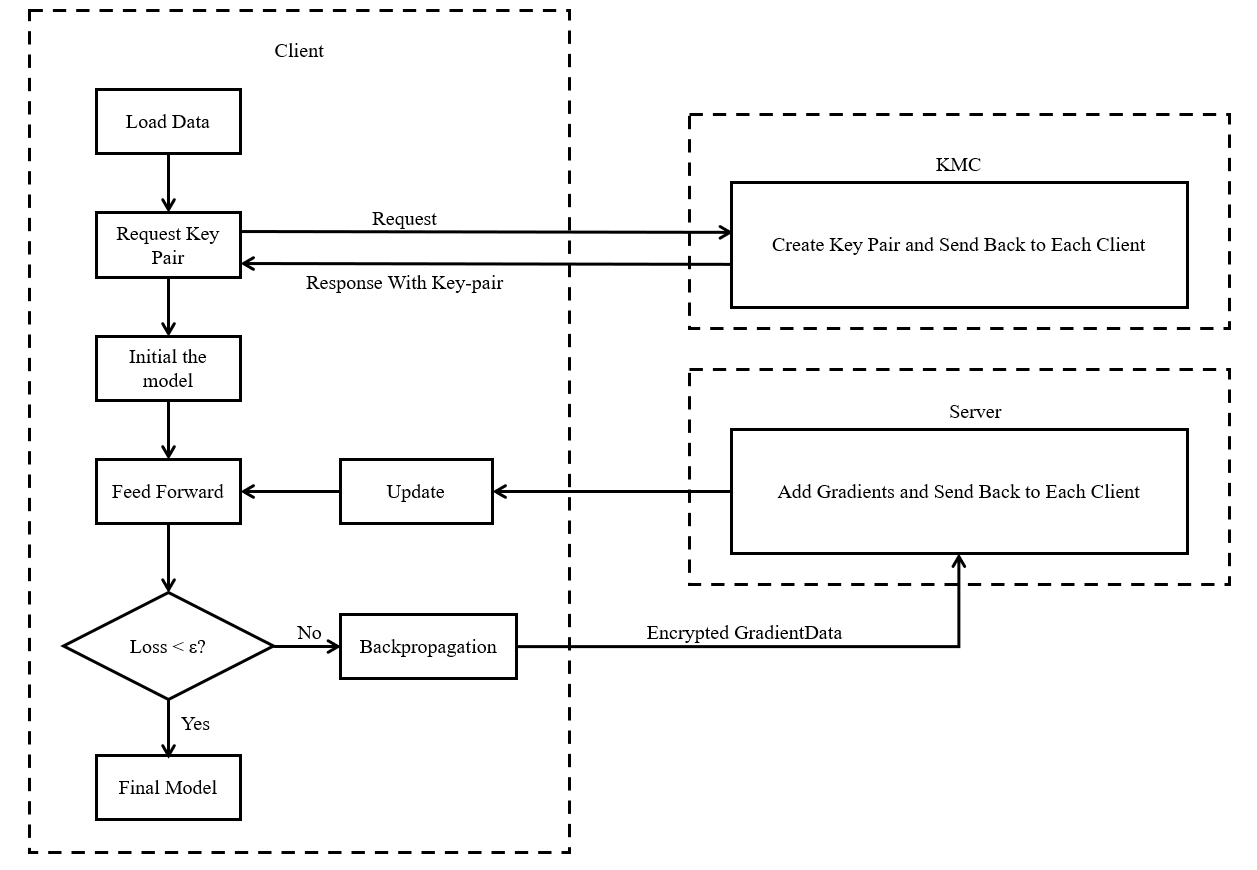}
\caption{The architecture of a Paillier federated network \cite {fang2021privacy},  where each client requests a key pair from the KMC, initializes the model, and performs forward propagation and backpropagation on encrypted data, ensuring the privacy of sensitive information. During the backpropagation step, the gradients are computed and encrypted before being sent back to the server. The server, which aggregates the gradients from all clients, sends the updated gradients back to each client, facilitating a secure and privacy-preserving training process.}
\label{fig461}
\end{figure*}

\textbf{PHE:} Phong et al. \cite{aono2017privacy} proposed a privacy-preserving DL framework using additive HE. By leveraging the asynchronous stochastic gradient descent algorithm and encrypting gradients with the Paillier system, the framework mitigates potential data leakage risks in cloud-based servers. Experiments showed that this approach achieved near 99\% accuracy on the MNIST dataset while effectively preventing data leakage between participants. This research highlights the potential of additive HE in achieving a balance between privacy and performance in distributed learning environments. In FL, Fang et al. \cite{fang2021privacy} introduced a framework combining additive HE and FL, whcih is showed in Fig. \ref{fig461}. This framework incorporates an optimized Paillier algorithm, significantly reducing encryption computation time while maintaining model accuracy deviations within 1\%. It is particularly well-suited for applications requiring high privacy standards, such as finance and healthcare. Their experiments further analyzed the impact of key length, network structure, and the number of clients on system performance, providing insights into balancing privacy and efficiency in complex environments. To address cross-silo FL requirements, Zhang et al. \cite{zhang2020batchcrypt} introduced the BatchCrypt system. The system employs a batch encryption method to compress gradient data, significantly reducing encryption computation and communication overhead. Experiments validated that BatchCrypt enhanced training speed by 23–93 times and reduced communication costs by 66–101 times compared to traditional methods, all while maintaining model accuracy. This system provides an efficient and secure encryption framework for cross-organizational data collaboration.

\textbf{FHE:} Lee et al. \cite{lee2022privacy} developed a DL inference model using FHE. Their research employed the CKKS scheme with bootstrapping technology to implement encrypted inference on the ResNet-20 model for the CIFAR-10 dataset, achieving an accuracy of 92.43\%, nearly matching the plaintext model's 91.89\%. To address challenges in non-linear activation functions within encrypted environments, they utilized high-precision polynomial approximations as substitutes and, for the first time, implemented Softmax encryption within the FHE framework to defend against model extraction attacks. This study underscores the practicality of FHE in privacy-preserving DL and efficient inference. Additionally, Sun et al. \cite{sun2018private} proposed an improved FHE scheme. Their work focused on enhancing computational efficiency and reducing noise accumulation through relinearization and modulus switching techniques. This scheme achieved significant efficiency improvements in private decision tree classification and other tasks like Naïve Bayes classification. Experimental results demonstrated its effectiveness, achieving high-performance homomorphic computations across multiple ML classification tasks.

\textbf{HE for SemCom:} Meng et al. \cite{meng2025secure} explored the feasibility of FHE in SemCom. They demonstrated that despite traditional encryption techniques disrupting semantic correlation, FHE preserves meaningful features in ciphertext, making it suitable for SemCom applications. Using Scale-Invariant Feature Transform (SIFT), their study validated that encrypted semantic features could still be extracted, ensuring that privacy is maintained while enabling effective communication. Furthermore, they proposed a homomorphic encrypted deep JSCC model, replacing ReLU activation with a square function and using average pooling to support homomorphic operations. Their experiments confirmed that the classification accuracy of their privacy-preserving model on encrypted data was nearly identical to that on plaintext data, demonstrating the potential of FHE in secure SemCom.

\subsubsection{Secure Multi-Party Computation}
\begin{figure*}[htbp]
\centering
\vspace{0mm}
\includegraphics[width=\linewidth]{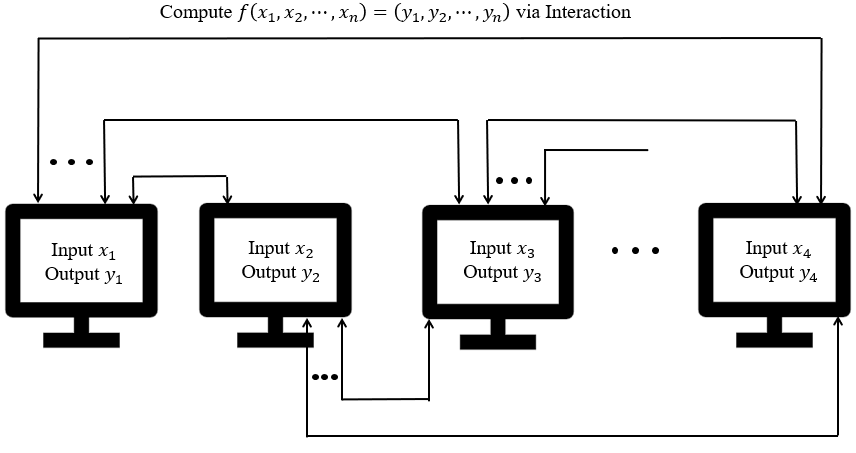}
\caption{The diagram of SMPC \cite{zhao2019secure}, where multiple parties collaboratively compute a function $f(x_1, x_2, \ldots, x_n)$ with inputs $x_1, x_2, \ldots, x_n$ and corresponding outputs $y_1, y_2, \ldots, y_n$ through secure interactions, ensuring data privacy. In this process, each party holds a portion of the data and performs computations on their own inputs, while the final result is shared and computed collaboratively without revealing any individual party's private input.}
\label{fig462}
\end{figure*}

SMPC serves as a core tool for privacy-preserving technologies, allowing multiple parties to jointly compute specific functions without revealing their private input data. It effectively resolves the conflict between data privacy and collaboration requirements. With the increasing reliance on distributed data in ML, SMPC provides robust privacy guarantees for secure model training and inference, striking a balance between data sensitivity and collaboration efficiency. Fig. \ref{fig462} illustrates a conceptual diagram of SMPC.

The core of SMPC lies in utilizing cryptographic techniques to ensure the privacy, correctness, robustness, and fairness of distributed computations. Key techniques include secret sharing, HE, and oblivious transfer. Secret sharing divides private data into multiple fragments distributed among the participants, ensuring that no single fragment reveals any meaningful information \cite{zhao2019secure}.
HE allows computations to be performed directly on encrypted data, protecting data privacy without requiring decryption during intermediate steps.

In SMPC security models, the semi-honest model assumes participants strictly follow the protocol but may try to infer additional information, while the malicious model allows participants to arbitrarily deviate from the protocol \cite{zhao2019secure}.
Foundational research by Yao and Goldreich \cite{goldreich1998secure} laid the theoretical groundwork for two-party and multi-party computations, establishing the basis for modern SMPC development
. In recent years, significant progress has been made in applying SMPC to ML, including advancements in asynchronous multi-party computation, the SPDZ framework, Falcon protocol, SecureNN, CRYPTEN framework, and applications in FL.


\textbf{General ML Frameworks:} These frameworks aim to provide secure, scalable, and efficient solutions for common ML algorithms, covering a wide range of tasks.
\begin{itemize}
 \item SPDZ Framework: Proposed by Chen et al. \cite{chen2019secure}, the SPDZ framework is a SMPC implementation that provides malicious security guarantees. It uses a preprocessing phase based on additive secret sharing to generate ``Beaver triples", significantly improving the efficiency of the online computation phase. The framework has been successfully applied to linear regression and logistic regression and can securely execute iterative algorithms like stochastic gradient descent on large-scale datasets.
  \item CRYPTEN Framework: Developed by Knott et al. \cite{knott2021crypten}, CRYPTEN integrates SMPC with ML by providing an API similar to PyTorch. It supports tensor computation, automatic differentiation, and modular neural networks, simplifying the implementation of privacy-preserving ML models. This design is particularly suitable for researchers and practitioners unfamiliar with cryptographic techniques, driving the wide application of SMPC in ML.
\end{itemize}

\textbf{Protocols for Neural Network Training and Inference :} SMPC protocols designed for DL address the challenges of secure training and inference in complex neural networks. These methods emphasize performance optimization in linear transformations and nonlinear activation functions.
\begin{itemize}
  \item SecureNN Protocol: Proposed by Wagh et al. \cite{wagh2019securenn}, SecureNN extends SMPC capabilities to neural network training. It employs a three-party computation model, avoiding costly cryptographic circuit operations and optimizing secure protocols for matrix multiplication, convolution, and nonlinear functions. Experiments on the MNIST dataset demonstrated that this protocol achieves over 99\%
  classification accuracy while significantly reducing communication complexity.
  \item Falcon Protocol: Developed by Wagh et al. \cite{wagh2020falcon}, the Falcon protocol is an end-to-end SMPC framework specifically designed for DL models.
  It supports high-capacity architectures such as VGG16 and AlexNet, incorporating batch normalization to enhance training stability. By adopting efficient secret-sharing-based protocols for nonlinear operations like ReLU and MaxPool, Falcon achieves an 8x performance improvement compared to previous methods. This makes it a practical choice for DL tasks and opens new possibilities for efficiently handling complex neural networks.
\end{itemize}

\textbf{Scalability and Communication Efficiency:} Research in SMPC has focused on improving scalability and reducing communication costs, which are crucial for distributed ML. Damgård et al. \cite{damgaard2009asynchronous} proposed an asynchronous multi-party computation protocol designed for large-scale distributed computation.
  Combining Shamir's secret sharing and preprocessing techniques, the protocol ensures security even when up to one-third of the participants are compromised. To enhance communication efficiency and scalability, they developed the Virtual Ideal Functionality Framework, which supports parallel secure multiplications. This demonstrates the practical potential of asynchronous protocols for large-scale data environments.

\textbf{Applications of SMPC in FL:} FL represents a key application area for SMPC, focusing on collaborative ML over distributed datasets while addressing privacy concerns. Mugunthan et al. \cite{mugunthan2019smpai} integrated SMPC into FL to address model parameter leakage. By employing cryptographic masking techniques to protect individual contributions and combining differential privacy with SMPC, they enhanced security during model updates. This framework is highly suitable for distributed application scenarios such as medical research, maintaining high computational accuracy while safeguarding data privacy.

\subsubsection{Trusted Execution Environment}
The TEE, as a hardware-level security technology, provides strong technical support for FL, showcasing significant advantages in areas such as privacy protection, secure model parameter aggregation, defense against malicious behavior, and side-channel attack prevention. TEE establishes trusted execution zones at the hardware level to ensure that sensitive information is not accessed or tampered with during operation. Below is an introduction to the mainstream TEE implementations in recent years and their applications in FL.

\begin{figure*}[htbp]
\centering
\vspace{0mm}
\includegraphics[width=\linewidth]{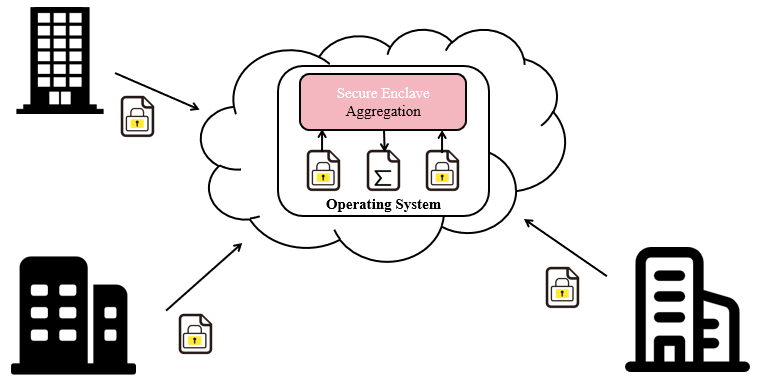}
\caption{SGX Diagram \cite{zhang2021shufflefl}, where secret shares and random masks are combined with careful zero padding to reconstruct missing data when some devices drop out, ensuring secure aggregation in federated learning scenarios. The process involves transforming the data using DFT, applying the peeling decoder to extract all shares, and using IDFT to recover the original data, while maintaining data privacy through the use of secret shares and random masks.}
\label{fig465}
\end{figure*}

\textbf{Mainstream TEE Implementations:} Intel SGX \cite{jauernig2020trusted} (Software Guard Extensions) utilizes "Enclave" isolation for user-space execution, making it suitable for privacy-protecting distributed computing. Intel SGX, which is shown in Fig. \ref{fig465}, provides hardware-level protection for applications, ensuring that sensitive data can be processed securely even in untrusted environments. This technology is critical in FL for safeguarding data during training, preventing leakage or unauthorized access even on client devices. ARM TrustZone \cite{mo2020darknetz} separates hardware resources into ``secure world" and ``normal world" and is widely used in IoT and mobile devices. TrustZone ensures robust isolation and security, allowing only certified code to execute in the secure world, thereby protecting local data privacy. AMD's SEV \cite{jauernig2020trusted} focuses on cloud computing environments, providing memory isolation for virtual machines and enhancing security in virtualized cloud environments. Unlike Intel SGX, SEV targets cloud service providers, ensuring that inter-virtual machine memory data remains protected even against malicious administrators.

\textbf{Applications in FL:} TEE ensures the privacy of training data on local devices. By executing training tasks within a client’s TEE, sensitive data remains in an isolated environment, preventing external access or leakage. For example, Fan et al. \cite{mo2021ppfl} introduced a TEE-based privacy protection mechanism in the PPFL framework, which significantly reduces the risks of data reconstruction and inference attacks by locally safeguarding client data and gradients. Additionally, Hamed Haddadi et al. \cite{mo2020darknetz} isolate the most sensitive layers of DNNs within TEE, effectively enhancing privacy protection. The global model aggregation process in FL is vulnerable to gradient inversion attacks. To address this issue, Yuhui Zhang et al. \cite{zhang2021shufflefl} proposed the ShuffleFL framework, which performs randomization and grouped aggregation within TEE to reduce the likelihood of attackers inferring original data from gradients. On the other hand, Aditya Pribadi Kalapaaking et al. \cite{kalapaaking2022blockchain} integrated blockchain technology to host model aggregation tasks within TEE, achieving tamper-proof and transparent model updates. Malicious clients in FL may attempt to degrade global model performance by uploading fabricated gradients or reusing outdated model updates. To counter this, Xiaoli Zhang et al. \cite{zhang2020enabling} developed the TrustFL framework, utilizing TEE to randomly verify the integrity of client training tasks and ensure adherence to predefined procedures. Furthermore, Arup Mondal et al. \cite{mondal2021poster} proposed FLATEE, which executes model updates in a distributed manner across multiple TEEs, significantly reducing the interference of malicious behavior on aggregation results. TEE can prevent side-channel attacks through hardware isolation and gradient obfuscation techniques. For instance, Yuhui Zhang et al.'s ShuffleFL framework \cite{zhang2021shufflefl} employs a random grouping strategy combined with secure processing within TEE to mitigate attacks through cache and timing analysis. Moreover, the PPFL framework \cite{mo2021ppfl} further reduces the risk of gradient leakage by protecting all layers of the model.

\subsubsection{Secure Aggregation}
Secure aggregation is a computational technology designed to protect data privacy, widely applied in distributed ML scenarios such as FL. In these scenarios, multiple participants (e.g., mobile devices or distributed nodes) collaborate to train a global model while ensuring data privacy. By using encryption, masking, or other privacy-preserving mechanisms, participants share only encrypted or perturbed local model updates with the server. As a result, the server can only access the aggregated result without revealing the individual updates from participants. Below, we discuss significant advancements in secure aggregation through detailed explanations of representative papers. 

\begin{figure*}[htbp]
\centering
\vspace{0mm}
\includegraphics[width=\linewidth]{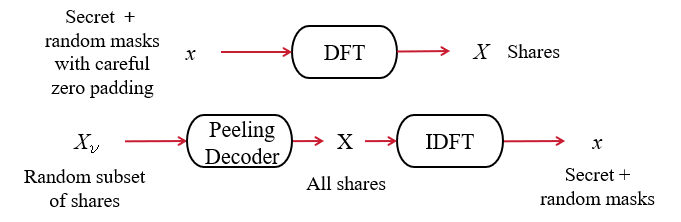}
\caption{Schematic diagram of the algorithm based on the FFT \cite{kadhe2020fastsecagg}, where secret shares and random masks are combined with careful zero padding to reconstruct missing data when some devices drop out, ensuring secure aggregation in federated learning scenarios.}
\label{fig464}
\end{figure*}

\textbf{Techniques Addressing Participant Dynamics:} Liu et al. \cite{liu2022efficient} proposed an innovative secure aggregation method that tackles the issue of device dropout. By leveraging a Homomorphic Pseudorandom Generator and Shamir's secret sharing, they developed an efficient and dropout-resilient aggregation mechanism. Unlike traditional methods that rely on expensive cryptographic primitives such as Diffie-Hellman key exchange, their solution utilizes this method for pre-key generation, significantly reducing communication overhead. Shamir's secret sharing ensures the system can recover global model updates even if some devices fail to participate, improving overall robustness. Experimental results demonstrated that this scheme outperformed existing methods by over six times in runtime efficiency, making it a practical choice for large-scale distributed systems. Kadhe et al. \cite{kadhe2020fastsecagg} designed a secure aggregation protocol based on the Fast Fourier Transform. As is showen in Fig. \ref{fig464}, this approach addresses the high computational complexity of traditional methods by introducing a multi-secret sharing mechanism optimized for FL. The FFT-based design reduces computational costs for both servers and clients while maintaining the capability to handle a fixed proportion of device dropouts. FastSecAgg showed exceptional scalability and adaptability in dynamic environments, making it ideal for large-scale distributed systems. Bonawitz et al. introduced a practical protocol based on one-time masks. The scheme ensures that individual model updates remain private by exchanging random pairwise masks among clients. Even if some devices drop out, the protocol reconstructs the missing masks using pre-distributed secret shares, enabling the server to compute the aggregated result securely. This design is particularly suitable for FL scenarios involving mobile devices, demonstrating effectiveness in handling high-dimensional data under unstable network conditions.

\textbf{Communication-Efficient Secure Aggregation Methods:} Bonawitz et al. \cite{bonawitz2016practical} explored methods to optimize communication costs in secure aggregation. They introduced a combination of random rotation and quantization techniques to compress model updates into smaller sizes while preserving essential information. Random rotation enhances privacy protection and significantly reduces communication overhead, making the method highly effective for transmitting high-dimensional model updates. Furthermore, the protocol employs adaptive tuning to adjust parameters dynamically, ensuring optimal performance across varying hardware configurations and network conditions. This approach effectively addresses the communication bottleneck in FL.

\textbf{Enhancing Privacy and Robustness:} Kairouz et al. \cite{kairouz2021distributed} presented a novel mechanism that combines discrete Gaussian noise with differential privacy to enhance secure aggregation. By locally adding discrete Gaussian noise to model updates, the approach ensures strong privacy guarantees while reducing the risk of data leakage. This mechanism is further integrated with secure aggregation techniques, allowing the server to access only the aggregated results. Their analysis quantified the trade-offs between communication efficiency, privacy protection, and model accuracy, offering an innovative solution for distributed learning systems that achieves near-centralized differential privacy performance at low communication costs. Zhao et al. \cite{zhao2021sear} proposed a hardware-based solution using Intel SGX to protect privacy and defend against Byzantine adversaries. The framework employs sampling-based anomaly detection to identify and filter malicious participants' updates by analyzing specific model features. Intel SGX ensures data confidentiality through hardware isolation, even in adversarial environments. This framework is well-suited for high-risk scenarios where robust privacy protection and resilience against malicious attacks are critical.

\subsubsection{Quantum Cryptography}
Quantum cryptography leverages the principles of quantum mechanics to provide enhanced security in communication systems. Unlike traditional cryptographic methods that rely on mathematical complexity, quantum cryptography uses the fundamental properties of quantum mechanics, such as superposition and entanglement, to ensure secure communication. This section discusses the integration of quantum cryptography with SemCom, focusing on how quantum techniques can enhance the security and efficiency of semantic data transmission.

\textbf{Quantum Key Distribution (QKD):} One of the most prominent applications of quantum cryptography is QKD. QKD allows two parties to generate a shared random secret key, which can be used for encrypting and decrypting messages. The security of QKD is based on the principles of quantum mechanics, making it immune to computational attacks that can compromise traditional cryptographic keys. In the context of  SemCom, QKD can be used to securely transmit semantic keys, ensuring that only authorized parties can access the semantic information. For instance, the study by Chehimi et al. \cite{chehimi2024quantum} proposes a quantum  SemCom framework that integrates QKD to enhance the security of semantic data transmission. The authors demonstrate that QKD can significantly reduce the risk of eavesdropping and ensure the confidentiality of semantic information.

\textbf{Quantum Digital Signatures(QDS):} QDS are another important application of quantum cryptography. QDS provides a way to sign digital messages using quantum states, ensuring the authenticity and integrity of the messages. In  SemCom, QDS can be used to sign semantic messages, preventing unauthorized parties from tampering with the semantic information. This is particularly important in scenarios where semantic data needs to be shared among multiple parties, such as in distributed SemCom networks. The work by Nunavath et al. \cite{nunavath2024towards} explores the use of QDS in quantum SemCom, highlighting its potential to enhance the security and reliability of semantic data transmission. The authors show that QDS can effectively prevent message tampering and ensure the integrity of semantic information.

\textbf{Quantum Secure Direct Communication (QSDC):} QSDC is a protocol that allows for the direct transmission of secret messages without the need for a pre-shared key. QSDC uses quantum states to encode the messages, ensuring that any eavesdropping attempt will be detected. In  SemCom, QSDC can be used to transmit semantic messages directly, providing an additional layer of security. This is particularly useful in scenarios where real-time  SemCom is required, such as in emergency response systems or military applications. The study by Khalid et al. \cite{khalid2023quantum} investigates the application of QSDC in quantum  SemCom, demonstrating its potential to enhance the security and efficiency of semantic data transmission. The authors highlight the advantages of QSDC in terms of reducing latency and improving the reliability of SemCom.

\begin{les}
In traditional communication systems, encryption technologies like AES and RSA ensure data confidentiality but limit semantic processing, posing challenges for SemCom. To address these challenges, emerging encryption technologies such as SMPC \cite{chen2019secure,knott2021crypten,wagh2019securenn,wagh2020falcon,damgaard2009asynchronous,mugunthan2019smpai}, HE \cite{aono2017privacy,fang2021privacy,lee2022privacy,sun2018private,zhang2020batchcrypt}, TEE \cite{jauernig2020trusted,mo2020darknetz,jauernig2020trusted,mo2021ppfl,mo2020darknetz,zhang2021shufflefl},  secure aggregation \cite{liu2022efficient,kadhe2020fastsecagg,bonawitz2016practical,kairouz2021distributed,zhao2021sear}, and Quantum Cryptography \cite{chehimi2024quantum,nunavath2024towards,khalid2023quantum} offer a balance between data protection and semantic processing, enhancing the efficiency and security of SemCom systems. Notably, PHE simplifies operations and is suitable for lightweight privacy-preserving scenarios, while FHE excels in supporting complex tasks with strong security guarantees. With ongoing advancements in algorithms and expanding applications, HE is set to play an increasingly vital role in privacy preservation and collaborative data computation in future SemCom.
\end{les}

\subsection{Block-Chain Technology}

\begin{figure*}[t]
\centering
\vspace{-10mm}
\includegraphics[width=\linewidth]{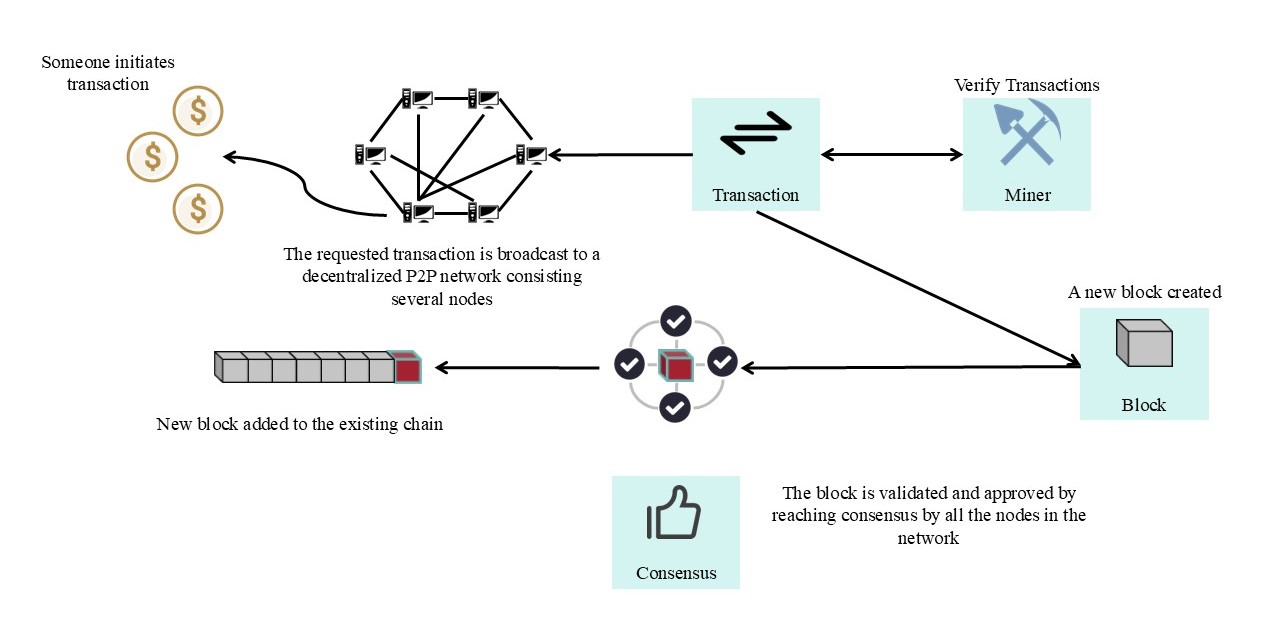}
\caption{Functional diagram of a blockchain network \cite{monrat2019survey}, where blockchain transaction handling starts with an individual initiating a transaction, which is then broadcasted across a decentralized peer-to-peer network of nodes. These nodes are individual computers maintaining a copy of the blockchain ledger, receive the transaction. Miners, specialized nodes equipped with enhanced capabilities, take on the task of verifying the transaction's legitimacy by checking the sender's balance, confirming the digital signature, and ensuring the transaction has not been previously spent, thus preventing fraud and maintaining the ledger's accuracy. Once verified, the miner includes the transaction in a newly created block, which is a collection of transactions, and solves a complex cryptographic puzzle to propose this block to the network.}
\label{fig471}
\end{figure*}

As shown in Fig. \ref{fig471}, block-chain technology is an undeniable ledger technology that stores transactions in high-security chains of blocks \cite{li2021blockchain}. In AI-enabled networks, block-chain technology has been widely applied to secure the model training in FL by sharing a shared transaction ledger without requiring a trusted third party \cite{kim2018device,peng2021vfchain,desai2021blockfla,8998397,9019859,8892848,8106871}. As in \cite{kim2018device,8998397}, each participant in SemCom system can attach the model during the model collaborative training process. After exchanging and verifying the received model updates, the miners compete to work on a block. The first miner who finishes the job records the generated block to the distributed ledger. Then the participants download the block for the local model update for model co-training. In addition to the model update process, given that it is impossible to change a single block without being detected \cite{kim2018device}, the node in SemCom system checks the block header to ensure the immutability and integrity of the received model and model slice during the model transmission process. Blockchain also enables SemCom system with audibility and traceability when a suspicious event occurs. As in \cite{desai2021blockfla,peng2021vfchain}, the clients' requests and the server's responses are recorded in an auditable manner. In such a way, the abnormal node can be found by evaluating the contribution of each node's model to the malicious event. 

The below, as shown in Table \ref{blockchain}, the following content, based on multiple academic studies, delves into various blockchain methods for different stages and their potential applications in SemCom.

\begin{longtable}{|A{2cm}|A{0.6cm}|P{9.5cm}|}  
\caption{\footnotesize{List of Representative Block-Chain Schemes}} 
\label{blockchain}\\
\hline
\textbf{Sub-class} & \textbf{Ref.} & \textbf{Descriptions, including} $\circledast$: \textbf{contributions; }

$\circleddash$: \textbf{trade-offs;} 
$\boxplus$: \textbf{performance metrics;}

$\Cap$: \textbf{applied in SemCom or promising for SemCom}
\\
\hline
\endfirsthead

\hline
\textbf{Sub-class} & \textbf{Ref.} & \textbf{Descriptions, including} $\circledast$: \textbf{contributions; }

$\circleddash$: \textbf{trade-offs;} 
$\boxplus$: \textbf{performance metrics;} 

$\Cap$: \textbf{applied in SemCom or promising for SemCom} \\
\hline
\endhead

\hline
\endfoot

\hline
\endlastfoot
Data Security & \cite{lin2023verifiable} &
$\circledast$: Integrates blockchain with semantic ecosystems for secure sharing of semantic information, utilizing a semantic consensus mechanism and semantic sharing methods.

$\circleddash$: Semantic redundancy and information consistency.

$\boxplus$: Accuracy, PSNR, and time.

$\Cap$: Applied in securing data sharing in SemCom.
\\
\hline

Data Security & \cite{lin2023blockchain} &
$\circledast$: Introduces a blockchain-assisted SemCom framework for optimizing knowledge bases in remote driving, reducing communication costs through semantic segmentation and edge computing.

$\circleddash$: Blockchain throughput and scalability.

$\boxplus$: Latency. PSNR and loss.

$\Cap$: Applied in knowledge management in SemCom.
\\
\hline

Prevent Node Leakage & \cite{liang2020intrusion} &
$\circledast$: Proposes an IoT-specific blockchain node design for intrusion detection, integrating smart contracts and encryption techniques to secure node communication.

$\circleddash$: Accuracy and number of parameters, measuring time and system performance.

$\boxplus$: Accuracy, precision and recall rate.

$\Cap$: Promising for securing nodes in decentralized SemCom systems.
\\
\hline

Prevent Node Leakage & \cite{li2020blockchain} &
$\circledast$: Introduces a blockchain-based FL framework with a Committee Consensus Mechanism, improving security and efficiency by restricting consensus to a selected group of nodes.

$\circleddash$: Model performance and resource consumption of deep learning models.

$\boxplus$: Accuracy.

$\Cap$: Promising for securing FL frameworks in SemCom.
\\
\hline

Model Framework Design & \cite{kim2019blockchained} &
$\circledast$: Proposes a decentralized blockchain framework for federated learning that eliminates central servers, ensuring data privacy and system robustness.

$\circleddash$: Challenges in ensuring global model consistency across decentralized nodes.

$\boxplus$: Average learning completion latency.

$\Cap$: Promising for decentralized ML applications in SemCom.
\\
\hline

Model Framework Design & \cite{rathore2019blockseciotnet} &
$\circledast$: Integrates blockchain, SDN, fog computing, and edge computing for IoT security in smart cities, enhancing attack detection and mitigation via smart contracts.

$\circleddash$: Complexity of integrating multiple technologies and ensuring seamless operation.

$\boxplus$: Accuracy, detection time, bandwidth.

$\Cap$: Promising for securing IoT-based SemCom systems.
\\
\hline

\end{longtable}

\subsubsection{Data Security}

A lot of researches have been done to prove that intermediate gradients can reveal sensitive information about training data, highlighting the limitations of FL frameworks in fully protecting data privacy, especially in the presence of malicious attackers. Moreover, the issue of data leakage during distributed transmission further exacerbates privacy concerns. Blockchain technology, with its decentralized architecture, cryptographic techniques, consensus algorithms, and distributed storage capabilities, offers a promising solution. By leveraging its unique data structure and integrating it with complementary technologies, blockchain can enhance the security of data transmission and effectively mitigate data leakage risks, providing a more robust and reliable framework for FL systems.

\textbf{Privacy-Enhanced FL:} Preuveneers et al. \cite{preuveneers2018chained} introduce Privacy-Enhanced FL , an innovative framework that leverages blockchain technology to enhance data security and privacy in FL. By recording iterative updates to eliminate the need for centralized training data. This approach upholds the decentralized nature of model training and effectively preserves data privacy. The proposed framework consists of three key components: the Key Generation Center, the Central Server, and multiple participants. It also integrates post-quantum security measures to protect the privacy of the aggregator and secure the data, even in cases where attackers may collude with multiple entities. Additionally, the auditing mechanism within the proposed framework monitors and validates blockchain transactions, enabling the detection of tampering or fraudulent activities related to model updates. This comprehensive solution significantly strengthens the security and integrity of FL systems.

\textbf{Blockchain-based FL:} Liu et al. \cite{liu2020secure} presents a blockchain-based FL framework aimed at safeguarding data security and privacy in 5G networks. The framework mitigates contamination attacks by automatically verifying model updates through smart contracts. It also incorporates local differential privacy techniques to protect against membership inference attacks. Leveraging the transparency and immutability of blockchain, the framework ensures robust data security, offering an effective solution for securing FL in 5G environments.

\textbf{Semantic information sharing blockchain framework:} Lin et al. \cite{lin2023verifiable} propose an innovative framework that combines blockchain technology with semantic ecosystems to facilitate the sharing of semantic information. The framework is composed of four main components: the semantic encoder, channel, semantic proof, and semantic decoder. To tackle the ``garbage-in, garbage-out" issue, the authors introduce a semantic consensus mechanism based on threshold signatures, which ensures that only accurate and relevant semantic information is recorded on the blockchain. Additionally, the framework employs a semantic sharing mechanism that utilizes state channels and task-related information bottleneck methods, significantly reducing information redundancy. By leveraging blockchain’s security and decentralization alongside the efficient information processing capabilities of SemCom, the framework enhances network resource utilization while ensuring data security and trust.

\begin{figure*}[t]
\centering
\vspace{-10mm}
\includegraphics[width=0.7\linewidth]{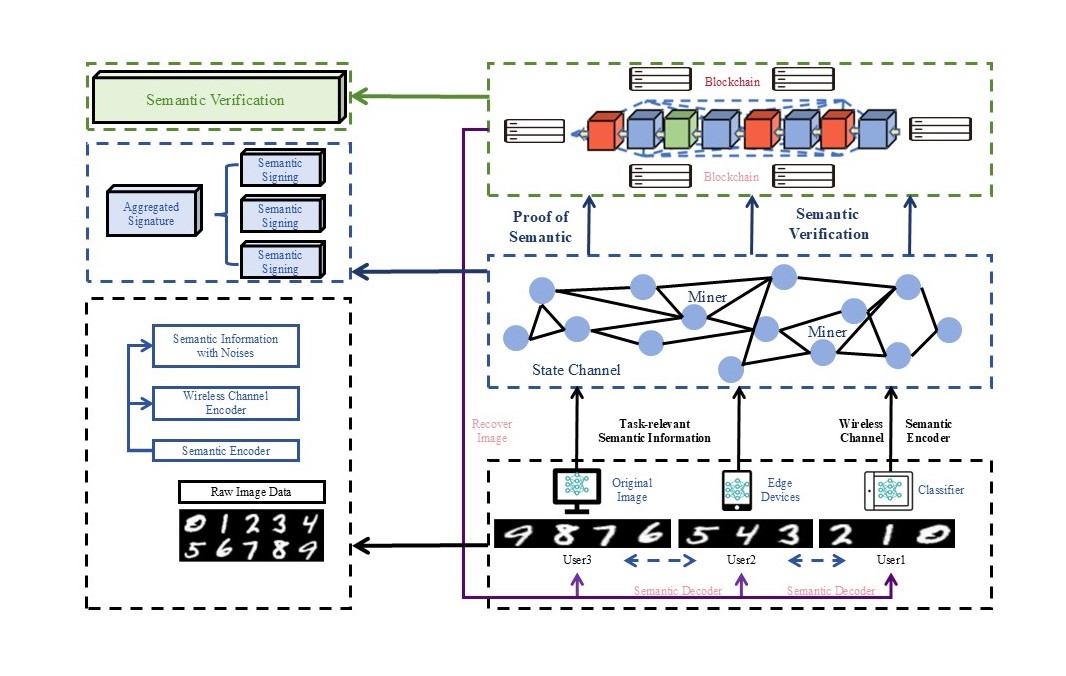}
\caption{An illustrative framework of Blockchain-Semantic ecosystems \cite{lin2023verifiable}, where begins with raw image data that undergoes semantic encoding to extract meaningful information, which is then transmitted through a wireless channel potentially affected by noise. At the receiving end, a semantic decoder retrieves the information, which is subsequently used for semantic analysis. This analysis involves an aggregated signature that encompasses semantic signatures, semantic strings, and semantic signatures to ensure data integrity and authenticity. Simultaneously, the system operates on a blockchain network where miners validate transactions through semantic verification, ensuring that the information recorded on the blockchain is accurate and tamper-proof. The proof of semantic is crucial here, as it serves as evidence of the semantic integrity of data. Once verified, the semantic information is used to update the blockchain, which is a series of blocks containing verified transactions. The process also includes a state channel that facilitates off-chain transactions for specific tasks, such as image recovery and task-relevant semantic information extraction, enhancing efficiency and reducing load on the main blockchain.}
\label{fig472}
\end{figure*}

\textbf{Optimize and maintain semantic knowledge base:} As illustrated in Fig. \ref{fig472}, Lin et al. \cite{lin2023blockchain} propose a blockchain and edge computing-assisted SemCom framework designed to optimize and maintain knowledge bases within the remote driving domain. The primary objective of the framework is to reduce communication costs by prioritizing the transmission of semantically meaningful information over simple bit-level data. The authors introduce a semantic segmentation strategy, leveraging geographically distributed edge nodes that can adapt to diverse contextual environments, enabling the integrated management of multiple knowledge bases. To ensure system security, the framework capitalizes on the tamper-proof characteristics of blockchain technology. Additionally, the architecture integrates task-specific sharding to enhance the blockchain network's ability to efficiently process knowledge base update transactions, significantly improving transaction throughput.

\textbf{Blockchain consensus mechanism based on DL:} Li et al. \cite{li2019dlbc} introduce a novel blockchain consensus mechanism called DL-Based Consensus to address the challenges posed by large models and training datasets. The proposed method harnesses the computational power of blockchain miners for DL training and replaces the traditional proof-of-work  hash computation with a Proof-of-Useful-Work approach, overcoming the limitations of conventional consensus mechanisms. To protect the ownership of DL models, they employ the DNN-watermark technique, which embeds a unique watermark within the model to verify ownership and prevent unauthorized use or modification. This approach not only enhances model security but also reduces the energy consumption and computational resource waste associated with blockchain operations, offering a more efficient solution for blockchain-based DL applications.

\subsubsection{Prevent Node Leakage}

Each node plays an important role in the blockchain network and is responsible for maintaining and running the blockchain, including validation, storage and dissemination of data. The large number of nodes in IoT is a key reason for security issues, as leakage of a single node reduces the security of the entire network.

\textbf{Secure blockchain node design scheme:} To address the security of nodes, Liang et al. \cite{liang2020intrusion} proposed a blockchain node design scheme, which is mainly centered on building a Smart Efficient Secure and Scalable intrusion detection system for the Internet of Things.The proposed system integrates five core components: a blockchain smart contract module, detection and analysis module, response module, data processing module, and collection module. Smart contracts define and enforce communication rules between nodes, ensuring compliance with predefined behaviors. By embedding databases and blockchain nodes, the system enhances data storage and retrieval efficiency. Localized data processing between nodes minimizes communication overhead while leveraging blockchain technology to maintain data consistency and security. The system employs the Elliptic Curve Digital Signature Algorithm to encrypt public and private keys, ensuring that only authorized nodes can generate or access specific communication messages. Encryption algorithms safeguard node communication against unauthorized access and tampering. All communication messages, including commands and feedback, must follow methods specified by smart contracts, with data integrity verified using the sender’s public key, the data itself, and its digital signature before being accessed by the receiver. These design features enhance the effectiveness and security of blockchain-enabled nodes within IoT intrusion detection systems.

\textbf{Committee consensus mechanism:} Li et al. \cite{li2020blockchain} present the Block-chain-based FL framework with Committee Consensus, an innovative solution to the security challenges in FL. This framework integrates blockchain technology for securely storing global models and local updates, safeguarding user data privacy while enabling collaborative model optimization. The federated blockchain structure features model blocks for storing global models and update blocks for recording local updates during each training round. The core of the proposed method is the Committee Consensus Mechanism, which improves security and efficiency by restricting the consensus process to a selected group of nodes forming a committee. This design ensures that malicious actors must control more than half of the committee to influence the global model, thereby significantly enhancing resilience to attacks while reducing the computational cost of consensus. Nodes within the system actively contribute by accessing the global model and uploading updates, fostering mutual cooperation. To reinforce security, the framework enforces strict node management via an initial manager node, permitting only authenticated devices to participate. A blacklist-based approach effectively blocks malicious nodes, preventing unauthorized access. By addressing vulnerabilities associated with malicious clients and central servers, the proposed framework provides a robust, secure, and efficient solution for FL in decentralized environments.

\textbf{Blockchain-based hybrid architecture:} Desai et al. \cite{desai2021blockfla} introduce BlockFLA, a blockchain-based hybrid architecture designed to enhance the security and efficiency of FL. BlockFLA achieves this by evaluating each node's model contributions to detect anomalous nodes involved in malicious activities. The architecture integrates a private chain and a public chain, each serving distinct purposes: the private chain functions as the aggregation server, while the public chain verifies node behavior and enforces penalties. In this system, nodes send hashes of their model updates to the public chain for verification. Smart contracts on the public chain automatically detect malicious activities and enforce penalties, such as forfeiting the deposit of the offending node and redistributing it to honest participants. Communication between nodes on the private chain is encrypted, ensuring the privacy of node data, while the tamper-proof transaction records on the public chain provide additional security and trust.By combining encrypted communication, tamper-resistant verification, and automated penalty mechanisms, BlockFLA establishes a secure and trustworthy FL environment, effectively mitigating the risks posed by malicious nodes and fostering reliable collaboration.

\textbf{Deep reinforcement learning-based node selection:} Lu et al. \cite{lu2020blockchain} propose a Deep Reinforcement Learning-based node selection algorithm to optimize the participation of nodes in a blockchain-enhanced asynchronous FL framework for IoV data sharing. This framework leverages asynchronous FL to allow nodes to train at varying speeds, eliminating delays caused by the slowest nodes and significantly enhancing system efficiency. It incorporates a hybrid blockchain architecture, PermiDAG, which combines a permissioned blockchain with a local Directed Acyclic Graph. In the local DAG, each vehicle node maintains its transaction records independently and synchronizes encrypted updates with neighboring vehicles, ensuring robust data privacy. This innovative approach improves the efficiency and security of FL in IoV environments, addressing challenges related to node participation, asynchronous training, and data privacy in decentralized systems.

\subsubsection{Model Framework Design}

In addition to starting at the data level and the node level, the overall architecture of the model can be designed to achieve the purpose of ensuring network security.The blockchain model typically follows a layered architecture, structured from the bottom up into the data layer, network layer, consensus layer, incentive layer, contract layer, and application layer. It enhances the flexibility and scalability of blockchain systems, allowing them to effectively meet the demands of diverse application scenarios.

\textbf{Decentralized block-chained FL framework:} Kim et al. \cite{kim2019blockchained} propose Blockchained FL, an innovative decentralized ML framework that leverages blockchain technology to facilitate the exchange and verification of local model updates between devices. This framework eliminates the need for a central server or centralized training data, enabling on-device ML. In BlockFL, each device computes its local model updates and uploads them to miners within the blockchain network. The miners use a Proof-of-Work consensus mechanism to exchange and verify these updates, recording the validated updates on the blockchain. Subsequently, devices download the new blocks to compute global model updates, completing the FL process. By design, BlockFL enhances system robustness, mitigates single points of failure, and incentivizes wider participation by rewarding devices in proportion to the size of their training samples. Furthermore, the framework ensures data privacy and security, making it a promising approach for decentralized ML.

\textbf{Block-chain secure IoT framework:} Rathore et al. \cite{rathore2019blockseciotnet} propose BlockSecIoTNet, a decentralized security architecture designed to enhance the security of IoT networks within smart city environments. This innovative architecture integrates key technologies such as software-defined networking, blockchain, fog computing, and mobile edge computing, forming a layered model with distinct functions at each layer: data collection, traffic analysis, and attack detection and mitigation. Attack detection models are shared and updated among fog nodes via smart contracts, thereby improving the accuracy of threat detection. Through this integrated approach, BlockSecIoTNet enables continuous monitoring and proactive analysis of potential security threats, providing optimized protection for IoT ecosystems in smart cities.

\textbf{Blockchain-based distributed learning framework:} Lugan et al. \cite{lugan2019secure} introduce TCLearn, a blockchain-based distributed learning security architecture that is applicable to consortia with varying levels of trust, supporting both public and permissioned blockchains. TCLearn offers tailored security and privacy mechanisms for public learning models, private learning models, and untrusted consortium members. The architecture utilizes the Federated Byzantine Protocol to ensure both model performance and data privacy by enabling the secure evaluation and logging of model updates via blockchain. This approach fosters enhanced trust and security in distributed learning environments, ensuring the integrity and confidentiality of the learning process.

\textbf{Blockchain-based distributed learning security framework}: Awan et al. \cite{awan2019poster} introduces the Blockchain-assisted Privacy-Preserving FL framework, which integrates blockchain technology with FL to enhance data privacy and security. By leveraging blockchain's immutability and decentralization, the framework ensures traceability and verification of model updates. It employs homomorphic encryption and proxy re-encryption techniques to securely aggregate client model updates without exposing local data. These updates are asynchronously recorded on the blockchain, addressing challenges such as random client dropouts. Additionally, the framework incorporates incentive mechanisms based on clients' contributions to global model optimization and introduces value premiums for model ownership, rewarding participating miners. This approach fosters a more transparent, equitable, and secure collaborative learning environment.

\textbf{Point-to-point ML framework:} Shayan et al. \cite{shayan2020biscotti} introduce Biscotti, a decentralized point-to-point ML framework designed to uphold data privacy and security in multi-party ML processes through the integration of blockchain technology and cryptographic techniques. The framework innovatively incorporates a Proof-of-Federation consensus protocol to filter and authorize key participants in the model update process. Furthermore, Biscotti employs differential privacy methods and secure multi-party computation techniques to mitigate data leakage risks and defend against model poisoning attacks. The system also adopts a verifiable secret sharing scheme to securely aggregate updates from multiple participants. By eliminating the reliance on a centralized server, Biscotti effectively establishes a FL environment that safeguards against known attacks, ensures data privacy, and enhances model robustness.



\begin{les}
In SemCom system, semantic information is inherently user-centric and user-generated, necessitating measures to preserve its value and safeguard it from unauthorized dissemination among participants during data mining processes. Integrating blockchain technology with SemCom can facilitate reliable updates to shared knowledge, foster trust among participants, and enhance the recognition of semantic information's intrinsic value. Blockchain-enabled SemCom can be achieved from the data \cite{preuveneers2018chained,liu2020secure,lin2023verifiable,lin2023blockchain,li2019dlbc}, node \cite{liang2020intrusion,li2020blockchain,desai2021blockfla,lu2020blockchain}, and model perspectives \cite{kim2019blockchained,rathore2019blockseciotnet,lugan2019secure,awan2019poster,shayan2020biscotti}.
\end{les}

\subsection{Model Compression}

Model compression aims at reducing the size of DL models and enhancing the operational efficiency. This is generally achieved by minimizing the model's computational requirements, such as the number of parameters and computational complexity, without compromising its accuracy significantly. The importance of model compression is paramount when deploying DL models on resource-constrained devices, such as mobile devices, embedded systems, or IoT devices, due to their limited memory, computational power, and battery life. By employing model compression techniques, devices can effectively decrease communication costs, conserve storage space, and eliminate parameter redundancy, thereby strengthening the SemCom system to defend against against multi-party cooperation attacks. Model compression techniques can be divided into the following sub-categories: model pruning, parameter quantization, low-rank decomposition, and knowledge distillation. The representative model compression schemes are listed in Tab. \ref{modelcompression}.

\begin{longtable}{|A{2cm}|A{0.6cm}|P{9.5cm}|}  
\caption{\footnotesize{List of Representative Model Compression Schemes}} 
\label{modelcompression}\\
\hline
\textbf{Sub-class} & \textbf{Ref.} & \textbf{Descriptions, including} $\circledast$: \textbf{contributions; }

$\circleddash$: \textbf{trade-offs;} 
$\boxplus$: \textbf{performance metrics;}

$\Cap$: \textbf{applied in SemCom or promising for SemCom}
\\
\hline
\endfirsthead

\hline
\textbf{Sub-class} & \textbf{Ref.} & \textbf{Descriptions, including} $\circledast$: \textbf{contributions; }

$\circleddash$: \textbf{trade-offs;} 
$\boxplus$: \textbf{performance metrics;} 

$\Cap$: \textbf{applied in SemCom or promising for SemCom} \\
\hline
\endhead

\hline
\endfoot

\hline
\endlastfoot
Model Pruning & \cite{jiang2022model} &
$\circledast$: Presents a novel FL approach called PruneFL, which adapts the model size during the training process to reduce communication and computation overhead while maintaining similar accuracy to the original model.

$\circleddash$: Communication and computation overhead and model accuracy.

$\boxplus$: Computation time, communication time, and test accuracy.

$\Cap$: Promising for FL training in SemCom.
\\

\hline

Model Pruning & \cite{ding2024patrol} &
$\circledast$: Uses privacy-oriented pruning to balance privacy, efficiency, and utility in collaborative inference. By deploying more layers at the edge and leveraging Lipschitz regularization and adversarial reconstruction training, the scheme reduces the risk of model inversion attacks and enhances the target inference model.

$\circleddash$: Accuracy, efficiency, and defense capability.

$\boxplus$: Prediction accuracy, PSNR, SSIM, and attack accuracy.

$\Cap$: Promising for SemCom to defend against inference attacks.
\\
\hline

Model Pruning   &  \cite{xie2020lite}  &
$\circledast$: Proposes a lightweight distributed SemCom system for text transmission with low complexity. By pruning the model redundancy and lowering weight resolution, the scheme significantly decreases the bandwidth required for model weight transmission.

$\circleddash$: Transmission accuracy and model complexity.

$\boxplus$: MSE and BLEU score.

$\Cap$: Applied in SemCom for text transmission.
\\
\hline

Parameter Quantization   &  \cite{nagel2021white}  &
$\circledast$: Discusses two main classes of algorithms: post-training quantization and quantization-aware-training. Post-training quantization is a lightweight approach that requires no re-training or labeled data. Quantization-aware-training requires fine-tuning and access to labeled training data but enables lower bit quantization with competitive results.

$\circleddash$: Complexity of layers and accuracy.

$\boxplus$: MSE and accuracy.

$\Cap$: Promising for parameter quantization in SemCom.
\\
\hline

Parameter Quantization   &  \cite{park2024vision}  &
$\circledast$: Introduces a vision transformer-based SemCom system with importance-aware quantization to facilitate SemComs without end-to-end training. Importance-aware quantization assigns varying quantization bits to patches, prioritizing those with higher importance.

$\circleddash$: Classification accuracy and quantization complexity.

$\boxplus$: Classification accuracy.

$\Cap$: Applied in SemCom without end-to-end training.
\\
\hline

Low-Rank Decomposition   &  \cite{swaminathan2020sparse}  &
$\circledast$: Introduces sparse low rank for compressing DNNs, which combines low-rank decomposition and neuron significance. The proposed method achieves better compression rates by sparsifying the singular value decomposition matrices and keeping lower rank for unimportant neurons.

$\circleddash$: Testing accuracy 

$\boxplus$: Compression rate and accuracy.

$\Cap$: Promising for model compression in SemCom.
\\
\hline

Low-Rank Decomposition   &  \cite{idelbayev2020low}  &
$\circledast$: Proposes a mixed discrete-continuous optimization approach to jointly optimize the ranks and matrix elements, resulting in better rank selection.

$\circleddash$: Classification accuracy and model size.

$\boxplus$: Test error, rank, and accuracy.

$\Cap$: Promising for model optimization in SemCom.
\\
\hline

Knowledge Distillation   &  \cite{gao2021residual}  &
$\circledast$: Proposes the residual error-based knowledge distillation to address the performance degradation in knowledge distillation due to the gap between the learning capacities of the student and teacher models.

$\circleddash$: Accuracy and robustness.

$\boxplus$: Top-1 accuracy, top-5 accuracy, and success rate of applying adversarial attacks 

$\Cap$: Promising for SemCom to defend against adversarial attacks.
\\
\hline

Knowledge Distillation   &  \cite{liu2023knowledge}  &
$\circledast$: Analyzes four types of knowledge transfer and model compression techniques to boost robustness and generalization in a multi-user SemCom system characterized by limited training samples and unforeseen interference.

$\circleddash$: Transmission performance and model complexity.

$\boxplus$: BLEU scores, parameters, size, training time, and inference time.

$\Cap$: Applied in multi-user SemCom.
\\
\hline

Knowledge Distillation   &  \cite{xing2023feddistillsc}  &
$\circledast$: Introduces a distributed SemCom system leveraging federated distillation for image classification tasks. This approach facilitates knowledge sharing without compromising private data, thereby enhancing each transmitter's semantic extraction capabilities.

$\circleddash$: Transmission performance and model complexity.

$\boxplus$: Top-1 accuracy.

$\Cap$: Applied in distributed SemCom.
\\
\hline

Knowledge Distillation   & \cite{nam2024language}   &
$\Delta$: Proposes the language-oriented SemCom by merging LLMs and generative models with SemCom.

$\Theta$: Transmission performance and model complexity.

$\Lambda$: Learned Perceptual Image Patch Similarity (LPIPS).

$\Xi$: Applied in LLM-aided SemCom for better robustness.
\\
\hline

Knowledge Distillation   &  \cite{albaseer2024tailoring}  &
$\circledast$: Introduces a SemCom framework tailored for heterogeneous, resource-constrained edge devices and computation-intensive servers. They leverages dynamic knowledge distillation to customize semantic models for each device, balancing computational and communication constraints while ensuring QoS.

$\circleddash$: Transmission performance and model complexity.

$\boxplus$: Inference accuracy, computational Complexity, and normalized transmit power consumption.

$\Cap$: Applied in SemCom with heterogeneous, resource-constrained edge devices and computation-intensive servers.
\\
\hline

\end{longtable}

\subsubsection{Model Pruning}

Model pruning diminishes the size of neural networks by eliminating unnecessary connections or parameters while striving to preserve their performance to the fullest extent. Model pruning primarily falls into two categories: unstructured pruning and structured pruning. Unstructured pruning involves directly removing individual parameters or connections, particularly those with smaller weight values or neuron nodes. While this method can drastically decrease the number of model parameters and theoretical computation, the resulting pruned model is typically sparse and disrupts the original model structure, necessitating special hardware for acceleration. Conversely, structured pruning operates at the level of filters or entire network layers. When a filter is pruned, the preceding and subsequent feature maps will undergo corresponding changes, yet the model's structure remains unaffected, enabling it to be accelerated by GPUs or other hardware. Structured pruning is more viable in practical applications.

\begin{figure*}[t]
\centering
\vspace{-10mm}
\includegraphics[width=\linewidth]{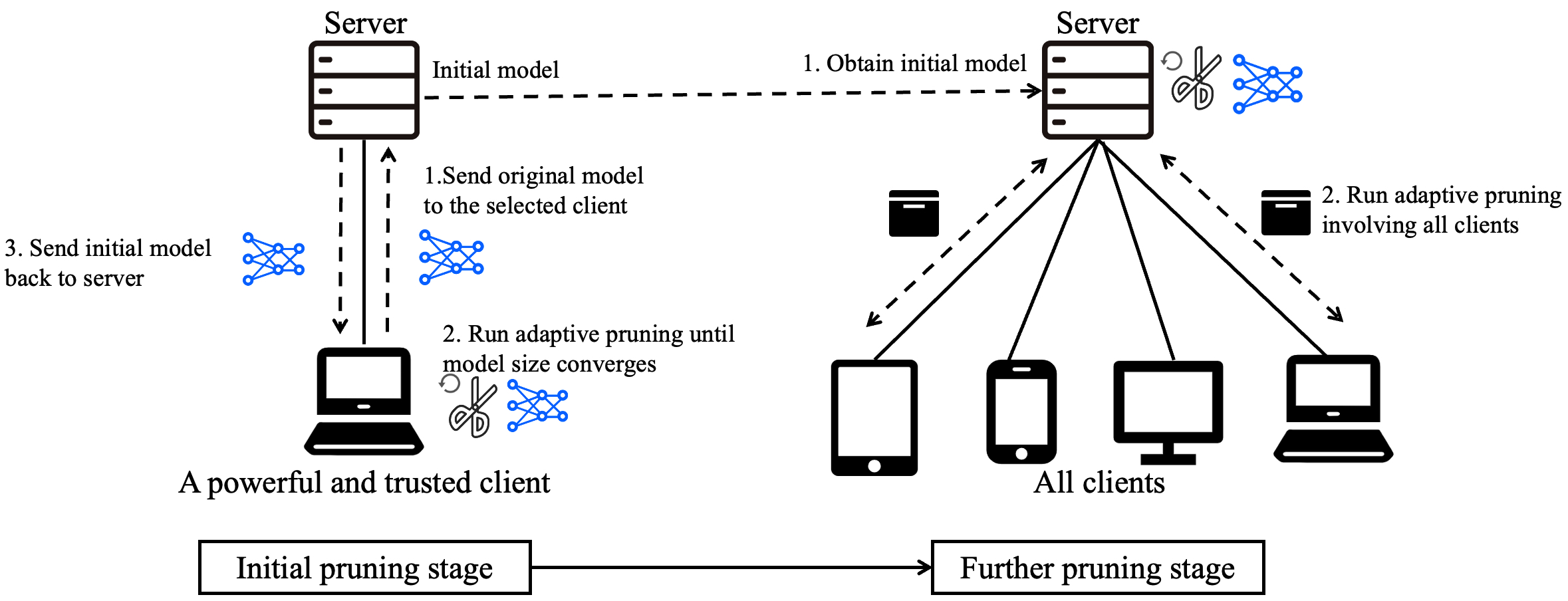}
\caption{Combination of model pruning and FL \cite{jiang2022model}, where two stages are involved, including initial pruning and further pruning stages. Initial pruning stage: Firstly, it facilitates the commencement of the FL process with a compact model, thereby substantially diminishing the computational and communication overhead per FL round. Secondly, when clients possess varied computational resources, the chosen client for pre-selection pruning can be a powerful and reliable one to expedite the pruning phase. Further pruning stage: The model refined through pre-selection pruning may not be fully optimized, as it is based solely on one client's data. However, it provides a solid foundation for the FL process encompassing all clients. Throughout the FL process, the standard FedAvg procedure with further adaptive pruning is augmented. This allows the model to dynamically expand or contract, optimizing training efficiency. In this phase, data from all participating clients contribute to the refinement.}
\label{fig481}
\end{figure*}

\textbf{Combination with FL:} As depicted in Fig. \ref{fig481}, Jiang et al. \cite{jiang2022model} present a novel FL approach called PruneFL, which adapts the model size during the training process to reduce communication and computation overhead while maintaining similar accuracy to the original model. The approach includes initial pruning at a selected client and further pruning as part of the FL process, and the model size is adapted to maximize the approximate empirical risk reduction divided by the time of one FL round. Experiments on edge devices show that PruneFL significantly reduces training time compared to conventional FL, and the pruned model converges to an accuracy similar to the original model, also acting as a lottery ticket of the original model.

\textbf{Defense against model inversion attacks:} Ding et al. \cite{ding2024patrol} introduce a Privacy-Oriented Pruning for Collaborative Inference (PATROL) scheme that uses privacy-oriented pruning to balance privacy, efficiency, and utility in collaborative inference. By deploying more layers at the edge and leveraging Lipschitz regularization and adversarial reconstruction training, PATROL reduces the risk of model inversion attacks and enhances the target inference model.

\textbf{Defense against membership inference attacks:} Wang et al. \cite{wang2020against} propose a pruning algorithm that reduces model storage and computational operations while also improving resistance to membership inference attacks. Experimental results demonstrate that the proposed algorithm can find a subnetwork that prevents privacy leakage and achieves competitive accuracy compared to the original DNNs, with attack accuracy reduced by up to 13.6\% and 10\% compared to the baseline and Min-Max game, respectively. Yuan et al. \cite{yuan2022membership} conducts the first analysis of privacy risks in the context of membership inference attacks and proposes a self-attention membership inference attack against pruned neural networks. Experiments show that the proposed attack performs better than existing attacks on pruned models. Additionally, the paper proposes a new defense mechanism to protect the pruning process by mitigating prediction divergence based on KL-divergence distance, which effectively mitigates privacy risks while maintaining sparsity and accuracy of the pruned models.

\textbf{Defense against adversarial attacks:} Ye et al. \cite{ye2019adversarial} propose a framework of concurrent adversarial training and weight pruning to enable model compression while preserving adversarial robustness, addressing the dilemma of adversarial training. The study finds that weight pruning is essential for reducing network model size in the adversarial setting, and training a small model from scratch cannot achieve either adversarial robustness or high standard accuracy. Wu et al. \cite{wu2021adversarial} combines adversarial training and model pruning in a joint formulation during training to address both model size and robustness against attacks. The method eliminates the need for heuristics and pre-trained models, enabling better compression and robustness.

\textbf{Pruning for SemCom:} 
Xie et al. \cite{xie2020lite} consider an IoT network scenario where the cloud/edge platform handles the training and updating of SemCom models, while IoT devices focus on data collection and transmission based on the trained model. To make it feasible for IoT devices, they propose a lightweight distributed SemCom system for text transmission with low complexity. By pruning the model redundancy and lowering weight resolution, the deep JSCC becomes affordable for IoT devices and significantly decreases the bandwidth required for model weight transmission.
Lee et al. \cite{lee2024asymmetric} introduce a resource-efficient asymmetric autoencoder framework that employs weight pruning and quantization to compress the encoder model on IoT devices. Simulation results show that the encoder-only pruning approach achieves a high compression rate of up to 98\% without sacrificing performance.

\subsubsection{Parameter Quantization}
Parameter quantization transforms the parameters within a neural network from high-precision formats (such as 32-bit floating-point numbers) to low-precision formats (such as 8-bit integers or even lower), with the objective of decreasing the model's computational and storage resource demands while preserving its performance levels \cite{gholami2022survey}.

\begin{figure*}[ht]
    \centering
		\subfigure[Quantization aware training.]{
			\includegraphics[width=0.4\textwidth]{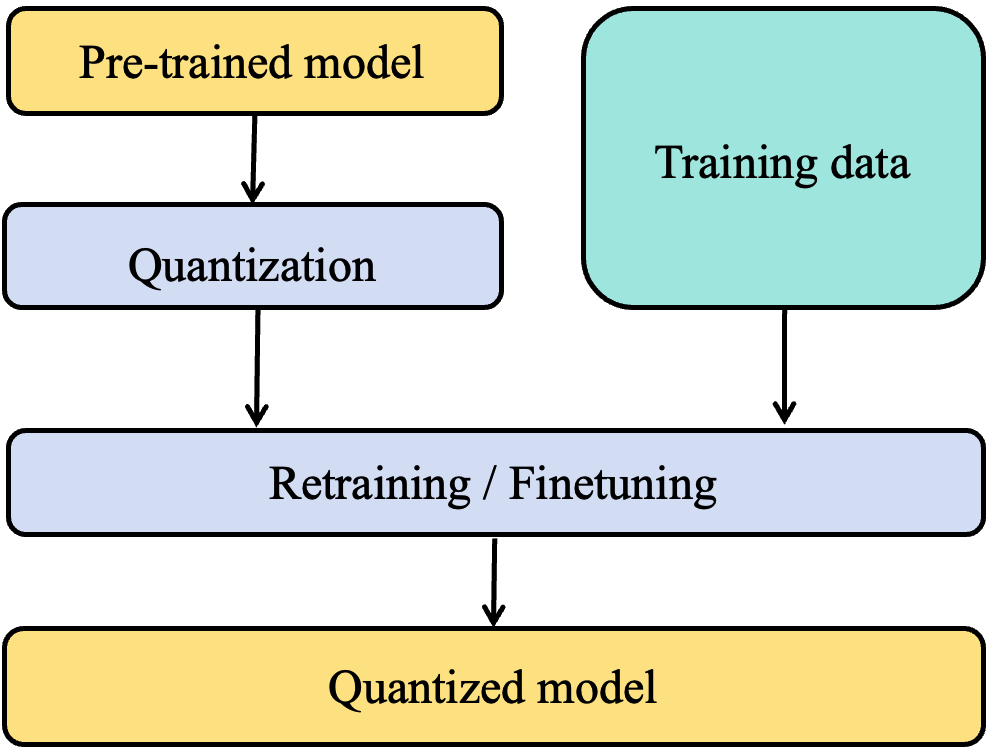}
			\label{fig482a}
		}
		\subfigure[Post training quantization.]{
			\includegraphics[width=0.4\textwidth]{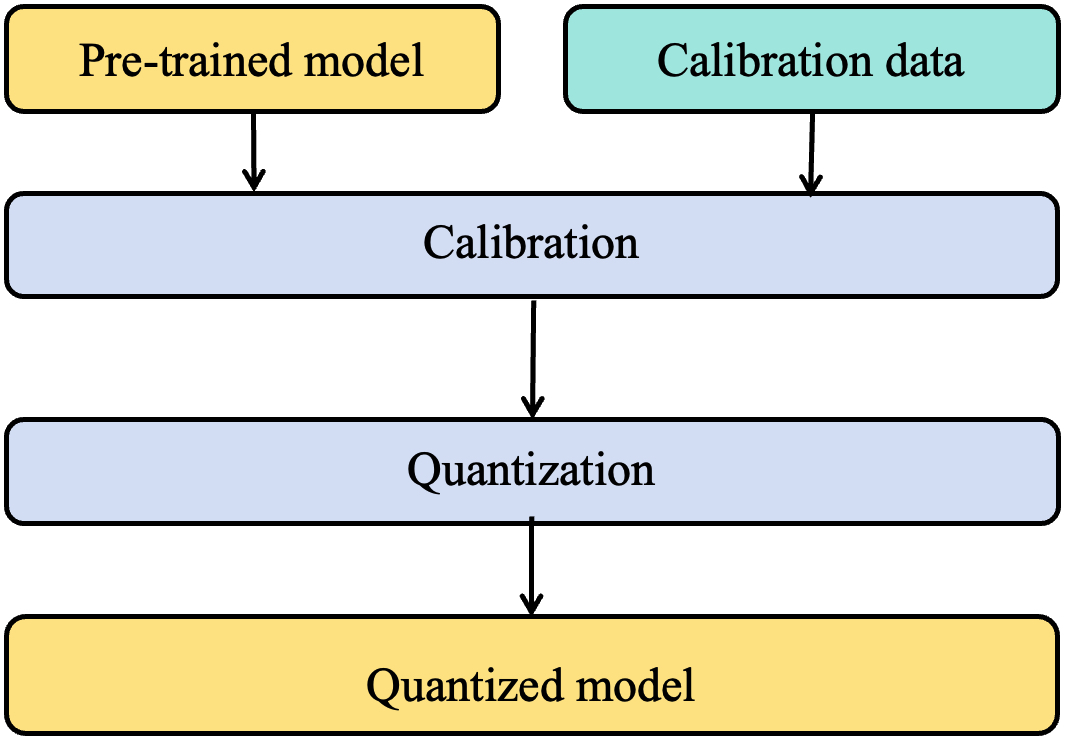}
			\label{fig482b}
		}
  \caption{Comparison between quantization aware training and post training quantization \cite{gholami2022survey}, where (a) Quantization aware training involves taking a pre-trained model, quantizing it, and subsequently fine-tuning it using training data. This fine-tuning helps adjust the parameters and mitigate any accuracy loss. (b): Post training quantization involves calibrating a pre-trained model using calibration data (such as a subset of the training data) to determine clipping ranges and scaling factors. Based on these calibration results, the model is then quantized.}
\label{fig482}
\end{figure*}

\textbf{Comparison between quantization aware training and post training quantization:} Nagel et al. \cite{nagel2021white} introduce advanced algorithms to mitigate the impact of quantization noise on neural network performance while maintaining low-bit weights and activations. They discuss two main classes of algorithms: post-training quantization and quantization-aware-training. As illustrated in Fig. \ref{fig482}, post-training quantization is a lightweight approach that requires no re-training or labeled data and is sufficient for achieving 8-bit quantization with close to floating-point accuracy. Quantization-aware-training requires fine-tuning and access to labeled training data but enables lower bit quantization with competitive results.

\textbf{Consideration of discretization errors between inputs and outputs of the discretizer:} Lee et al. \cite{lee2021network} consider discretization errors and propose the element-wise gradient scaling method, which adaptively scales each gradient element based on its sign and the error between the continuous input and discrete output of the discretizer. They demonstrate the effectiveness of the proposed method through extensive experimental results on image classification datasets with diverse network architectures and bit-width settings.

\textbf{Asymmetric quantization:} Bhalgat et al. \cite{bhalgat2020lsq+} propose LSQ+, an extension of LSQ that introduces a general asymmetric quantization scheme with trainable scale and offset parameters to accommodate negative activations. LSQ+ also alleviates the problem of high instability or variance in final training performance commonly seen in gradient-based learnable quantization schemes by using an MSE-based initialization scheme for the quantization parameters.

\textbf{Quantization for large language models:} Shao et al. \cite{shao2023omniquant} introduce OmniQuant, a technique that achieves good performance in diverse quantization settings while maintaining the computational efficiency of post-training quantization. OmniQuant includes two innovative components, learnable weight clipping and learnable equivalent transformation, which optimize various quantization parameters efficiently within a differentiable framework using block-wise error minimization. Xu et al. \cite{xu2023qa} propose a quantization-aware low-rank adaptation algorithm that balances the degrees of freedom of quantization and adaptation using group-wise operators. The designed algorithm can quantize weights during fine-tuning to reduce time and memory usage, and naturally integrates large language models and auxiliary weights into a quantized model without loss of accuracy.

\textbf{Quantization for SemCom:} To facilitate SemComs without end-to-end training, Park et al. \cite{park2024vision} introduce a vision transformer-based SemCom system with importance-aware quantization for wireless image transmission. The system leverages the attention scores of a pretrained vision transformer model to assess the importance of image patches. Based on these scores, importance-aware quantization assigns varying quantization bits to patches, prioritizing those with higher importance. This is achieved by solving a weighted quantization error minimization problem, where weights increase with attention scores.

\subsubsection{Low-Rank Decomposition}
Low-rank decomposition breaks down a large matrix into the product of two or more smaller, simpler matrices that generally possess lower ranks, subsequently decreasing the number of model parameters and mitigating the computational burden.

\textbf{Combination with sparsity:} Swaminathan et al. \cite{swaminathan2020sparse} introduce a new method called sparse low rank for compressing DNNs, which combines the ideas of low-rank decomposition and neuron significance. The proposed method achieves better compression rates by sparsifying the singular value decomposition matrices and keeping lower rank for unimportant neurons. Experimental results show that the proposed approach outperforms vanilla truncated SVD and a pruning baseline, achieving better compression rates with minimal or no loss in accuracy.
Xue et al. \cite{xue2021multilayer} propose a multilayer sparsity-based tensor decomposition method for low-rank tensor completion. The method aims to depict complex hierarchical knowledge with implicit sparsity attributes hidden in a tensor by encoding structured sparsity through multiple-layer representations. Specifically, it uses the CANDECOMP/PARAFAC model for tensor decomposition and introduces a new sparsity insight of subspace smoothness as the third-layer sparsity.

\textbf{Degeneracy in the tensor decomposition of convolutional kernels:} Phan et al. \cite{phan2020stable} introduce a novel method to stabilize the low-rank approximation of convolutional kernels in DNNs, addressing the issue of degeneracy in tensor decomposition. The proposed method ensures efficient compression while preserving the high-quality performance of the neural networks. This study is the first to address degeneracy in the tensor decomposition of convolutional kernels.

\textbf{Selection of optimal rank in each layer:} Idelbayev et al. \cite{idelbayev2020low} propose a mixed discrete-continuous optimization approach to jointly optimize the ranks and matrix elements, resulting in better rank selection. This makes low-rank compression more attractive and demonstrates its effectiveness by achieving similar classification error with faster inference times compared to ResNet using a VGG network.

\textbf{Domain generalization:} Piratla et al. \cite{piratla2020efficient} propose a common specific decomposition method for domain generalization, which jointly learns a common component that generalizes to new domains and a domain-specific component that overfits on training domains. The domain-specific components are discarded after training, and only the common component is used for prediction.

\subsubsection{Knowledge Distillation}
Knowledge distillation leverages a pre-trained, more sophisticated teacher model to instruct the training of a streamlined student model. Throughout the training phase, the student model mimics the predictive behavior of the teacher model, effectively decreasing the model's size and parameter count while striving to retain the teacher model's accuracy levels. Based on whether the teacher model is updated concurrently with the student model, the learning frameworks can be categorically divided into offline distillation, online distillation, and self-distillation. Knowledge distillation-enabled SemCom has been studied for multi-user scenarios and edge devices.

\textbf{Offline distillation:} Considering that he performance of the smaller student network can degrade when the gap between the student and teacher network sizes is too large, Mirzadeh et al. \cite{mirzadeh2020improved} introduce the multi-step knowledge distillation, which uses an intermediate-sized network to bridge the gap between the student and teacher. Theoretical analysis and experiments on various datasets and architectures demonstrate the effectiveness of this approach. Asif et al. \cite{asif2020ensemble} present a framework for learning compact CNN models with improved classification performance and generalization using a student model with parallel branches trained using ground truth labels and information from high-capacity teacher networks. The framework provides two benefits: promoting heterogeneity in learning features and encouraging collaboration among branches to improve prediction quality.

\textbf{Online distillation:} Wu et al. \cite{wu2021peer} propose a peer collaborative learning method for online knowledge distillation, which integrates online ensembling and network collaboration into a unified framework. By constructing a multi-branch network for training, where each branch is a peer, and using random augmentation and an additional classifier to assemble feature representations as the peer ensemble teacher, the method transfers knowledge from a high-capacity teacher to the peers and optimizes the ensemble teacher. Additionally, the temporal mean model of each peer is used as the peer mean teacher to facilitate collaboration among peers, leading to richer knowledge transfer and better generalization. Zhang et al. \cite{zhang2021adversarial} propose the adversarial co-distillation network to enhance the ``dark knowledge" in co-distillation by generating extra divergent examples using only the standard training set. The designed end-to-end approach consists of an adversarial phase with GANs to generate divergent examples and a co-distillation phase with multiple classifiers to learn from these examples. The two phases are learned in an iterative and adversarial way, and additional modules are designed to ensure the quality of divergent examples.

\begin{figure*}[t]
\centering
\vspace{-10mm}
\includegraphics[width=0.6\linewidth]{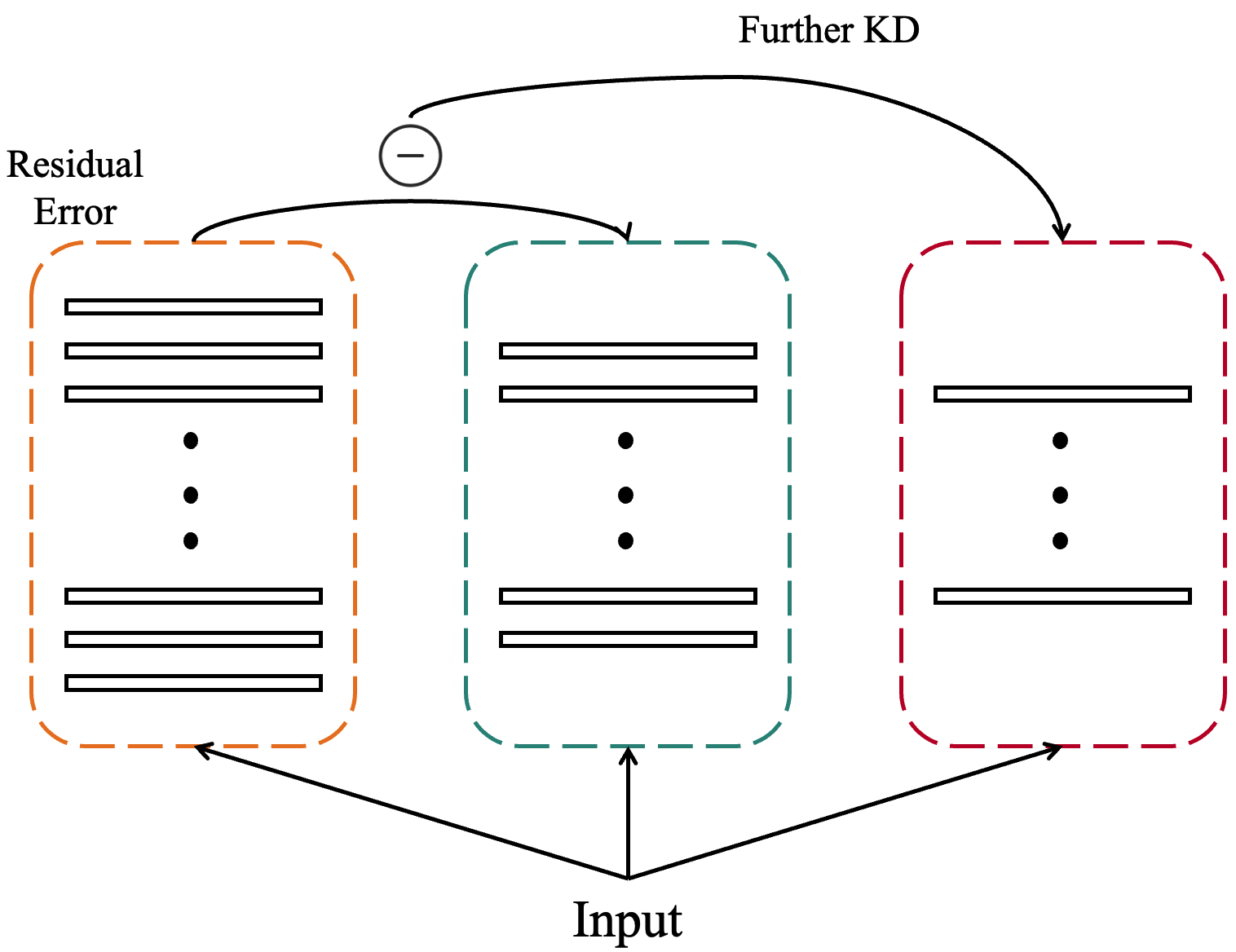}
\caption{Residual error based knowledge distillation \cite{gao2021residual}, where the Assistant (middle) is employed to enhance the distillation of knowledge from the Teacher (left). The assistant aids in optimizing the transfer of knowledge, with the goal of capturing the residual error between the Student's (right) and Teacher's feature maps.}
\label{fig483}
\end{figure*}

\textbf{Self-distillation:} Zhang et al. \cite{zhang2020self} offer a new interpretation of teacher-student training as amortized MAP estimation and relate self-distillation to label smoothing, emphasizing the importance of predictive diversity in addition to predictive uncertainty. Experimental results demonstrate the utility of predictive diversity across multiple datasets and neural network architectures. They propose a novel instance-specific label smoothing technique that promotes predictive diversity without a separately trained teacher model and find that it often outperforms classical label smoothing. As illustrated in Fig. \ref{fig483}, Gao et al. \cite{gao2021residual} propose the residual error based knowledge distillation to address the performance degradation in knowledge distillation due to the gap between the learning capacities of the student and teacher models. The proposed method introduces an assistant model to learn the residual error between the feature maps of the student and teacher, allowing them to complement each other and better transfer knowledge from the teacher. An effective method is further devised to derive the student and assistant models from a given model without increasing the total computational cost.

\textbf{Defense against adversarial attacks:} Papernot et al. \cite{papernot2016distillation} propose the defensive distillation to reduce the effectiveness of adversarial samples on DNNs. Analytical investigation shows that defensive distillation grants generalizability and robustness properties when training DNNs. Empirical studies demonstrate that defensive distillation can significantly reduce the effectiveness of adversarial sample creation. 
Ham et al. \cite{ham2024neo} propose the NEO-KD, a knowledge distillation-based adversarial training strategy, to improve the robustness of multi-exit neural networks against adversarial attacks. NEO-KD uses neighbor knowledge distillation to guide the output of adversarial examples to tend to the ensemble outputs of neighbor exits of clean data, and employs exit-wise orthogonal knowledge distillation to reduce adversarial transferability across different sub-models.

\textbf{Knowledge distillation for Multi-user SemCom:} Liu et al. \cite{liu2023knowledge} focus on a multi-user SemCom system characterized by limited training samples and unforeseen interference. To enhance model generalization and reduce its size, they introduce a system leveraging knowledge distillation, where a Transformer-based encoder-decoder serves as the semantic encoder-decoder, and fully connected neural networks function as the channel encoder-decoder. They analyze four types of knowledge transfer and model compression techniques. Key system and model parameters, such as noise and interference levels, the number of interfering users, and the dimensions of the encoder and decoder, are taken into account. Numerical results indicate that knowledge distillation significantly boosts robustness and generalization against unexpected interference and mitigates performance degradation when the model size is compressed. Nguyen et al. \cite{nguyen2024optimizing} study the downlink communication from a base station to users with varying computational capacities. The users utilize different versions of Swin transformer models for source decoding and a streamlined channel decoding architecture. They introduce an innovative training approach that integrates transfer learning and knowledge distillation to enhance the performance of users with limited computing power.

\textbf{Federated knowledge distillation for SemCom:} Xing et al. \cite{xing2023feddistillsc} introduce FedDistillSC, a distributed SemCom system leveraging federated distillation for image classification tasks. Following local training with both public and private datasets, the system refines the local models of each transmitter through weighted aggregation based on receiver-generated classification probabilities. This approach facilitates knowledge sharing without compromising private data, thereby enhancing each transmitter's semantic extraction capabilities. Notably, our method avoids the need for uploading or downloading model parameters, conserving bandwidth resources and mitigating data security risks associated with parameter leakage. Considering that mismatched knowledge bases disrupt semantic alignment between transmitters and receivers, leading to significant semantic errors, Lu et al. \cite{lu2024efficient} propose a semantic knowledge base synchronization framework leveraging federated knowledge distillation for knowledge base establishment and dynamic updates. The framework employs a mutual distillation mechanism to extract knowledge from diverse local knowledge bases. Additionally, they compress the global knowledge base to boost synchronization efficiency.

\textbf{Combination of knowledge distillation and LLMs for SemCom:} Nam et al. \cite{nam2024language} propose the language-oriented SemCom by merging LLMs and generative models with the SemCom paradigm. They introduce three innovative algorithms: 1) Semantic source compression, which compresses text prompts into key headwords preserving syntactic essence and context; 2) Semantic error mitigation, which enhances robustness by substituting headwords with synonyms to counter errors; and 3) Listener-adaptive prompt generation, which creates customized prompts tailored to the listener's language style using in-context learning.

\textbf{Knowledge distillation-enabled SemCom for edge devices:} Albaseer et al. \cite{albaseer2024tailoring} introduce a SemCom framework tailored for heterogeneous, resource-constrained edge devices and computation-intensive servers. They leverages dynamic knowledge distillation to customize semantic models for each device, balancing computational and communication constraints while ensuring QoS. They formulate an optimization problem and develop an adaptive algorithm that iteratively refines semantic knowledge on edge devices, resulting in tailored models suited to their resource profiles.

\begin{les}
Model compression reduces the size and enhances the efficiency of DL models by minimizing computational requirements without compromising accuracy significantly. Its importance is crucial for deploying Deep JSCC models on resource-constrained devices, as it helps decrease communication costs, conserve storage, and strengthen SemCom systems against multi-party cooperation attacks. Model compression can be achieved through model pruning, parameter quantization, low-rank decomposition, or knowledge distillation. Specifically, model pruning can be designed to defense against model inversion attacks \cite{ding2024patrol}, membership inference attacks \cite{wang2020against,yuan2022membership}, and adversarial attacks \cite{ye2019adversarial,wu2021adversarial}. Model pruning for SemCom has also been studied \cite{xie2020lite,lee2024asymmetric}. Parameter quantization can be divided into quantization aware training and post training quantization \cite{nagel2021white,gholami2022survey}, and the applications can be extended to task oriented SemCom \cite{lee2021network}, large knowledge base \cite{shao2023omniquant,xu2023qa}, and lightweight SemCom without end-to-end training \cite{park2024vision}. Low-rank decomposition can be divided into the following sub-problems: sparsity-enabled decomposition \cite{swaminathan2020sparse,xue2021multilayer}, decomposition of convolutional kernels \cite{phan2020stable}, decomposition of each layer in deep JSCC models \cite{idelbayev2020low}, and task-oriented SemCom \cite{piratla2020efficient}. Knowledge distillation can be devided into offline \cite{mirzadeh2020improved,asif2020ensemble}, online \cite{wu2021peer,zhang2021adversarial}, and self distillation \cite{zhang2020self,gao2021residual}, and it is effective to defend against adversarial attacks \cite{papernot2016distillation,ham2024neo}. Knowledge distillation has also been studied for multi-user SemCom \cite{liu2023knowledge,nguyen2024optimizing,xing2023feddistillsc,lu2024efficient}, combination with LLMs and SemCom \cite{nam2024language}, and SemCom edge devices \cite{albaseer2024tailoring}.
\end{les}

\subsection{Physical-Layer Security}

Physical-layer security (PLS) involves utilizing the randomness of wireless channels, such as interference, fading, and noise, to ensure secure transmission. Shannon proposed the concept of ``perfect secret", which can be extended to ``perfect semantic secret" in SemCom as
\begin{equation}
I(M_S;C_S)=0
\end{equation}
where $M_S$ is the semantic information at the transmitter. $C_S$ represents the encrypted semantic features, that is, the observation of semantic eavesdroppers. Perfect semantic secret necessitates that the mutual information between $M$ and $C$ is zero. If the condition of perfect semantic secret is met, semantic eavesdroppers can only recover the original semantic information from the encrypted semantic characteristics through random guessing, as it represents the optimal strategy. The various random factors present in wireless communications can be effectively harnessed to establish the comparative advantages of legitimate semantic transmission and ensure the security of SemCom. The related PLS techniques include beamforming, artificial noise, relay cooperation, Intelligent reflecting Surface (IRS), physical-layer key generation (PLKG), and physical-layer authentication (PLA), which are descried in detail as follows. The representative PLS schemes are listed in Tab. \ref{physicallayersecurity}.

\begin{longtable}{|A{2cm}|A{0.6cm}|P{9.5cm}|}  
\caption{\footnotesize{List of Representative Physical-Layer Security (PLS) Schemes}}
\label{physicallayersecurity}\\
\hline
\textbf{Sub-class} & \textbf{Ref.} & \textbf{Descriptions, including} $\circledast$: \textbf{contributions; }

$\circleddash$: \textbf{trade-offs;} 
$\boxplus$: \textbf{performance metrics;}

$\Cap$: \textbf{applied in SemCom or promising for SemCom}
\\
\hline
\endfirsthead

\hline
\textbf{Sub-class} & \textbf{Ref.} & \textbf{Descriptions, including} $\circledast$: \textbf{contributions; }

$\circleddash$: \textbf{trade-offs;} 
$\boxplus$: \textbf{performance metrics;} 

$\Cap$: \textbf{applied in SemCom or promising for SemCom} \\
\hline
\endhead

\hline
\endfoot

\hline
\endlastfoot
Beamforming & \cite{zhang2024beamforming} &
$\circledast$: Introduces a semantic-bit coexisting communication framework and proposes a spatial beamforming scheme to address the heterogeneous inter-user interference issue.

$\circleddash$: Qos for semantic-bit users and algorithm complexity.

$\boxplus$: SSIM.

$\Cap$: Applied in semantic-bit coexisting communications.
\\

\hline

Beamforming & \cite{wu2024deep2} &
$\circledast$: Proposes a deep joint semantic coding and beamforming scheme for airship-based massive MIMO image transmission networks in near-field communication networks.

$\circleddash$: Transmission performance and network parameter complexity.

$\boxplus$: PSNR, MS-SSIM, and LPIPS.

$\Cap$: Applied in SemCom for massive MIMO image transmission.
\\
\hline

Beamforming & \cite{raha2024towards} &
$\circledast$: Develops a semantic-based method to enhance the robustness of beamforming in 6G wireless communication, particularly for high-mobility applications like intelligent transportation systems and virtual reality platforms.

$\circleddash$: Transmission performance and beamforming complexity.

$\boxplus$: Accuracy and received power.

$\Cap$: Applied in SemCom with high-mobility applications.
\\
\hline

Beamforming & \cite{yang2024secure2} &
$\circledast$: Explores the allocation of secure resources in a downlink system, accommodating multiple authorized users and potential eavesdroppers. 

$\circleddash$: Communication rate and semantic secrecy performance.

$\boxplus$: Worst-case semantic secrecy rate and Multiple SIgnal Classification (MUSIC) Spectrum.

$\Cap$: Applied in SemCom to defend against eavesdroppers.
\\
\hline

Beamforming & \cite{dai2024secure} &
$\circledast$: Focuses on scenarios involving multiple eavesdroppers. For the beamforming subproblem, it employs the Dinkelbach algorithm with predetermined semantic parameters to simplify the objective function and proposes an iterative alternating algorithm to solve a series of convex subproblems.

$\circleddash$: Communication rate and semantic secrecy performance.

$\boxplus$: Secure semantic efficiency.

$\Cap$: Applied in SemCom to defend against eavesdroppers.
\\
\hline

Artificial Noise & \cite{zhang2018artificial} &
$\circledast$: Provides an artificial-noise-aided optimal beamforming scheme for a two-layer unicast system to maximize high-level information security while adhering to low-level secrecy constraints.

$\circleddash$: Secrecy performance and algorithm complexity.

$\boxplus$: Secrecy rate.

$\Cap$: Promising for secure SemCom.
\\
\hline

Artificial Noise & \cite{jiang2018secrecy} &
$\circledast$: Proposes two secrecy energy efficiency maximization schemes for energy-efficient secure downlink communication in OFDM-based cognitive radio networks with an eavesdropper having multiple antennas.

$\circleddash$: Secrecy performance and power consumption.

$\boxplus$: Average secrecy energy efficiency.

$\Cap$: Promising for secure SemCom.
\\
\hline

Relay Cooperation & \cite{luo2022autoencoder} &
$\circledast$: Introduces a SemCom scheme for wireless relay channels, which uses an Autoencoder for encoding and decoding sentences from the semantic dimension. A novel semantic forward mode is designed for the relay node to forward semantic information at the semantic level, particularly in scenarios where the source and destination nodes do not share common knowledge.

$\circleddash$: Transmission performance and network complexity.

$\boxplus$: BLEU scores.

$\Cap$: Applied in SemCom where the source and destination nodes do not share common knowledge.
\\
\hline

Relay Cooperation & \cite{tang2023cooperative} &
$\circledast$: Proposes a DL-based cooperative SemCom system on relay channels that enhances reliability and adaptability to varying channel conditions through an on-demand semantic forwarding framework.

$\circleddash$: The degree of semantic information recovery and the transmit energy consumption.

$\boxplus$: BLUE score and sentence similarity.

$\Cap$: Applied in SemCom with varying channel conditions.
\\
\hline

Relay Cooperation & \cite{guo2024distributed} &
$\circledast$: Introduces the distributed task-oriented communication networks that leverages multimodal semantic transmission and edge intelligence to improve task performance. The framework integrates multimodal knowledge of semantic relays and adaptive adjustment capability of edge intelligence.

$\circleddash$: Transmission performance and model complexity.

$\boxplus$: Classification accuracy.

$\Cap$: Applied in multimodal SemCom.
\\
\hline

Intelligent Reflecting Surface & \cite{du2023semantic} &
$\circledast$: Proposes the inverse SemCom, which encodes task-related source messages into a hyper-source message for data transmission or storage, instead of extracting semantic information from messages. The proposed framework includes three algorithms for data sampling, IRS-aided encoding, and self-supervised decoding.

$\circleddash$: Multi-task performance and multi-task complexity.

$\boxplus$: PSNR.

$\Cap$: Applied in multi-task SemCom.
\\
\hline

Intelligent Reflecting Surface & \cite{huang2024joint2} &
$\circledast$: Proposes a unified design framework for IRS-aided SemCom systems tailored for classification tasks. The framework is based on the Infomax principle, which aims to maximize the mutual information between the received feature and the label of the input data sample.

$\circleddash$: Transmission performance and IRS complexity.

$\boxplus$: Inference accuracy. 

$\Cap$: Applied in tasks-oriented SemCom.
\\
\hline

Intelligent Reflecting Surface & \cite{chen2024ris} &
$\circledast$: Offers the paradigm of IRS-based on-the-air SemCom to enable SemComs in the wave domain, leveraging IRSs and on-the-air diffractional DNNs. This scheme discusses data and control flow issues, and provides a performance analysis with image transmission as an example.

$\circleddash$: Transmission performance and digital hardware.

$\boxplus$: PSRN, MS-SSIM, and computation time.

$\Cap$: Applied in IRS-aided SemCom.
\\
\hline

Intelligent Reflecting Surface & \cite{sun2024multi} &
$\circledast$: introduces a novel concept utilizing multi-functional IRSs to facilitate semantic anti-jamming communication and computing within MEC-supported integrated aerial-ground networks. This framework leverages a semantic transceiver's inherent resilience and data compression abilities, while multi-functional-IRSs customize the wireless environment through signal reflection, refraction, amplification, and energy harvesting, achieving extensive global coverage, reliable connectivity, and high-speed computing.

$\circleddash$: Transmission performance, security, and complexity

$\boxplus$: Semantic computation rate.

$\Cap$: Applied in SemCom to defend against semantic jamming attacks.
\\
\hline

Physical-Layer Key Generation & \cite{zhang2021deep} &
$\circledast$: Develops a PLKG scheme for frequency-division duplexing systems in IoT. The scheme uses DL to establish a feature mapping function between different frequency bands, enabling two users to generate highly similar channel features.

$\circleddash$: Key generation performance and model energy consumption.

$\boxplus$: NMSE, key error rate, and key generation ratio.

$\Cap$: Promising for SemCom to defend against eavesdroppers.
\\
\hline

Physical-Layer Key Generation & \cite{zhao2022semkey} &
$\circledast$: Develops the SemKey to significantly improve the secret key generation rate for SemCom by exploring the underlying randomness of the system and utilizing IRS.

$\circleddash$: Key generation performance and network complexity.

$\boxplus$: MSE, key generation rate, and P-value.

$\Cap$: Applied in SemCom to defend against eavesdroppers.
\\
\hline

Physical-Layer Authentication & \cite{gao2023esanet} &
$\circledast$: Proposes an environment semantics enabled PLA network for 6G endogenous security. The network extracts a frequency independent wireless channel fingerprint from CSI in a massive MIMO system using environment semantics knowledge.

$\circleddash$: Detection performance and model complexity.

$\boxplus$: True positive rate, false positive rate, and detection accuracy.

$\Cap$: Promising for secure SemCom.
\\
\hline

Physical-Layer Authentication & \cite{ma2024physical} &
$\circledast$: Proposes a pseudo-random watermark hopping-based PLA scheme to enhance both security and communication performance. By generating a pseudo-random sequence and designing a watermark hopping mechanism, the scheme increases randomness and security while also reducing authentication latency and improving communication performance.

$\circleddash$: False alarm rate and miss detection rate.

$\boxplus$: False alarm rate, miss detection rate, achievable rate, and outage probability.

$\Cap$: Promising for secure SemCom.
\\
\hline

\end{longtable}

\subsubsection{Beamforming}
Beamforming technology primarily utilizes multi-antenna systems to form beams with specific directionality by precisely controlling the phase and amplitude of the transmission signals from each antenna in the array. This technology can substantially enhance the signal strength for the intended receiver while suppressing signal transmission in other directions, thus diminishing the received signal quality for eavesdroppers.

\begin{figure*}[t]
\centering
\vspace{-10mm}
\includegraphics[width=\linewidth]{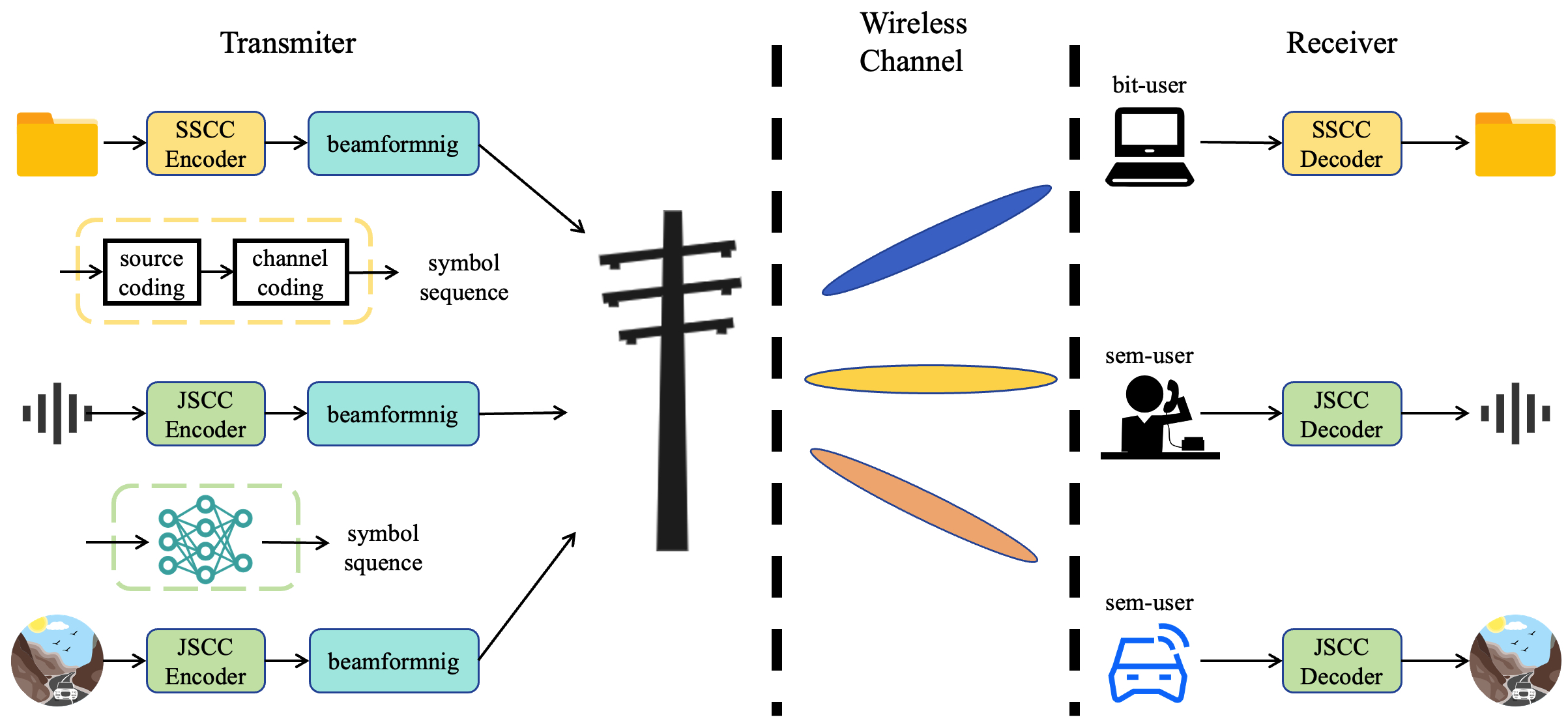}
\caption{Semantic-bit coexisting communication framework \cite{zhang2024beamforming}, where a single-cell downlink MU-MISO system is considered. The base station has multiple transmit antennas, and each user possesses a single antenna. The users are categorized into two groups: bit-users utilizing BitCom, and sem-users utilizing SemCom.}
\label{fig491}
\end{figure*}

\textbf{Semantic-bit coexisting environment:} As illustrated in Fig. \ref{fig491}, Zhang et al. \cite{zhang2024beamforming} introduces a semantic-bit coexisting communication framework and proposes a spatial beamforming scheme to address the heterogeneous inter-user interference issue. To maximize the semantic rate for SemCom users while ensuring QoS for BitCom users, a data-driven method and majorization minimization are employed to derive a semi-closed form solution. Additionally, a low-complexity version of the MM-FP algorithm is proposed to reduce computational complexity, and simulation results demonstrate the superiority of both algorithms compared to conventional beamforming techniques.

\textbf{Massive MIMO applications:} Wu et al. \cite{wu2024deep2} proposes a deep joint semantic coding and beamforming scheme for airship-based massive MIMO image transmission networks in near-field communication networks. By integrating SemCom with massive MIMO technology, the scheme aims to improve transmission efficiency and capacity. The proposed scheme extracts semantics from both the image source and channel state information, fuses them into complex-valued semantic features, and then employs hybrid data and model-driven semantic-aware beamforming networks for efficient transmission. At the receiver, a semantic decoding network reconstructs the transmitted images.

\textbf{Varying environmental conditions:} Raha et al. \cite{raha2024towards} develops a semantic-based method to enhance the robustness of beamforming in 6G wireless communication, particularly for high-mobility applications like intelligent transportation systems and virtual reality platforms. The proposed method utilizes the YOLOv8 algorithm to extract semantic data from RGB camera images and K-means clustering with GPS data to identify target vehicles. A lightweight model is used to predict the optimal beamforming index to maintain ultra-reliable low-latency communication.
Raha et al. \cite{raha2024advancing} further proposes a robust beamforming technique to ensure consistent Quality of Service (QoS) under varying environmental conditions in 6G. The proposed technique leverages semantic localization and optimal beam selection problems to maximize users' data rates. A novel method using K-means clustering and YOLOv8 model is proposed to solve the semantic localization problem, while a lightweight hybrid architecture combining a transformer and CNN is proposed to solve the beam selection problem. A novel metric, Accuracy-Complexity Efficiency, is introduced to quantify the performance of the proposed model.

\textbf{Integrated sensing and SemCom systems:}
Yang et al. \cite{yang2024secure2} and Dai et al. \cite{dai2024secure,dai2025joint} study the security issues of beamforming technology in integrated sensing and SemCom systems. Yang et al. \cite{yang2024secure2} explore the allocation of secure resources in a downlink system, accommodating multiple authorized users and potential eavesdroppers. To augment security at the semantic level, semantic information is extracted from the original data prior to transmission. Successful recovery of the received information at the user end necessitates the use of a knowledge base shared with the base station, which is pre-stored. The objective is to maximize the cumulative semantic secrecy rate for all users while ensuring a minimum quality of service for each and maintaining overall sensing performance. They propose an iterative algorithm utilizing the alternating optimization technique. Simulation outcomes highlight the superiority of the proposed algorithm in terms of secure SemCom and reliable detection capabilities. Dai et al. \cite{dai2024secure} focus on scenarios involving multiple eavesdroppers. Initially, we introduce the concept of secure semantic rate to quantify the dependability of user information acquisition. Subsequently, they formulate an optimization problem aimed at maximizing secure semantic efficiency. To address this problem, they decompose it into two subproblems: beamforming optimization and semantic parameter optimization. For the beamforming subproblem, they employ the Dinkelbach algorithm with predetermined semantic parameters to simplify the objective function and propose an iterative alternating algorithm to solve a series of convex subproblems. For the semantic parameter subproblem, they mathematically demonstrate the monotonic decrease of the objective function in relation to the semantic parameter, given a fixed beamforming vector. Dai et al. \cite{dai2025joint} focus on scenarios with multiple eavesdroppers and accounting for both perfect and imperfect CSI. Similarly, they split this problem into two subproblems: beamforming optimization and semantic parameter optimization.

\subsubsection{Artificial Noise}
Artificial noise technology involves adding intentional and pseudo-random noise to the transmission signals to interfere with eavesdroppers. This noise has minimal impact on legitimate receivers but can drastically reduce the received signal-to-noise ratio for eavesdroppers, thereby safeguarding the security of information.

\textbf{Optimal beamforming:} Zhang et al. \cite{zhang2018artificial} provide a layered PLS model to secure multiple messages simultaneously with a cascade security structure. An artificial-noise-aided optimal beamforming scheme is proposed for a two-layer unicast system to maximize high-level information security while adhering to low-level secrecy constraints. A successive convex approximation-based algorithm is used to address the nonconvexity of the problem, and a low-complexity zero-forcing beamforming scheme is introduced for efficiency.

\textbf{Cognitive radio network applications:} Jiang et al. \cite{jiang2018secrecy} propose two secrecy energy efficiency maximization schemes for energy-efficient secure downlink communication in OFDM-based cognitive radio networks with an eavesdropper having multiple antennas. The proposed schemes exploit the instantaneous and statistical CSI of the eavesdropper, respectively, to defend against eavesdropping and improve energy efficiency. The problems are transformed into equivalent subtractive problems and then approximate convex problems using the difference of two convex functions approximation method. Two-tier power allocation algorithms are proposed to achieve $\epsilon$-optimal solutions.

\textbf{Cellular vehicle-to-everything network applications:} Wang et al. \cite{wang2020physical} investigates the potential of artificial noise and secure beamforming for enhancing the security of cellular vehicle-to-everything networks. Using stochastic geometry, the PLS is studied, with locations of nodes modeled by Cox processes and Poisson point processes. The coverage probability and bounds on the secrecy probability are calculated and validated by simulation results. Analytical results on effective secrecy throughput are also obtained, demonstrating the reliability and security of wiretap channels.

\textbf{Visible light communication applications:} Pham et al. \cite{pham2024design} explores energy-efficient artificial noise schemes for enhancing PLS in visible light communications. Two transmission schemes, selective artificial noise-aided single-input single-output and artificial noise-aided multiple-input single-output, are compared in terms of secrecy energy efficiency. In the former, the closest LED luminaire to the legitimate user transmits the information-bearing signal, while the rest transmit artificial noise to degrade eavesdroppers' channels. In the latter scheme, all luminaries transmit a combination of the information-bearing signal and artificial noise. When eavesdroppers' CSI is unknown, an indirect design to maximize the legitimate user's channel energy efficiency is proposed, along with a low-complexity zero-forcing design. When eavesdroppers' CSI is known, the design maximizes the minimum secrecy energy efficiency among all eavesdroppers.

\subsubsection{Relay Cooperation}

Relay cooperation enhances SemCom performance by facilitating multiple nodes to share and utilize each other's resources during semantic information transmission. This technology holds a pivotal role in ensuring PLS. By harnessing the intricate spatial domain characteristics formed by multiple relay nodes, the received signals at these nodes can be regulated through adjustments to the antenna transmission weights. The legitimate receiver consolidates all the content relayed to obtain the original information and achieve diversity gain, whereas the signal captured by a solitary relay node incorporates random weighting, rendering it infeasible to accurately reconstruct the semantic information content.

\textbf{Semantic text transmission:} Luo et al. \cite{luo2022autoencoder} introduce a SemCom scheme for wireless relay channels, which uses an Autoencoder for encoding and decoding sentences from the semantic dimension. The Autoencoder provides anti-noise performance for the system. Additionally, a novel semantic forward mode is designed for the relay node to forward semantic information at the semantic level, particularly in scenarios where the source and destination nodes do not share common knowledge. Hu et al. \cite{hu2024semantic} propose the semantic relay to address the challenges of implementing resource-demanding DL-based SemCom on mobile devices with limited computing and storage resources. The semantic relay serves as an edge server to provide SemCom services for both semantic users with rich resources and conventional users with limited resources. Two new transmission protocols are proposed for text transmissions via the semantic relay, and an optimization problem is formulated to maximize the weighted sum-rate of all users. An efficient algorithm is proposed to solve this non-convex problem, and numerical results demonstrate the effectiveness of the proposed algorithm and the superior performance of the semantic relay compared to traditional decode-and-forward relays, especially in the small bandwidth regime.

\begin{figure*}[t]
\centering
\vspace{-10mm}
\includegraphics[width=0.9\linewidth]{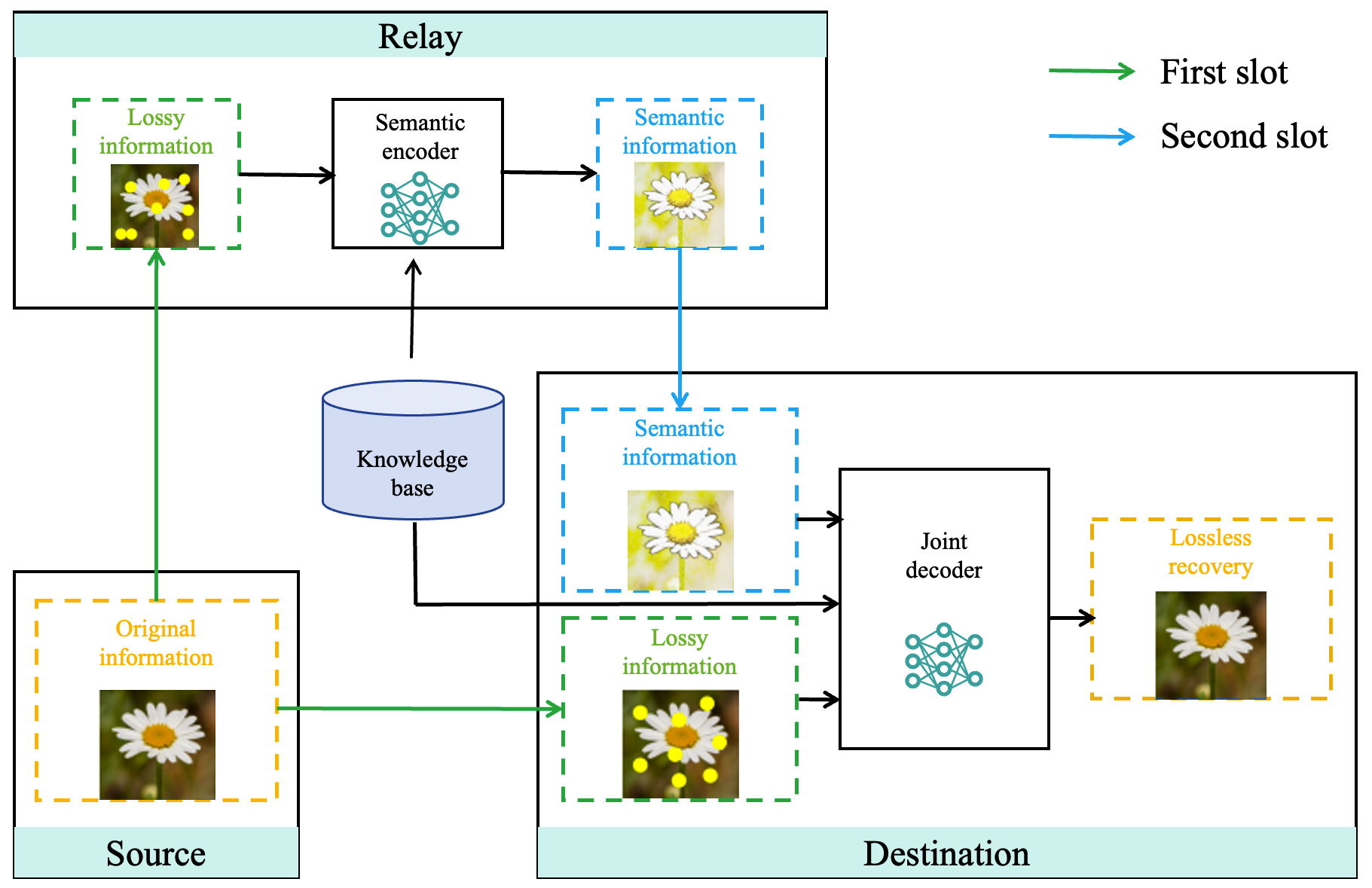}
\caption{Semantic-forward for cooperative communications \cite{lin2024semantic}, where SemComs are employed at both the relay and destination to minimize the payload of the relay-destination link. The deep JSCC algorithm for this relaying systems is applied to enable the destination to recover the source information accurately with the help of semantic information provided by the relay.}
\label{fig492}
\end{figure*}

\textbf{Cooperative SemCom:} Tang et al. \cite{tang2023cooperative} proposes a DL-based cooperative SemCom system on relay channels that enhances reliability and adaptability to varying channel conditions through an on-demand semantic forwarding framework. The system uses a semantic similarity check to determine whether semantic forwarding is needed from the relay and a semantic combining method to effectively merge semantic information received through different paths. Additionally, a new metric called semantic energy efficiency is proposed to balance the degree of semantic information recovery and transmit energy consumption. As depicted in Fig. \ref{fig492}, Lin et al. \cite{lin2024semantic} introduce a relaying framework called semantic-forward for cooperative communications in 6G wireless networks. The semantic-forward relay reduces payload and enhances network robustness by extracting and transmitting semantic features. A joint source-channel coding algorithm is designed based on cooperative communications with side information and the turbo principle to improve decoding gains.

\textbf{Multimodal SemCom:} Guo et al. \cite{guo2024distributed} introduce the distributed task-oriented communication networks that leverages multimodal semantic transmission and edge intelligence to improve task performance. The framework integrates multimodal knowledge of semantic relays and adaptive adjustment capability of edge intelligence. Key techniques in the framework include semantic alignment and complement, a semantic relay scheme for deep joint source-channel relay coding, and collaborative device-server optimization and inference.

\subsubsection{Intelligent Reflecting Surface}
IRS constitutes a planar structure comprised of numerous programmable reflecting units, capable of dynamically tuning the phase, amplitude, and frequency attributes of reflected electromagnetic waves. Through intelligent control of these units, the IRS attains precise management of incident electromagnetic waves, thereby modifying the transmission properties of the wireless channel. This functionality allows the IRS to emerge as a potent tool for PLS, bolstering the confidentiality of communications by fashioning a programmable wireless environment. In scenarios involving eavesdroppers, the IRS can seamlessly collaborate with the base station's beamformer and adjust its own reflection phase shifts to maximize the system's minimum secrecy rate. By effectively reducing the SNR of eavesdroppers while enhancing the SNR of legitimate receivers, the IRS notably elevates the confidentiality performance of wireless systems.

\begin{figure*}[t]
\centering
\vspace{-10mm}
\includegraphics[width=\linewidth]{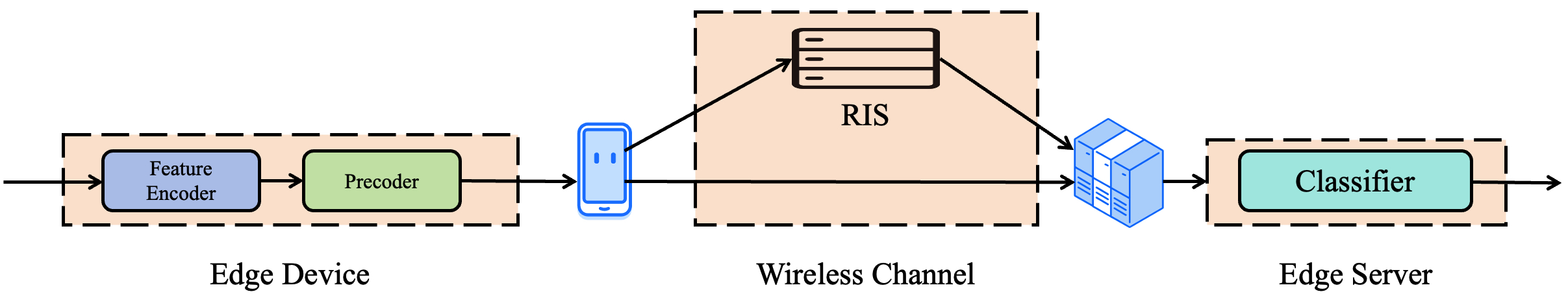}
\caption{Unified framework for IRS-aided SemCom systems \cite{huang2024joint2}, where the edge device hosts the feature encoder, while the edge server hosts the classifier. For classification, the edge device uses a DNN-based feature encoder. This feature is then precoded using a linear precoder. Subsequently, it is transmitted over the IRS-assisted wireless channel, resulting in a noisy signal. The edge server employs its classifier determine the estimated classification outcome.}
\label{fig493}
\end{figure*}

\textbf{Task-oriented SemCom:} Du et al. \cite{du2023semantic} propose a new paradigm called inverse SemCom, which encodes task-related source messages into a hyper-source message for data transmission or storage, instead of extracting semantic information from messages. The proposed framework includes three algorithms for data sampling, IRS-aided encoding, and self-supervised decoding. Wang et al. \cite{wang2024irs} propose an IRS-enhanced secure SemCom to ensure PLS from a task-oriented semantic perspective. The paper introduces a multi-layer codebook for hierarchical semantic representation and defines novel semantic security metrics, secure semantic rate and secure semantic spectrum efficiency. Additionally, a noise disturbance enhanced hybrid deep reinforcement learning-based resource allocation scheme is proposed to maximize secure semantic spectrum efficiency by optimizing semantic representation bits, IRS reflective coefficients, and subchannel assignment. A semantic context awared state space is also introduced to solve the dimensional catastrophe problem. As illustrated in Fig. \ref{fig493}, Huang et al. \cite{huang2024joint2} propose a unified design framework for IRS-aided SemCom systems tailored for classification tasks. The framework is based on the Infomax principle, which aims to maximize the mutual information between the received feature and the label of the input data sample. Unlike end-to-end training, the proposed approach decouples the learning and communication modules, allowing for a more efficient design of feature encoding, joint active and passive beamforming, and classification.

\textbf{Discussion on data and control flow:} Chen et al. \cite{chen2024ris} offer the paradigm of IRS-based on-the-air SemCom to enable SemComs in the wave domain, leveraging IRSs and on-the-air diffractional DNNs. The computations in this scheme occur inherently as wireless signals pass through IRSs, offering light-speed computation, low power consumption, and the ability to handle multiple tasks simultaneously. This scheme presents a system model, discusses data and control flow issues, and provides a performance analysis with image transmission as an example. Overall, IRS-based SemCom offer appealing characteristics compared to traditional digital hardware-based approaches.

\textbf{Defense against semantic eavesdroppers:} Wang et al. \cite{wang2024star} propose the use of a simultaneous transmitting and reflecting reconfigurable intelligent surface (STAR-RIS) to achieve privacy protection in a SemCom system. The STAR-RIS is used to enhance signal transmission between a base station and a destination user, while converting the signal to interference for an eavesdropper. Simulation results show that the proposed method outperforms other benchmarks in protecting SemCom privacy, as evidenced by a significantly lower task success rate achieved by the eavesdropper.

\textbf{Defense against semantic jamming attacks:} Sun et al. \cite{sun2024multi} introduce a novel concept utilizing multi-functional IRSs to facilitate semantic anti-jamming communication and computing within MEC-supported integrated aerial-ground networks. This framework leverages a semantic transceiver's inherent resilience and data compression abilities, while multi-functional-IRSs customize the wireless environment through signal reflection, refraction, amplification, and energy harvesting, achieving extensive global coverage, reliable connectivity, and high-speed computing. They formulate a semantic computation rate maximization problem, considering imperfect jammer's CSI, energy partition constraints for computation offloading, semantic similarity requirements, semantic computation rate targets, and multi-functional-IRS self-sustainability. By transforming the imperfect CSI into a worst-case scenario using a discretization method, They propose a fast-converging monotonic optimization algorithm combined with decoupled second-order cone programming to achieve a globally optimal solution with minimal feasibility evaluations.

\subsubsection{Physical-Layer Key Generation}
PLKG hinges on leveraging the reciprocity, temporal variability, and spatial decorrelation properties of wireless channels. In the realm of SemCom, the inherent complexity and randomness of the channel environment enable both communicating parties to capture highly correlated channel characteristics through a shared channel. These channel characteristics, which fluctuate with time and space, introduce a level of randomness and enhance the eavesdropping resistance of the physical layer keys.

\textbf{Static environment applications:} Aldaghri et al. \cite{aldaghri2020physical} introduce a low-complexity method called induced randomness for high-rate secret key generation in wireless communication systems, even in static environments. This method leverages the uniqueness of wireless channel coefficients and locally generated randomness by the communicating parties, Alice and Bob. The work considers two scenarios: direct communication between Alice and Bob, and communication through an untrusted relay. Post-processing is done to generate highly-correlated samples that are converted into bits, and disparities are mitigated. Hashing is used to compensate for information leakage to an eavesdropper and to ensure consistency of the generated key bit sequences.

\textbf{Frequency-division duplexing system applications:} Zhang et al. \cite{zhang2021deep} develop a PLKG scheme for frequency-division duplexing systems in IoT. The scheme uses DL to establish a feature mapping function between different frequency bands, enabling two users to generate highly similar channel features. The proposed key generation neural network demonstrates excellent performance in terms of randomness, key generation ratio, and key error rate, and is suitable for resource-constrained IoT devices in frequency-division duplexing systems.

\textbf{IRS-assisted PLKG:} Jin et al. \cite{jin2022ris} propose a IRS-assisted PLKG scheme to significantly enhance key generation performance. By optimizing reflecting coefficients and formulating a power minimization problem, the scheme achieves a higher key generation rate and reduced transmit power compared to existing relay-assisted schemes. Gao et al. \cite{gao2024physical} presents an overview of IRS-aided PLKG in static indoor environments, highlighting its channel model, hardware architectures, potential application scenarios, and design challenges. Experimental results demonstrate that the key generation rate is significantly enhanced by using IRS in a static indoor environment. Zhao et al. \cite{zhao2022semkey} develop the SemKey to significantly improve the secret key generation rate for SemCom by exploring the underlying randomness of the system and utilizing IRS. This can potentially pave the way for PLS, addressing the vulnerability of these systems to various attacks due to the openness of wireless channels and the fragility of neural models.

\subsubsection{Physical-Layer Authentication}
PLA utilizes the unique propagation characteristics of the wireless environment or the inherent hardware attributes of access devices to realize the verification of the transmitter's identity \cite{meng2024survey}. In comparison to upper-layer authentication methods, PLA offers notable advantages, including reduced complexity, accelerated authentication processes, minimized latency, and increased resistance to forgery. This approach significantly enhances the reliability of information transmission and introduces innovative perspectives for achieving secure SemCom.

\begin{figure*}[t]
\centering
\vspace{-10mm}
\includegraphics[width=\linewidth]{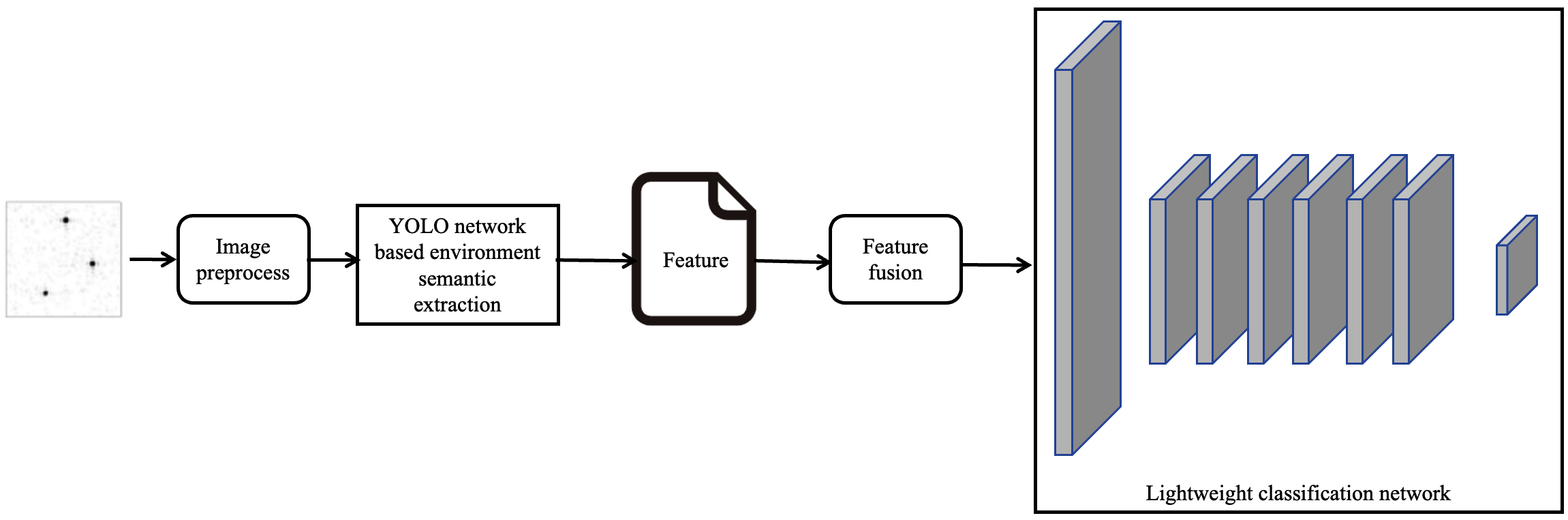}
\caption{Environment semantics enabled PLA \cite{gao2023esanet}, where angle-delay features are extracted for higher interpretability and efficient neural network training. Using the trained EsaNet, physical layer spoofing attacks can be promptly detected in time-varying wireless environments. You Only Look Once (YOLO), an advanced single-stage object detection network, is also employed to swiftly capture angle-delay features from received signals, minimizing data processing overhead and authentication latency. Subsequently, a lightweight neural network is designed to classify without a detection threshold, making it suitable for unknown wireless environments.}
\label{fig494}
\end{figure*}

\textbf{Environment semantics-enabled PLA:} As illustrated in Fig. \ref{fig494}, Gao et al. \cite{gao2023esanet} propose an environment semantics enabled PLA network, EsaNet, for 6G endogenous security. The network extracts a frequency independent wireless channel fingerprint from CSI in a massive MIMO system using environment semantics knowledge. The received signal is transformed into an RGB image and processed by a YOLO network to quickly capture the frequency independent wireless channel fingerprint. A lightweight classification network is then used to distinguish legitimate from illegitimate users. Experimental results demonstrate that EsaNet can effectively detect physical layer spoofing attacks and is robust in time-varying wireless environments.

\textbf{Generative AI-enabled PLA:} Meng et al. \cite{meng2022physical} present a PLA scheme named hierarchical variational autoencoder (HVAE) for IIoT, which achieves high authentication performance without requiring attackers' prior channel information. The HVAE consists of an AE module for CIR characteristics extraction and a VAE module for improving representation ability and outputting authentication results. A new objective function considering both single-peak and double-peak Gaussian distributions is constructed in the VAE module. Simulations under static and mobile IIoT scenarios demonstrate the superiority of the proposed HVAE over other PLA schemes, even with limited training data.

\textbf{Multiuser PLA:} Meng et al. \cite{meng2023multiobservation} present a multiuser authentication architecture named multiobservation-multichannel-attribute-based authentication for IIoT that enhances wireless security by considering both multi-receiver observations and multiple channel attributes. The proposed architecture provides additional spatial recognition characteristics for multi-users and two gradient boosting optimization-based schemes are proposed to better fit the channel features of multi-observations. Meng et al. \cite{meng2023multi} further propose a multiattacker detection architecture using multidimensional fingerprints for more robust device identification. Four clustering-based PLA schemes that do not require training fingerprint sets are introduced. To enhance precision, graph learning-based PLA approaches with minimal labeled fingerprints are also proposed.

\textbf{Watermark-based active PLA:} Ma et al. \cite{ma2024physical} propose a pseudo-random watermark hopping-based PLA scheme to enhance both security and communication performance. By generating a pseudo-random sequence and designing a watermark hopping mechanism, the scheme increases randomness and security while also reducing authentication latency and improving communication performance. The scheme superimposes the tag on pilot signals, avoiding the need for message recovery before authentication. Theoretical and experimental results show that the proposed scheme decreases the bit error rate, outage probability, and increases the achievable rate of the system while significantly improving security performance.

\begin{les}
PLS leverages the randomness of wireless channels to ensure secure transmission. Shannon's concept of ``perfect secret" extends to ``perfect semantic secret" in SemCom, ensuring eavesdroppers can only recover original information through random guessing. Various PLS techniques, including beamforming, artificial noise, relay cooperation, IRS, PLKG, and PLA, can be harnessed to establish secure semantic transmission. Beamforming can be leveraged in semantic-bit coexisting environments \cite{zhang2024beamforming}, massive MIMO applications \cite{wu2024deep2}, varying environmental conditions \cite{raha2024towards,raha2024advancing}, and integrated sensing and SemCom systems \cite{yang2024secure2,dai2024secure,dai2025joint}. Artificial noises can be used to optimal beamforming \cite{zhang2018artificial}, and are extended to cognitive radio network \cite{jiang2018secrecy}, cellular vehicle-to-everything network \cite{wang2020physical}, and visible light communication applications \cite{pham2024design}. Relay cooperation is effective for semantic text transmission \cite{luo2022autoencoder}, Cooperative SemCom \cite{tang2023cooperative,lin2024semantic}, and Multimodal SemCom \cite{guo2024distributed}. IRS can be combined with other PLS techniques, such as artificial noise \cite{arzykulov2023artificial}, PLKG \cite{jin2022ris,gao2024physical,zhao2022semkey}, and PLA \cite{meng2023efficient}.
\end{les}

\section{Future Research Directions}
\label{section 5}

\subsection{Dynamic and Intelligent Data Cleaning}
The future of data cleaning technique will evolve from static preprocessing towards dynamic and intelligent methods. The primary objective is to ensure semantic consistency and contextual adaptability of data within dynamic, multimodal, and semantic-driven environments. Below are the pivotal challenges and their potential solutions.

\textit{Research Challenges:} (1) Complexity of semantic noise: In the realm of SemCom, noise is not limited to numerical errors caused by sensor inaccuracies, but also encompasses semantic ambiguities, such as vague textual descriptions or erroneous image annotations, and contextual conflicts, like mismatched video actions and audio descriptions. For instance, in autonomous driving, a camera may incorrectly identify a ``stop sign" as a ``speed limit sign," resulting in the SemCom system issuing erroneous instructions. (2) Real-time adaptation to dynamic semantics: Semantic environments evolve over time, such as real-time updates of a patient's condition in medical diagnosis, rendering traditional static cleaning rules inadequate for adapting to dynamic semantic relationships. For example, in remote surgery, sensor data cleaning must identify and correct physiological signal offsets caused by patient position changes in real-time. (3) Heterogeneous fusion of multimodal data: Multimodal data sources, including texts, images, and sensors, require uniform alignment at the semantic level. However, noise characteristics vary significantly across different modalities. Image noise is pixel-based, while text noise is grammatical. A relevant example is in smart cities, where the semantic correlation between traffic camera data (images) and social media texts (such as ``congestion on a certain road segment") necessitates dynamic verification.

\textit{Potential Solutions:} (1) Multimodal semantic alignment and cleaning technology: Develop cross-modal attention mechanisms to facilitate joint learning of semantic consistency across multimodal data. For instance, in smart healthcare, aligning pathological images with diagnostic reports can help identify and rectify discrepancies in descriptions, such as unreported tumor areas. (2) Lightweight real-time cleaning framework technology: Implement a hierarchical cleaning architecture leveraging edge computing. Lightweight rule filtering, such as outlier detection is executed on terminal devices, while complex semantic analysis is performed in the cloud. For example, in IIoT, sensor endpoints filter noise data in real-time (like sudden temperature spikes), while the cloud analyzes equipment failure patterns using knowledge graphs.

\subsection{Explainable Robust Learning}

Explainable robust learning for SemCom emerges as a pivotal research direction for the future. The endeavor to collaboratively optimize robustness and explainability in deep JSCC encounters numerous hurdles, yet promising avenues exist to overcome them.

\textit{Research Challenges:} (1) Insufficient model robustness: Traditional JSCC models demonstrate robust resilience against physical channel noise, such as additive white Gaussian noise. However, they lack defensive mechanisms against semantic-level adversarial samples, like semantic perturbations or malicious interference. In real-world communication scenarios, the generalization capability of low-frequency semantic samples is limited, causing model performance to plummet when confronted with outliers, such as rare semantic expressions. (2) Trade-off between explainability and performance: The intricate structure of DNNs obscures the decision-making logic, making it challenging to identify specific issues in semantic understanding, such as encoder bottlenecks or decoder preferences. Furthermore, existing explainability tools, including attention visualization and feature attribution, struggle to quantify the dependency of explanation results on channel conditions in communication contexts.

\textit{Potential Solutions:} (1) Explainability-driven model optimization: Utilize a phased explainable architecture, such as encoding-transmission-decoding decoupling, combined with attention visualization at the semantic encoding layer, to pinpoint bottlenecks in semantic understanding. Incorporate causal graph models or counterfactual analysis to uncover spurious correlations in semantic decisions, such as pseudo-associations between channel noise and semantic labels. Additionally, construct a semantic verification module based on knowledge graphs to validate the rationality of the model's decision-making logic through external knowledge bases. (2) Joint optimization framework: Develop a multi-objective optimization framework that dynamically adjusts robust training strategies based on explainability analysis results, for instance, by enhancing the protection of attention regions. During deployment, employ an online learning mechanism coupled with lightweight fine-tuning techniques, to adapt in real-time to variations in channel conditions and shifts in semantic distributions.

\subsection{Multi-Strategy Combined Backdoor Defense}

With the progression of technology, the methodologies and techniques employed in backdoor attacks are continually undergoing evolution. Attackers can utilize diverse trigger mechanisms, injection methods, and attack objectives to carry out these attacks, rendering them exceptionally varied and intricate. Consequently, a solitary defense strategy frequently falls short in addressing the multifaceted characteristics of backdoor attacks. A defensive methodology that amalgamates multiple strategies can holistically consider security requirements from various viewpoints and defend against backdoor attacks from diverse angles.

\textit{Research Challenges:} Attackers have the capability to launch backdoor attacks through a multitude of avenues, including semantic deception and malicious code injection. These diversified attack vectors render it challenging for a single defense strategy to fully counteract. Moreover, despite the exceptional performance of deep JSCC models in semantic comprehension and generation, their intricacy and dependencies introduce security vulnerabilities. Attackers might exploit the model's internal structure or training data for backdoor attacks, or disrupt its normal functionality by tampering with model parameters.

\textit{Potential Solutions:} Training the model with a diverse array of datasets can bolster its resilience to data, thereby mitigating the impact of tampered or injected malicious data on model performance. Concurrently, data cleaning techniques can proficiently eliminate or flag malicious data, ensuring the authenticity and integrity of the training data. Additionally, pruning and compressing the deep JSCC model can decrease its complexity and redundancy, thereby fortifying its resistance to malicious attacks. Pruning techniques can eliminate parameters that contribute minimally to the training data, while compression techniques can reduce the model's size and diminish the risk of it being compromised.

\subsection{Differential Privacy-based Deep JSCC}
While differential privacy offers robust mathematical foundations and the capability to quantify and analyze privacy leakage risks, ensuring the accuracy and efficiency of SemCom while safeguarding privacy presents a formidable challenge. The detailed discussions are as follows.

\textit{Research Challenges:}(1) Balancing privacy and accuracy: Differential privacy effectively safeguards user privacy, yet the accompanying data noise may undermine the accuracy and efficiency of SemCom. SemCom systems hinge on precise data interpretation and processing. Compromised data quality can lead to erroneous extraction of semantic information, thereby impacting the overall system performance.
(2) Elevated computational complexity: As datasets grow larger and queries become more intricate, the computational load on SemCom systems intensifies. In this context, implementing differential privacy can significantly escalate computational resource consumption, potentially slowing the system's response and detracting from user experience.
(3) Necessity for privacy protection mechanism optimization: Differential privacy technology in SemCom is still nascent. Optimizing these mechanisms to mitigate their impact on system performance is a crucial area for future exploration. Examples include refining adaptive and local differential privacy technologies to better align with diverse data attributes and system demands.

\textit{Potential Solutions:}
(1) Optimizing differential privacy algorithms: To minimize data noise and ensure semantic information accuracy, differential privacy algorithms must be refined. Techniques like Gaussian Differential Privacy offer robust privacy protection without compromising accuracy. Furthermore, research should focus on enhancing the parallelization and distributed processing capabilities of these algorithms to handle large-scale data more efficiently.
(2) Adaptive differential privacy: Developing adaptive differential privacy algorithms that dynamically adjust protection levels based on data characteristics and query requirements is essential. In high-risk situations, protection should be heightened; in low-risk contexts, it can be appropriately lowered to balance privacy and system performance. Real-time monitoring of data sensitivity and usage scenarios allows for dynamic adjustment of privacy protection parameters, informed by users' historical behaviors and current requests.
(3) Local differential privacy: Embracing local differential privacy technology distributes privacy protection across various nodes or devices. This reduces reliance on central servers and enhances system fault tolerance and scalability. In SemCom systems, nodes collaborate, mutually monitor, and protect privacy, fostering a multi-layered privacy protection mechanism.

\subsection{Efficient Homomorphic Encrypted SemCom}

Homomorphic encryption offers privacy protection capabilities for SemCom, facilitating direct computation on encrypted data to safeguard against sensitive information leakage. Nevertheless, its practical deployment is hindered by high computational complexity and substantial resource demands.

\textit{Research Challenges:} The complexity of ciphertext operations in homomorphic encryption, including addition and multiplication, surpasses that of plaintext operations by a wide margin, leading to considerable delays in semantic encoding and decoding, and thus posing a challenge to meeting real-time communication requirements. For example, a single multiplication operation in FHE can take several seconds. Mobile terminals and IoT devices, constrained by limited computing power and memory, struggle to handle the intensive computational tasks associated with homomorphic encryption, thereby impeding the universal applicability of SemCom.

\textit{Potential Solutions:} (1) Algorithm optimization and lightweight design: Dynamically adjust the encryption level according to the complexity of semantic tasks. For instance, employ FHE solely for key semantic features while opting for lightweight encryption for other components. Leverage semantic sparsity to devise sparse ciphertext computation protocols, thereby omitting encryption operations in non-critical areas. (2) Hardware-algorithm co-acceleration: Leverage FPGA/ASIC acceleration and design hardware that incorporates instruction sets tailored for homomorphic encryption, thereby hardware-accelerating core operations. Furthermore, harness the parallel computing prowess of GPUs to expedite large-scale ciphertext matrix operations.

\subsection{Smart Contract-Enabled SemCom}

Blockchain smart contracts, by leveraging their decentralized nature, transparent execution, and tamper-proof characteristics, offer data credibility assurance, automated protocol management, and robust anomaly response capabilities for SemCom. Nevertheless, their seamless integration with SemCom is hindered by various challenges as follows.

\textit{Research Challenges:} (1) Execution efficiency and real-time constraints of smart contracts: The blockchain consensus mechanism and contract execution necessitate multi-node verification, which increases end-to-end latency in SemCom and poses difficulties in meeting real-time interaction demands, such as those in autonomous driving or industrial control. (2) Accuracy and efficiency challenges in anomaly detection: Complex anomaly detection algorithms cannot be directly implemented within smart contracts, and relying on off-chain computing might introduce trust risks. Furthermore, rule-based simple anomaly detection methods (like threshold determination) lack the capability to adequately identify attacks.

\textit{Potential Solutions:} (1) Hierarchical architecture and on-chain/off-chain collaboration: Deploy core logic on-chain while offloading complex semantic processing tasks to a trusted off-chain execution environment. For instance, in IoV, real-time semantic interactions between vehicles are facilitated through state channels, with hash values of collision warning results stored on-chain for evidentiary purposes. (2) Efficient anomaly detection and response framework: Develop a lightweight rule base utilizing a finite state machine to identify common semantic attack patterns. AI anomaly detection models should be run on edge nodes, with only summaries of suspicious events submitted to on-chain contracts for verification.

\subsection{Semantic Channel Fingerprint Database-Enabled PLS}

The semantic channel fingerprints in the semantic channel fingerprint database can serve as a basis for authentication. When a device attempts to access the network, it only needs to compare its semantic channel characteristics with the information in the database to quickly determine its legitimacy. This significantly improves the efficiency and speed of identity authentication.

\textit{Research Challenges:} (1) Environmental dynamic interference affecting channel fingerprints: The wireless channel fingerprints undergo substantial alterations due to multipath effects and moving obstructions. In IIoT, the movement of robotic arms leads to significant fluctuations in CSI, posing a challenge for traditional static channel fingerprint databases to manage effectively.
(2) Risks of tampering in semantic channel fingerprint repositories: Attackers have the potential to inject forged fingerprints by generating CSI perturbations using GANs, which can impair the semantic extraction capability of channel fingerprints and mislead the authentication results.

\textit{Potential Solutions:} (1) Implement spatial-temporal joint modeling: Achieve intelligent integration of dynamic CSI matrices and semantic features.
(2) Enhance PLS with cryptographic techniques: Leverage blockchain-based distributed fingerprint evidence storage, utilizing an advanced Practical Byzantine Fault Tolerance (PBFT) consensus mechanism combined with Pedersen commitments to conceal the original features.

\section{Conclusion}
\label{section 6}
When SemCom converges with 6G, it not only enhances communication efficiency but also introduces opportunities alongside challenges related to security and privacy issues. This article initially introduces the concept, architecture, and lifecycle of SemComs. Subsequently, it delves into the security and privacy concerns that arise during each phase of SemCom systems, encompassing knowledge base poisoning attacks, gradient leakage, server vulnerabilities, attacks targeting communication bottlenecks, model slice attacks, semantic adversarial attacks, semantic eavesdropping, semantic inference attacks, and semantic jamming attacks. Furthermore, we propose diverse techniques to counteract these security and privacy threats, such as data cleaning, robust learning, defensive strategies against backdoor attacks, adversarial training, differential privacy, cryptography, blockchain technology, model compression, and physical-layer security. Ultimately, we conclude the paper with recommendations and future research directions for these defense methods.

\section{Acknowledgements}
The work presented in this paper was supported by the National Key R\&D Program of China No. 2020YFB1806905, the National Natural Science Foundation of China No. 61871045, No. 62401074, and No. 61932005, Beijing Natural Science Foundation No. L242012 and No. L232051, the research foundation of Ministry of Education-China Mobile under Grant MCM20180101, and the Joint Research Fund for Beijing Natural Science Foundation and Haidian Original Innovation under Grant L232001.

\section{CRediT authorship contribution statement}
\textbf{Rui Meng:} investigation, methodology, writing, and supervision. 
\textbf{Song Gao:} investigation and writing. 
\textbf{Dayu Fan:} investigation and writing.
\textbf{Haixiao Gao:} investigation and writing. 
\textbf{Yining Wang:} investigation and writing. 
\textbf{Xiaodong Xu:} writing (review and editing), funding acquisition, and supervision. 
\textbf{Bizhu Wang:} writing (review and editing) and funding acquisition.
\textbf{Suyu Lv:} investigation and writing. 
\textbf{Zhidi Zhang:} investigation and writing. 
\textbf{Mengying Sun:} writing (review and editing) and funding acquisition. 
\textbf{Shujun Han:} writing (review and editing) and funding acquisition. 
\textbf{Chen Dong:} writing (review and editing) and funding acquisition. 
\textbf{Xiaofeng Tao:} writing (review and editing) and funding acquisition.
\textbf{Ping Zhang:} funding acquisition and supervision.

\section{Data availability}
No data was used for the research described in the article.

\bibliographystyle{elsarticle-num}
\bibliography{ref.bib}



\end{document}